\documentstyle[epsfig,aps,pre,twocolumn,floats,eqsecnum]{revtex}

\begin{document}

\newcommand{\gsim}{
\,\raisebox{0.35ex}{$>$}
\hspace{-1.7ex}\raisebox{-0.65ex}{$\sim$}\,
}

\newcommand{\lsim}{
\,\raisebox{0.35ex}{$<$}
\hspace{-1.7ex}\raisebox{-0.65ex}{$\sim$}\,
}

\newcommand{\const}{
{\rm const}
}

\newcommand{\onehalf}{\mbox{\scriptsize 
\raisebox{1.5mm}{1}\hspace{-2.7mm}
\raisebox{0.5mm}{$-$}\hspace{-2.8mm}
\raisebox{-0.9mm}{2}\hspace{-0.7mm}
\normalsize }}

\newcommand{\fivehalf}{\mbox{\scriptsize 
\raisebox{1.5mm}{5}\hspace{-2.7mm}
\raisebox{0.5mm}{$-$}\hspace{-2.8mm}
\raisebox{-0.9mm}{2}\hspace{-0.7mm}
\normalsize }}

\newcommand{\onethird}{\mbox{\scriptsize 
\raisebox{1.5mm}{1}\hspace{-2.7mm}
\raisebox{0.5mm}{$-$}\hspace{-2.8mm}
\raisebox{-0.9mm}{3}\hspace{-0.7mm}
\normalsize }}

\newcommand{\twothird}{\mbox{\scriptsize 
\raisebox{1.5mm}{2}\hspace{-2.7mm}
\raisebox{0.5mm}{$-$}\hspace{-2.8mm}
\raisebox{-0.9mm}{3}\hspace{-0.7mm}
\normalsize }}

\newcommand{\fourthird}{\mbox{\scriptsize 
\raisebox{1.5mm}{4}\hspace{-2.7mm}
\raisebox{0.5mm}{$-$}\hspace{-2.8mm}
\raisebox{-0.9mm}{3}\hspace{-0.7mm}
\normalsize }}

\newcommand{\onefourth}{\mbox{\scriptsize 
\raisebox{1.5mm}{1}\hspace{-2.7mm}
\raisebox{0.5mm}{$-$}\hspace{-2.8mm}
\raisebox{-0.9mm}{4}\hspace{-0.7mm}
\normalsize }}

\newcommand{\oneeights}{\mbox{\scriptsize 
\raisebox{1.5mm}{1}\hspace{-2.7mm}
\raisebox{0.5mm}{$-$}\hspace{-2.8mm}
\raisebox{-0.9mm}{8}\hspace{-0.7mm}
\normalsize }}

\bibliographystyle{prsty}

\title{ \begin{flushleft}
%{\small \em submitted to}\\
{\small 
PHYSICAL REVIEW E 
\hfill
VOLUME  
{\normalsize 58}, 
NUMBER 
{\normalsize 1}
\hfill 
%MONTH XXX
$\qquad\qquad\qquad\quad$
JULY {\normalsize 1998}
%, 3250-3256}
}\\
\end{flushleft}  
Semi-infinite anisotropic spherical model: Correlations at $T\geq T_c$
}      

\author{
D.~A. Garanin$^*$
%\renewcommand{\thefootnote}{\fnsymbol{footnote}}
%\footnotemark[1]
}

\address{
Max-Planck-Institut f\"ur Physik komplexer Systeme, N\"othnitzer Strasse 38,
D-01187 Dresden, Germany\\
%}
%\date{\today}
%\maketitle
%\abstract{
\smallskip
{\rm (Received 12 February 1998)}
\bigskip\\
\parbox{14.2cm}
{\rm
The ordinary surface magnetic phase transition is studied for the exactly
solvable anisotropic spherical model (ASM), which is the limit $D\to\infty$ 
of the $D$-component uniaxially anisotropic classical vector model.
The bulk limit of the ASM is similar to that of the spherical model, apart
from the role of the anisotropy stabilizing ordering for low lattice
dimensionalities, $d\leq 2$,
at finite temperatures.
The correlation functions and the energy density profile in the semi-infinite
ASM are calculated analytically and numerically for $T\geq T_c$  and 
$ 1\leq d\leq \infty$.     
Since the lattice dimensionalities $d=1,\,2,\,3$, and 4 are special, a continuous
spatial dimensionality $d'=d-1$ has been introduced for dimensions parallel to
the surface.
However, preserving a discrete layer structure perpendicular to the surface
avoids unphysical surface singularities and allows numerical solitions that
reveal significant short-range features near the surface. 
The results obtained generalize the isotropic-criticality results for $2<d<4$ 
of Bray and Moore 
[Phys. Rev. Lett. {\bf 38},  735  (1977);
J. Phys. A {\bf 10},  1927  (1977)].
\smallskip
\begin{flushleft}
PACS numbers: 64.60.Cn, 75.10.Hk, 75.30.Pd
\end{flushleft}
} 
} 
\maketitle

\renewcommand{\thefootnote}{\fnsymbol{footnote}}
\footnotetext{ 
$^*$Permanent address: I. Institut f\"ur Theoretische Physik, Universit\"at
Hamburg,
Jungius Strasse 9, D-20355 Hamburg, Germany. 

Electronic addresses: garanin@mpipks-dresden.mpg.de\\
garanin@physnet.uni-hamburg.de \\
http://www.mpipks-dresden.mpg.de/$\sim$garanin/
 }

\section{Introduction}
\label{introduction}

Magnetic ordering in semi-infinite and film geometries is an old problem currently 
receiving increasing attention because of enormous progress in fabrication of magnetic  
structures on the atomic scale.
Theoretical methods using the mean field approximation (MFA) or the phenomenological 
Landau theory \cite{mil71,woldewhalpal71,kagome71,binhoh7274,lubrub75mfa}
provided classification and description of the
qualitative features of different types of surface phase transitions. 
High-temperature series expansions \cite{binhoh7274}
and Monte Carlo simulations \cite{bin72}, as well as the scaling analysis 
\cite{binhoh7274,fisbar72} and the $\epsilon$ expansion 
\cite{lubrub7375eps,diedie81,diedie81energy}, 
shed light on the details of the surface critical behavior.
 A general review of these approaches can be found in 
Refs. \cite{bin83ptcp,bar83ptcp}.
Examples of recent work in film geometry based on $\epsilon$ expansion
are Refs. \cite{kreeisdie95,eiskredie96}.
A special case is the confined two-dimensional Ising ($S=\onehalf$) model, for which exact
solutions have been found \cite{mccoywu67,fisfer67,auyfis80,mikfis9394}. 
 
The ordinary surface phase transition of the semi-infinite ferromagnet occurs at the bulk
critical temperature $T_c$.
It is characterized by a number of surface critical exponents the definition of which can be
found in Ref. \cite{bin83ptcp}.
In particular, the susceptibilities at the surface with respect to the fields applied either in the bulk
or at the surface are described by the exponents $\gamma_1$ and $\gamma_{11}$,
respectively, which in the MFA are given by $\gamma_1=\onehalf$ and
$\gamma_{11}=-\onehalf$ (no divergence), 
in contrast to the bulk exponent $\gamma=1$.
In thin films, which are more important for applications and more interesting for the
experiment, there are additional effects, such as the lowering of $T_c$ in comparison to its bulk 
value and the crossover between the three- and two-dimensional behavior as a function of
the film thickness $L$ \cite{libab92,schsierie96,bab97}.

The main body of the theoretical work on surface phase transitions is being done, with an 
exclusion of the 2$d$ Ising model, starting from the field-theoretical continuous Hamiltonians 
or free energies.
Such an approach has proven to be very useful for establishing the universality classes
and critical laws but, on the other hand, the (nonuniversal) absolute values of  observables, such
as critical amplitudes, remain undetermined.
In addition to the restrictions of the field-theoretical 
methods that are well known from the bulk physics, there are more specific problems related to the role of the lattice discreteness in
confined geometries.
One can question how the continuous approach can be
applied to thin films consisting of a {\em mesoscopic} number of layers.
A similar question can be addressed to the semi-infinite ferromagnets as well --- does the 
continuous approximation apply in the region near the boundary, $n\sim 1$, where 
$n=1,\,2,\ldots$ is the layer number? 
The MFA, or the Landau theory, gives a positive answer to this question near criticality, where
the correlation length $\xi_c$ is much larger than the lattice spacing $a_0$ and the order 
parameter --- the magnetization --- cannot change at distances smaller than $\xi_c$.
This is, however, not the case if one goes beyond the MFA and considers 
spin-spin correlation functions.
If the temperature is high enough, or the system is classical, spin waves with the
wave vectors up to the edge of the Brillouin zone $k\sim k_{\rm max} =\pi/a_0$ are excited.
This means that correlation functions comprise $a_0$ as the length parameter,
additionally to $\xi_c$, and thus there can be inhomogeneities near the
surface of a ferromagnet on the scale of several atomic layers, even near criticality.
This boundary region, $n\sim 1$, is that which can be locally probed in experiments, and here 
the continuous approximation may become, at least quantitatively, wrong.

Spin-wave effects in weakly anisotropic systems drastically change their behavior in 
comparison to the MFA predictions.
Magnetic models with continuous symmetry in low dimensions, $d\leq 2$, cannot order at
finite temperatures, and for models with $d>2$ the correlation length is infinite in the
whole region below $T_c$. 
These effects are not less important than the critical coupling of fluctuations giving rise to 
nonclassical critical indices.
In  Heisenberg systems the linear spin-wave theory satisfactorily describes the above
mentioned effects well below $T_c$ but breaks down at elevated temperatures.
There is, however, a model where a kind of spin-wave description is valid in the whole
temperature range, whereas the critical fluctuation coupling vanishes.
This is the $D$-component classical spin vector model proposed by Stanley 
\cite{sta68prl,sta74} in the limit $D\to\infty$.
Stanley has shown \cite{sta68pr} that in this limit the partition function of a homogeneous
ferromagnet coincides with that of the spherical model (SM).
The latter was advanced by Berlin and Kac \cite{berkac52} as an exactly solvable substitute 
for the Ising model.
The formalism contains, however, the spin-wave integral over the Brillouin 
zone and describes rather the properties of the isotropic Heisenberg model.
The spherical model in its traditional form was extended to inhomogeneous systems by 
Barber and Fisher \cite{barfis73}, who found a nonmonotonic dependence 
$T_c(L)$ for thin films.
This unexpected feature was attributed to the failure of the global spin constraint in the 
inhomogeneous case.
Later, an improved version of the SM was proposed, which used spin constraints
on each lattice site \cite{kno73}, and it was shown that this version is equivalent to the
$D=\infty$ model in the inhomogeneous case.
Application of the SM with constraints in each layer \cite{cosmazmih76} yielded a 
reasonable monotonic dependence of the numerically calculated 
$T_c(L)$ for the films in four dimensions.
Because of the complexity of models of this type, most researchers still use  the 
more convenient global-constraint SM in confined geometries
 (see, e.g., Refs. \cite{allpat93,pat94,all95}).
Very recently, a compromise model was proposed \cite{danbraami97b}, which uses
a constraint for the spins at the surface in addition to that for the bulk ones.
The properties of such a model are closer, in a sense,  to those of the SM with the local spin 
constraint and, hence, to those of the $D=\infty$, or $O(\infty)$, model.

A remarkable property of the Stanley model is that it can be easily extended to the
anisotropic case, whereas the traditional SM cannot.
This is very important because the anisotropy breaking a continuous symmetry plays a
crucial role in low-dimensional systems, where it stabilizes  ordering at finite
temperatures.
The limit $D\to\infty$ of the uniaxial $D$-component vector model determines the so-called
anisotropic spherical model (ASM), which is described in the inhomogeneous case by a 
closed set of equations for the variables on the lattice sites obtained in Ref. \cite{gar96jpa}.
The ASM was applied in Ref. \cite{gar96jpal} to study the dependence $T_c(L)$ for 
ferromagnetic films in three dimensions.
Here, for all fixed values of the film thickness $L$ one has $T_c\to 0$ in the isotropic limit
because of the two-dimensionality of the film, and such a behavior is
pertinent for any Heisenberg film with finite $D$ as well.
Thus, allowing for the anisotropy is relevant here, and it cannot be done within the traditional SM.
An interesting feature of the solution obtained in Ref. \cite{gar96jpal} is the role of the
correlation length of the transverse spin components, $\xi_{c\alpha}$, in the crossover from
the three- to two-dimensional behavior of the film, which takes place for 
$L \lsim \xi_{c\alpha}$.
Note that in the finite-size scaling analysis (see, e.g., Ref. \cite{bin83ptcp}) only the diverging
longitudinal correlation length $\xi_{cz}$ is used, whereas the noncritical 
transverse correlation length is disregarded as an irrelevant variable.

Another application of the ASM is to the temperature-driven phase transition between 
the Bloch and linear (Ising-like) domain walls in uniaxial ferromagnets at
some $T_B<T_c$ \cite{gar96jpa}.
This phase transition was studied within the framework of the MFA and the Landau theory in 
Refs. \cite{bulgin63,brozij71,sartrubis76,nie76,lajnie79} and with a field-theoretical method
in Ref. \cite{lawlow81}.
A low minimum of the domain-wall mobility at $T_B$ predicted in 
Refs. \cite{gar91llb,gar91edw} was used to identify the domain-wall phase transition in the 
dynamic susceptibility experiments on Ba and Sr hexaferrites 
\cite{koegarharjah93,harkoegar95}.
It is clear that the anisotropy is an important characteristic of the model, giving rise to the 
very existence of domain walls of {\em finite width} separating the ``up'' and ``down'' domains.
For this reason, the attempts to describe domain walls with the traditional SM
in both versions with
 \cite{abrrob80} and without \cite{angbuncos81} the global spin constraint could
not yield relevant results.  

In the recent work of Ref. \cite{gar97zpb} it was shown that the ASM and SM are 
{\em not} equivalent, even in the isotropic homogeneous case, if the longitudinal correlation
function (CF) $S_{zz}({\bf k})$ below $T_c$ is involved.
Whereas in the traditional SM the CF is proportional to $1/k^2$ at small wave vectors, 
the behavior of  $S_{zz}({\bf k})$ in the ASM shows a more complicated behavior which is
governed by the spin-wave effects and is sensitive to the dimensionality.
In three dimensions $S_{zz}({\bf k})\propto 1/k$, which is familiar from the linear spin-wave
theory.
The above law holds for $k \lsim \kappa_m\propto T_c-T$, i.e.,
 there is a critical length scale $\xi_m = 1/\kappa_m$ in the theory.
The length $\xi_m$ is analogous to the ``bare'', or the mean-field,
correlation length below $T_c$, whereas the true longitudinal correlation
length $\xi_{cz}$ remains infinite in the isotropic model below $T_c$.
The former is responsible for the crossover of the real-space CF from 
$S_{zz}({\bf r})\propto 1/r$ for $r\lsim\xi_m$ to $S_{zz}({\bf r})\propto 1/r^2$ for $r\gsim\xi_m$.

The ASM equations of Ref. \cite{gar96jpa}, as well as those for the SM without the global 
spin constraint \cite{cosmazmih76}, are rather complicated, strongly nonlinear equations for 
the variables on a lattice.
In the latter case some researchers termed them analytically intractable. 
Nevertheless, for the weakly anisotropic ASM in the domain-wall geometry it was possible to
guess the solution \cite{gar96jpa}, which yielded an example of the phase
transition of an interface that is
analytically tractable beyond the MFA .
Surfaces with free boundary conditions make the problem much more complicated.
In Ref. \cite{gar96jpal} only the most important and partially rough asymptotes for the
$T_c$ shift in films could be obtained, and numerical calculations have not been performed.
The aim of this work is to investigate the influence of surfaces on magnetic
ordering in more detail for the ordinary phase transition in the semi-infinite
ASM in the temperature range $T\geq T_c$.
As we shall see, analytical solutions are available in the dimensionality ranges 
$1\leq d \leq 2$ and $d\geq 4$, as well as for $2<d<4$, both at and away from
the isotropic criticality.
In addition, the problem will be solved numerically in all the relevant cases.

The same problem presented here was addressed in the seminal work by Bray and Moore 
\cite{bramoo77prl77jpa}, who considered a field theory with the 
volume and surface Hamiltonian densities of the type $(\nabla\phi)^2 + \tau\phi^2 + u\phi^4$
and $c\phi^2$, respectively, for the $n$-component vector order parameter
$\phi$ in the limit $n\to \infty$. 
Since this model is isotropic, the range $1\leq d \leq 2$ is excluded from the outset due to
the absence of ordering in the bulk.
For $2<d<4$ a very important solution for the correlation function at criticality was obtained,
however, only for the ``magic'' $d$-dependent values $u^*(d)$ of the coupling constant $u$. 
This solution yielded the anomalous dimensions $\eta_\|=d-2$ and $\eta_\perp=(d-2)/2$,
with which all other surface critical indices could be derived from the scaling arguments 
(see, e.g., Refs. \cite{bin83ptcp,bar83ptcp}). 
These critical indices depend on $d$ and differ from the mean-field ones, as well as from
those of the global-constraint semi-infinite SM.
In the solution of Bray and Moore the required coupling constant $u^*(d)$ vanishes for 
$d=4$, i.e., the model simplifies to the exactly solvable and trivial Gaussian model showing
the mean-field behavior.
On one hand, it seems reasonable, because the critical indices indeed simplify to their
mean-field values for $d>4$.
On the other hand, one could desire more detailed information about the critical behavior
of the models with $u\neq u^*$, in particular for $u\neq 0$ and $d>4$.
Unfortunately, no extension of the analytical results of Bray and Moore in
this direction or for the off-criticality case, is possible.
Also a numerical solution  is hampered for this model by insurmountable difficulties.
Apart from the obvious impossibility of handling the inhomogeneities on the scale of the lattice
spacing with a continuous field theory, it turns out that this model cannot be solved
numerically  at all because near the surface, where the boundary condition is set, the continuous
approximation does not apply.
In fact, this is an example of a situation in which using a field-theoretical
approach in statistical mechanics brings only disadvantages.
By contrast, the ASM formulated from the beginning in its true form on a lattice 
leads to the ASM equations which are well defined and suitable for numerical
solution, and also can be considered continuously far from the surface.

The main body of the article is organized as follows.
In Sec. \ref{basic} the system of equations describing the ASM in zero field is
written down.
Its bulk solution, which differs from the well known solution for the spherical
model by the incorporating the uniaxial anisotropy, is studied for all lattice
dimensions.
The quantity playing the central role in the theory, the gap parameter
$G_n$, is related to the reduced energy density $\tilde U_n$.
The continued-fraction formalism, which is mainly needed for the numerical
solution of the ASM system of equations, is described.
Section \ref{basic} concludes with the results for the variation of $G_n$ far above
and far below $T_c$.
In Sec. \ref{lowhigh} the ASM is solved analytically in low ($1\leq d \leq 2$) and high
($4\leq d$) lattice dimensions, starting from the exact solution for the one-dimensional
``toy'' model.
The energy density profiles and spin correlation functions  
are calculated analytically in all possible cases.
In Sec. \ref{middle} the most interesting case $2< d < 4$ is investigated.
Analytical solutions are obtained for the isotropic model at criticality and 
away from the isotropic criticality.
In Sec. \ref{numerical} the semi-infinite ASM is numerically solved in the
whole range of dimensions at  $T\geq T_c$.
The results for the energy density profile and correlation functions are presented.
In Sec. \ref{disc} the main results of the paper are summarized and
compared with the results of other approaches.

\section{Basic relations}
\label{basic}

\subsection{ASM equations}
\label{ASM}

We start from the Hamiltonian of the uniaxially anisotropic classical $D$-vector model,
which, in the absence of the magnetic field,  can be written in the form 
%\marginpar{dham}
%
\begin{equation}\label{dham}
{\cal H} = 
- \frac{1}{2}\sum_{ij}J_{ij}
\left(
m_{zi}m_{zj}  
+ \eta \sum_{\alpha=2}^D m_{\alpha i} m_{\alpha j}
\right) ,
\end{equation}
where ${\bf m}_i$ is the normalized $D$-component vector, 
$|{\bf m}_i|=1$, and $\eta \leq 1$ is the dimensionless anisotropy factor. 
In the mean-field approximation the Curie temperature of 
this model is $T_c^{\rm MFA}=J_0/D$, where $J_0$ is the zero Fourier 
component of the exchange interaction. 
It is convenient to use $T_c^{\rm MFA}$ as the energy scale and introduce
the dimensionless temperature variable $\theta\equiv T/T_c^{\rm MFA}$.
Using the diagram technique for classical spin systems \cite{garlut84d,gar94jsp}, 
 in the limit $D\to\infty$ one arrives at the closed system of equations \cite{gar96jpa} for the average magnetization $m_i\equiv \langle m_{zi}\rangle$ and the correlation function
of transverse ($\alpha\geq 2$) spin components: 
$s_{ij}\equiv D\langle m_{\alpha i}m_{\alpha j}\rangle$.
This system of equations describing the anisotropic spherical model consists of the magnetization equation
%\marginpar{mageq}
%
\begin{equation}\label{mageq}
m_i = G_i \sum_j \lambda_{ij}m_j ,
\end{equation}
the Dyson equation for the transverse CF 
%\marginpar{cf2}
%
\begin{equation}\label{cf2}
s_{ii'} 
= 
\theta G_i \delta_{ii'}
+ 
\eta G_i \sum_j \lambda_{ij}s_{ji'} ,
\end{equation}
and the kinematic relation playing the role of the spin constraint on a lattice site $i$,
%\marginpar{sconstr}
%
\begin{equation}\label{sconstr}
s_{ii} + m_i^2 = 1 .
\end{equation}
Here $\delta_{ij}$ is the site Kronecker symbol, $\lambda_{ij}\equiv J_{ij}/J_0$, and the so-called gap parameter $G_i$ is the one-site cumulant spin average 
\mbox{$D\langle m_{\alpha i}m_{\alpha i}\rangle_{\rm cum}/\theta$} 
renormalized by Gaussian fluctuations. 
We will see below that $G_i$ can be related to the energy density at the site
$i$.

The ASM system of equations is self-consistent.
It is instructive to compare it with the MFA equations that
can be recovered via the following steps.
(i)  Ignore correlations in (\ref{cf2}), which leads to $s_{ii}=\theta G_i$.
(ii) Express $G_i$ through $m_i$ with the help of (\ref{sconstr}) to get  
 $G_i=(1-m_i^2)/\theta$.
(iii) Insert the latter into (\ref{mageq}) to obtain the closed equation for
magnetization.
The form of the latter is simplified with respect to the general-$D$ case
because of the simplification of the Langevin function in the limit $D\to \infty$. 
The resulting $m_i$ is zero above $\theta_c=1$ and nonzero below $\theta_c$.
(iv) With $G_i$ determined, which is simply $G_i=1/\theta$ for $\theta\geq
\theta_c$,  return to (\ref{cf2}) to find the improved correlation function.
It is clear in step (iv) that the MFA is not self consistent, even in the simpler case above $\theta_c$.
This is the reason why the MFA value of $\theta_c$ found from the CF equation (\ref{cf2})
with $G_i=1/\theta$ for the spatially homogeneous isotropic low-dimensional
magnets is nonzero, in contradiction with the result of more rigorous
approaches.
By contrast, the ASM equations are free from such an inconsistency and
they correctly describe the dimensional effects in isotropic and
weakly-anisotropic systems.
It should be noted that in the ``Ising limit'' $\eta=0$ all the steps above
leading to the MFA equations are exact, i.e., the Ising model in the limit
$D\to\infty$ is exactly described by the mean-field approximation.

For the model with the nearest-neighbor interaction $J_{ij}$ on the 
$d$-dimensional hypercubic lattice in the semi-infinite geometry, it is 
convenient to use the Fourier representation in $d'=d-1$ translationally invariant dimensions parallel to the surface and the site representation in the $d$th dimension. 
The Dyson equation (\ref{cf2}) for the Fourier-transformed CF $\sigma_{nn'}({\bf q})$ 
then takes the form of a system of the second-order finite-difference
equations in the set of layers
 $n=1,\,2,...,\infty$,
%\marginpar{cffd}
%
\begin{equation}\label{cffd}
2b_n\sigma_{nn'} - \sigma_{n+1,n'} - \sigma_{n-1,n'}  = (2d\theta/\eta)\delta_{nn'}. 
\end{equation}
$b_n$ here is given by
%\marginpar{bn}
%
\begin{equation}\label{bn}
b_n = 1 + d[(\eta G_n)^{-1}-1] + d'(1-\lambda'_{\bf q}) ,
\end{equation}
where $\lambda'_{\bf q}$ is given by 
%\marginpar{lampr}
%
\begin{equation}\label{lampr}
\lambda'_{\bf q} = \frac{1}{d'}\sum_{i=1}^{d'} \cos(q_i) 
\end{equation}
and the lattice spacing $a_0$ is set to unity.
The magnetization equation (\ref{mageq}) takes the form 
%\marginpar{mageqfd}
%
\begin{equation}\label{mageqfd}
2\bar b_n m_n  - m_{n+1} - m_{n-1} = 0 ,
\end{equation}
with $\bar b_n  = 1 + d[ G_n^{-1}-1]$.
Since the layer with $n=0$ is absent, one can use
%\marginpar{bc}
%
\begin{equation}\label{bc}
\sigma_{0n'} = 0, \qquad m_0 = 0
\end{equation}
as the free boundary conditions to (\ref{cffd}) and (\ref{mageqfd}).
If the interaction in the boundary layer differs from that in the bulk 
(see, e.g., Ref.  \cite{bin83ptcp}) the form of the boundary conditions changes.  
The constraint equations (\ref{sconstr}) can now be written as
%\marginpar{sconstrfd}
%
\begin{equation}\label{sconstrfd}
s_{nn} + m_n^2 = 1, \qquad s_{nn} = \int\!\!\!\frac{d^{d'}{\bf q}}{(2\pi)^{d'}} \sigma_{nn}({\bf q}) . 
\end{equation}
For $n\gg 1$, $q\ll 1$, and $\kappa\ll 1$, where $\kappa$ is the inverse
transverse correlation length defined by (\ref{kapdef}) below, the second-order finite-difference CF equation 
(\ref{cffd}) simplifies to the differential equation for the Green's function 
%\marginpar{cfdiffeq}
%
\begin{equation}\label{cfdiffeq}
\left( \frac{d^2}{dn^2} - \tilde q^2 + 2dG_{1n} \right) \sigma_{nn'} = - 2d\theta\delta(n-n') ,
\end{equation}
where $n$ is considered as a continuous variable, $\tilde q \equiv\sqrt{\kappa^2+q^2}$,
and 
%\marginpar{g1ndef}
%
\begin{equation}\label{g1ndef}
G_{1n} \equiv G_n-G \ll 1
\end{equation}
is the deviation of the gap parameter from its bulk value $G$.
The magnetization equation (\ref{mageqfd}) takes on a similar form with
$\tilde q \Rightarrow 0$ and without the inhomogeneous term.

Before proceeding, let us consider the solution of equations (\ref{cffd}) for the important special
 variation of $G_n$ near the surface, in which only $G_1$ may differ from the bulk value
$G$.
In this case one can solve (\ref{cffd}) directly with the result 
%\marginpar{signnbc}
%
\begin{equation}\label{signnbc}
\sigma_{nn'} = \frac{ d\theta/\eta }{ \sqrt{b^2-1} }
\left[
\alpha^{|n-n'|} - \alpha^{n+n'-2} \frac{ \alpha - 2b_{11} }{ \alpha^{-1} - 2b_{11} } 
\right],
\end{equation}
where $\alpha \equiv b-\sqrt{b^2-1}$, $b$ is the bulk value of $b_n$, and  $b_{11}\equiv b-b_1$.
One of the particular cases of (\ref{signnbc}) is $b_{11}=0$, which corresponds to the MFA or to 
the high-dimensional ($d\geq 4$) lattices for $n\gg 1$ (see Sec. \ref{solutiond4}).
Here (\ref{signnbc}) simplifies to
%\marginpar{signnbchi}
%
\begin{equation}\label{signnbchi}
\sigma_{nn'} = \frac{ d\theta/\eta }{ \sqrt{b^2-1} }
\left[
\alpha^{|n-n'|} - \alpha^{n+n'} 
\right].
\end{equation}
Another particular case is $G_1=[2d/(2d-1)]G$, i.e., $2b_{11}=1$.
As we shall see below, a variation of $G_n$ close to this one is realized for low-dimensional ($d\leq 2$)
lattices in the weakly anisotropic case at low temperatures.
Here (\ref{signnbc}) reduces to
%\marginpar{signnbclo}
%
\begin{equation}\label{signnbclo}
\sigma_{nn'} = \frac{ d\theta/\eta }{ \sqrt{b^2-1} }
\left[
\alpha^{|n-n'|} + \alpha^{n+n'-1} 
\right].
\end{equation}
For $1-\eta \ll 1$ and $\tilde q \ll 1$  one can define $2b_{11}\cong 1 - c$
and simplify  (\ref{signnbc}) to
%\marginpar{signnbc'}
%
\begin{equation}\label{signnbc'}
 \sigma_{nn'} \cong \frac{ d\theta }{ \tilde q } \left[ e^{-\tilde q|n-n'|} + 
f(\tilde q) e^{-\tilde q(n+n'-2)} \right] ,
\end{equation}
with $f(\tilde q) = ( \tilde q - c )/( \tilde q + c)$.
This result could also be obtained solving the differential equation
(\ref{cfdiffeq}) with the boundary condition
%\marginpar{bclo}
%
\begin{equation}\label{bclo}
 \left [ c \sigma_{nn'} - \frac{ d }{ dn }\sigma_{nn'}  \right] _{n=1} = 0
\end{equation}
following from (\ref{bc}).
The remarkable feature of this solution is that it becomes insensitive to the
exact form of the boundary condition if $\tilde q \ll c$ or $\tilde q \gg c$.
In fact, as will be argued below, the limiting forms of $\sigma_{nn'}$ with both signs of the
surface-induced term, (\ref{signnbchi}) and (\ref{signnbclo}), are realized
for more general variations of $G_n$, in which  $G_n$ differs from the bulk
value in some localized region near the surface, $n\lsim n^*$.

Note that the quantity $c$ above is similar to the coefficient in the
surface-energy term that is introduced in the phenomenological field theory
of phase transitions and it defines the extrapolation length
$c=\lambda_e^{-1}$ (see, e.g.,  Ref. \cite{bin83ptcp}).
This term was used, in particular, by Bray and Moore 
\cite{bramoo77prl77jpa}, who have set $c=\infty$ for the ordinary phase transition
to remove the uncertainty.
We shall see, however, that the weakly anisotropic models with $d\le4$ the
 microscopic solution  is characterized by effective values of $c$ of order $\kappa \ll 1$, 
i.e., by large extrapolation lengths.

The equation for the longitudinal correlation function, $\sigma_{nn'}^{zz}$, is not coupled to the ASM system of equations, since fluctuations of the (only one) longitudinal spin component make contributions of order $1/D$ to the physical quantities, which disappear in the spherical limit.
Above $T_c$ in zero field this equation has the form (\ref{cffd}) with $\eta=1$.
The latter amounts to replacing 
$\tilde q \Rightarrow \tilde q_z \equiv \sqrt{\kappa_z^2 + q^2}$ in (\ref{cfdiffeq}),
where $\kappa_z$ is the inverse longitudinal correlation length defined by (\ref{kapzdef})
below.
Thus both $\sigma_{nn'}$ and $\sigma_{nn'}^{zz}$ are given by the same
function of different arguments, and the latter is more convenient, since its
argument spans the wider range, starting from zero at criticality ($\kappa_z=0$).
In this limit, $\tilde q_z =0$, formula (\ref{signnbc'}) for
$\sigma_{nn}^{zz}$ reduces to the expression
%\marginpar{sigzzext}
%
\begin{equation}\label{sigzzext}
\sigma_{nn}^{zz} \cong 2d\theta ( n - 1 + \lambda_e) ,
\end{equation}
where the extrapolation length $\lambda_e$ is given by $\lambda_e = 1/c$.

\subsection{ Energy and susceptibilities }
\label{energy}

For the ferromagnetic model described by the Hamiltonian (\ref{dham}),
the energy corresponding to the $i$th site, $U_i$, is determined in the spherical limit 
$D\to\infty$ by the spontaneous magnetization $m_i\equiv\langle m_{zi}\rangle$ 
and the transverse CF $s_{ij}$:
%\marginpar{ui}
%
\begin{equation}\label{ui}
U_i = - \frac{1}{2} \sum_j J_{ij} m_i m_j  - \frac{\eta}{2} \sum_j J_{ij} s_{ij} .
\end{equation}
It is convenient to consider the reduced energies $\tilde U \equiv U/|U_0|$ where 
$U_0=-J_0/2$ is the ground-state energy.
In the semi-infinite geometry the reduced energy corresponding to any site in the $n$th layer can be 
written as
%\marginpar{un}
%
\begin{equation}\label{un}
\tilde U_n =  2d'\tilde U_{nn} + \tilde U_{n,n-1} + \tilde U_{n,n+1},
\end{equation}
where $\tilde U_{nn}$ is due to the interaction with one of the neighbors in the same layer and 
$\tilde U_{n,n\pm 1}$ is due to that with the neighbors in the adjacent layers. 
The terms of (\ref{un}) can be represented through magnetization and the layer CF 
$\sigma_{nn'}({\bf q})$ as
%\marginpar{unn}
%
\begin{eqnarray}\label{unn}
&&
\tilde U_{nn} = -\frac{1}{2d} \left[ 
m_n^2 + \int\!\!\!\frac{d^{d'}{\bf q}}{(2\pi)^{d'}} \eta \lambda'_{\bf q} \sigma_{nn}({\bf q})
\right] ,
\\
&&
\tilde U_{n,n\pm 1} = -\frac{1}{2d} \left[
m_n m_{n\pm 1} + \int\!\!\!\frac{d^{d'}{\bf q}}{(2\pi)^{d'}} \eta \sigma_{n,n\pm 1}({\bf q}) 
\right]. \nonumber
\end{eqnarray}
In practice, only the total energy of the site  $\tilde U_n$ is needed.
Using the CF equation  (\ref{cffd}) to eliminate $\sigma_{n,n\pm 1}$ 
and the magnetization equation   (\ref{mageqfd}) to eliminate $m_{n\pm 1} $ in (\ref{unn}), one comes to the remarkably simple result
%\marginpar{utiln}
%
\begin{equation}\label{utiln}
\tilde U_n = \theta - 1/G_n  .
\end{equation}
The deviation of the energy density from the bulk density $\tilde U$ is given by
%\marginpar{util1n}
%
\begin{equation}\label{util1n}
\tilde U_{1n}\equiv \tilde U_n - \tilde U \cong  G_{1n}/G^2  ,
\end{equation}
where $G_{1n}$ is defined by  (\ref{g1ndef}).
This formula provides additional physical interpretation of $G_{1n}$, besides
that following from the role it plays in the correlation functions (see Sec. \ref{bulk}).

The susceptibilities of a ferromagnet are related to the correlation functions.
In the semi-infinite geometry the generic susceptibility is that describing the response of the 
spin polarization in the $n$th layer to the dimensionless magnetic field 
${\bf h}\equiv J_0 {\bf H}$ in the $n'$th one.
In the region above $T_c$ which is considered throughout the paper, it can be written as
%\marginpar{chinn1}
%
\begin{equation}\label{chinn1}
\chi_{\alpha nn'} =  \partial \langle m_{\alpha n} \rangle / \partial h_{\alpha n'}  =
 \sigma_{nn'}^{\alpha\alpha}(0) / \theta .
\end{equation}
Here, in the anisotropic case the longitudinal ($\alpha=1=z$) and transverse ($\alpha\ne 1= z$) 
susceptibilities are different.
For the transverse susceptibility the corresponding layer correlation function $\sigma_{nn'}(0)$ (the
index $\alpha$ is dropped for convenience) is determined by the system of linear equations 
(\ref{cffd}).
The longitudinal layer CF $\sigma_{nn'}^{zz}(0)$ satisfies the same system of equations with 
$\eta \Rightarrow 1$ in (\ref{bn}).
The most important of the susceptibilities (\ref{chinn1}) is $\chi_{\alpha 11}$, corresponding to the
boundary layer.
Whereas for $\eta<1$ the transverse susceptibility $\chi_{11}$ is noncritical, the longitudinal
one $\chi_{z11}$ shows critical behavior with the critical index $\gamma_{11}$.
One can also consider the response in the $n$th layer to the homogeneous field.
The appropriate susceptibilities are given by
%\marginpar{chin}
%
\begin{equation}\label{chin}
\chi_{\alpha n} = \frac{ \partial \langle m_{\alpha n} \rangle }
{ \partial h_{\alpha } } =
\sum_{n'=1}^\infty \chi_{\alpha nn'}  =
\frac 1\theta \sum_{n'=1}^\infty \sigma_{nn'}^{\alpha\alpha}(0) .
\end{equation}

\subsection{Bulk limit and continuous dimensions}
\label{bulk}

In the homogeneous case, $m_i=m$ and $G_i=G$ are constants, and the equation (\ref{cf2}) can be easily solved with the help of the Fourier transformation.
This yields the Fourier-transformed transverse CF
%
%\marginpar{cfhomo0}
%
\begin{equation}\label{cfhomo0}
s({\bf k}) = \frac{ \theta G }{ 1-\eta G \lambda_{\bf k} } .
\end{equation}
The longitudinal CF $s_{zz}({\bf k})$ is given above $T_c$ by the same expression with $\eta=1$. 
Now the autocorrelation function $s_{ii} $ can be expressed as
%\marginpar{cfhomo}
%
\begin{equation}\label{cfhomo}
s_{ii} = \int\!\!\!\frac{d^d{\bf k}}{(2\pi)^d} s({\bf k}) = \theta G P(\eta G) = 1-m^2,
\end{equation}
where 
%\marginpar{px}
%
\begin{equation}\label{px}
P(X) \equiv \int\!\!\!\frac{d^d{\bf k}}{(2\pi)^d}\frac{1}{1-X\lambda_{\bf k}} 
\end{equation}
is the lattice Green's function.
The quantity $\lambda_{\bf k} \equiv J_{\bf k}/J_0$ is given for the
nearest-neighbor (nn) interaction by
(\ref{lampr}) with $d'\Rightarrow d$.
The total wavevector $\bf k$ is related to $\bf q$ above by 
${\bf k} = k_z{\bf e}_z + {\bf q}$, where ${\bf q\cdot e}_z=0$.
The last equation in (\ref{cfhomo}) together with the equation
%\marginpar{mageqb}
%
\begin{equation}\label{mageqb}
m(1-G) = 0
\end{equation}
following from (\ref{mageq}) in the homogeneous case completely describes the
ASM in zero magnetic field $H$.
The homogeneous ASM equations for $H\neq 0$ can be found in 
Refs. \cite{garlut84d,gar96prb}.
The lattice integral $P(X)$ has the following properties:
%\marginpar{plims}
%
\begin{equation}\label{plims}
\renewcommand{\arraystretch}{1.2}
P(X) \cong
\left\{
\begin{array}{lll}
1 + X^2/(2d),                      & X \ll 1                           \\
1/(1-X^2)^{1/2},		  & d=1		\\
(1/\pi)\ln(8/X_1),                & X_1 \ll 1,   & d=2 \\
W_3 - cX_1^{1/2},              & X_1 \ll 1,   & d=3 \\
W_4 - cX_1\ln(c'/ X_1),     & X_1 \ll 1,   & d=4 \\     
W_d - cX_1,                   & X_1 \ll 1,   & d>4,       
\end{array}
\right. 
\end{equation}
where $X_1 \equiv 1-X$ and 
$W_3=1.51639$, $W_4=1.23947$, $W_5=1.15631$, etc., are the Watson integrals.
Since above $T_c$ the constraint equation (\ref{sconstr}) yields $s_{ii}=1$, equation (\ref{cfhomo}) determines the value of $G$ which increases with decreasing temperature.
The high-temperature asymptote of $G$ is $G \cong \theta^{-1} [ 1 - \eta^2 \theta^{-2}/(2d) ]$,
$\theta \gg 1$.
This results in $\tilde U \cong -\eta^2/(2d\theta)$ for the energy in the bulk, which is given by 
(\ref{utiln}) with $n \to \infty$.
The criticality is determined by $G=1$, which corresponds to closing the gap in the longitudinal correlation function $s_{zz}({\bf k})$ [see (\ref{cfhomo0})].
This is the reason for calling $G$ the ``gap parameter''.
Below $\theta_c$ one obtains $G=1$ from (\ref{mageqb}) and then
%\marginpar{tc} 
%
\begin{equation}\label{tc}
m = \sqrt{1-\theta/\theta_c}, \qquad \theta_c = 1/P(\eta) 
\end{equation}
from (\ref{cfhomo}).
Here the value of the Curie temperature $\theta_c$ \cite{garlut84d}
generalizes the well-known result of the spherical model $\theta_c=1/W$ \cite{berkac52} for the anisotropic case.

The influence of the anisotropy on the ordering in the ASM is rather essential.
The anisotropic gap in the transverse CF $s({\bf k})$ prevents long-wavelength excitations (transverse fluctuations) from destroying the long-range order in two dimensions, and $\theta_c$ determined by (\ref{tc}) is finite for $\eta<1$.
Moreover, the phase transition at finite temperature occurs even in the {\em one-dimensional}
ASM.
This surprising result is due to the switching off of the {\em longitudinal} fluctuations in the limit $D\to\infty$, which are responsible for the breakdown of the long-range order in one dimension.

The bulk solution of the linear system of equations (\ref{cffd}) has the form
%\marginpar{sigbulk}
%
\begin{equation}\label{sigbulk}
\sigma_{nn'}^{\rm bulk}({\bf q}) = 
\frac{d\theta}{\eta} \frac{\alpha^{|n-n'|}}{\sqrt{b^2-1}},
\qquad \alpha \equiv b-\sqrt{b^2-1} ,
\end{equation}
where $b$ is given by (\ref{bn}) with $G_n\Rightarrow G$.
This result could also be obtained by the integration of the bulk transverse CF $s({\bf k})$ given by (\ref{cfhomo0}) over $k_z$.
For the weakly anisotropic ASM, $1-\eta \ll 1$, at small wave vectors the transverse 
correlation functions in the bulk, (\ref{cfhomo0}) and (\ref{sigbulk}),  have the form
%\marginpar{cfbulk}
%
\begin{equation}\label{cfbulk}
s({\bf k}) \cong \frac{ 2d\theta }{\kappa^2 + k^2},
\qquad
\sigma_{nn}^{\rm bulk}({\bf q}) = \frac{ d\theta}{ \sqrt{\kappa^2 + q^2} },
\end{equation}
where $\kappa$ is defined by 
%\marginpar{kapdef}
%
\begin{equation}\label{kapdef}
\kappa^2 \equiv 2d[1/(\eta G) -1] \cong 2d[1-\eta G] \ll 1 .
\end{equation}
One can see that the transverse correlation length $\xi_{c\alpha}\equiv 1/\kappa$ increases 
without diverging with decreasing temperature down to $\theta_c$ and remains constant below 
$\theta_c$, in accordance with the behavior of $G$.

The field-theoretical multiple-component $\phi^4$ model used by Bray and Moore \cite{bramoo77prl77jpa} extends in a natural way for arbitrary noninteger lattice dimensions $d$.
The discrete structure of the lattice which is important near the surface is, however, lost in such a 
model.
A better way to get a continuous-dimension model to study crossover between
different lattice dimensions is to consider the $d'$ translationally invariant dimensions as continuous, preserving
the dimension $z$ perpendicular to the surface as discrete.
This amounts to making the long-wavelength approximation
%\marginpar{lamcomb}
%
\begin{equation}\label{lamcomb}
\lambda_{\bf k} = \frac 1d \cos k_z + \frac{d'}{d}\lambda'_{\bf q},
\qquad  \lambda'_{\bf q} \Rightarrow 1-\frac{q^2}{2d'}
\end{equation}
{\em in the whole Brillouin zone} for the part of the expression for the exchange integral 
$J_{\bf k} = J_0 \lambda_{\bf k} $.
The natural hypercubic cutoff $|k_i|\leq \pi$ and the corresponding density of states are modified for the  ${\bf q}$ components according to
%\marginpar{contint}
%
\begin{equation}\label{contint}
\int\!\!\!\frac{d^{d'}{\bf q}}{(2\pi)^{d'}} \ldots \Rightarrow 
\frac{d'}{\Lambda^{d'}}\int_0^\Lambda dq q^{d'-1}\ldots
\end{equation}
with $\Lambda = \sqrt{2(d+1)}$ .
One can check that the sum rules
%\marginpar{sumrules}
%
\begin{equation}\label{sumrules}
\int_{-\pi}^\pi\!\!\!\frac{dk_z}{2\pi} \frac{d'}{\Lambda^{d'}}\int_0^\Lambda dq q^{d'-1}
\left\{ 1 \atop \lambda_{\bf k} \right\} = \left\{ 1 \atop 0 \right\}
\end{equation}
are satisfied.
Now using (\ref{sigbulk}), one can, instead of (\ref{cfhomo}), write
%\marginpar{cfhomo1}
%
\begin{equation}\label{cfhomo1}
\frac{d'}{\Lambda^{d'}}\int_0^\Lambda dq q^{d'-1}\frac{d\theta}{\eta} \frac{1}{\sqrt{b^2-1}} = 
\theta G P(\eta G),
\end{equation}
which is the definition of $P(X)$ in our continuous-dimension model.
The resulting $P(X)$ posesses the same general properties (\ref{plims}).
The Watson integrals $W$ for some values of $d$ are
$W_{2.5}=2.527059$, $W_{3.0}=1.719324$, $W_{4.0}=1.321825$, and 
$W_{5.0}=1.192848$.

For both hypercubic and continuous-dimension lattices the singular behavior of the
integral $P(X)$ for $\kappa \ll 1$ is described by
%\marginpar{plimsd}
%
\begin{equation}\label{plimsd}
\renewcommand{\arraystretch}{1.2}
P(X) \cong
\left\{
\begin{array}{ll}
C_d/ \kappa^{2-d},		  & 1 \leq d \leq 2	\\
W - C_d\kappa^{d-2},                      & 2< d<4 .
\end{array}
\right. 
\end{equation}
Here $C_d=A_d\times dM_d$, where the non-universal factor $A_d$ reads
%\marginpar{Addef}
%
\begin{equation}\label{Addef}
A_d \equiv \left\{
\begin{array}{ll}
S_{d'}/(2\pi)^{d'}, 	& \mbox{hypercubic lattice} \\
d'/\Lambda^{d'},		& \mbox{continuous dimensions} 
\end{array}
\right.
\end{equation}
and $S_d = 2\pi^{d/2}/\Gamma(d/2)$ is the surface of the $d$-dimensional unit sphere.
The universal quantity $M_d$ which will be needed below is given by
%\marginpar{Mddef1}
%
\begin{equation}\label{Mddef1}
M_d \equiv \int_0^{\Lambda/\kappa} \frac{ dy y^{d'-1} }{ \sqrt{1+y^2} }
= \frac{ 1 - \kappa^{2-d} }{ 2\pi^{1/2} } \Gamma \left(\frac{ d-1 }{ 2 }\right) 
\Gamma \left(\frac{ 2-d }{ 2 }\right)
\end{equation}
for $d\leq2$ and
%\marginpar{Mddef2}
%
\begin{eqnarray}\label{Mddef2}
&&
M_d \equiv \int_0^\infty \!\! dy \, y^{d'-1} \left[ \frac 1y - \frac{ 1 }{ \sqrt{1+y^2} } \right]
\nonumber\\
&&\qquad
= \frac{ \pi  }{ \cos(\pi\mu) } \frac{ \Gamma(d-1) }{ 2^{d-1} \Gamma^2(d/2) },
\qquad \mu\equiv\frac{ d-3 }{ 2}  
\end{eqnarray}
for $2<d<4$.
The factor $1 - \kappa^{2-d}$ in  (\ref{Mddef1}) is for $\kappa\ll 1$ close to unity if $d$ is
not close to 2.
It is needed for $2-d \ll 1$ to give 
%\marginpar{Mddef1'}
%
\begin{equation}\label{Mddef1'}
M_d \cong \frac{ 1 - \kappa^{2-d} }{ 2-d }\cong  \ln(1/\kappa)
\end{equation}
 with logarithmic accuracy.
For  $d \to 1$ one obtains $C_d\to 1 $, in accordance with (\ref{plims}).

In the anisotropic case $\eta<1$, the value of $G$ determined from the equation 
$\theta G P(\eta G) = 1$ approaches 1 linearly just above the Curie temperature 
$\theta_c$ given by (\ref{tc}):
%\marginpar{glin}
%
\begin{equation}\label{glin}
1 - G \cong \tau / I(\eta),   \qquad   \tau \equiv \theta/\theta_c - 1 ,
\end{equation}
where
%\marginpar{ieta}
%
\begin{equation}\label{ieta}
I(X) \equiv 1 + \frac{XP'(X)}{P(X)} ,  \qquad P'(X) \equiv \frac{dP(X)}{dX}.
\end{equation}
For the weakly anisotropic model this solution is valid in the narrow region defined by 
$1-G\ll 1-\eta$, i.e., below the crossover temperature $\tau^* = (1-\eta)I(\eta)$.
For different lattice dimensions $\tau^*$ reads
%\marginpar{taustar}
%
\begin{equation}\label{taustar}
\renewcommand{\arraystretch}{1.2}
\tau^* \sim
\left\{
\begin{array}{ll}
1,	  		&  d<2	\\
1/\ln[1/(1-\eta)],               & d=2 \\
(1-\eta)^{(d-2)/2},	& 2 < d < 4 \\
(1-\eta)	\ln[1/(1-\eta)],	& d=4 \\
1-\eta,			& d>4.
\end{array}
\right. 
\end{equation}
For $\tau\gg\tau^*$ one has
%\marginpar{gtc}
%
\begin{equation}\label{gtc}
\renewcommand{\arraystretch}{1.2}
1-G \sim
\left\{
\begin{array}{ll}
\theta^{2/(2-d)},	 	& d<2	\\
\exp(-A_d^{-1}/\theta),                  & d=2, \, 2.0\\
\tau^{2/(d-2)},		& 2<d<4\\
\tau/\ln\tau,		& d= 4\\
\tau,			& d>4,
\end{array}
\right. 
\end{equation}
where, according to (\ref{Addef}),  $A_2^{-1} = \pi$ and $A_{2.0}^{-1} = \sqrt{6}$.
Here the result for $d\leq 2$ is valid for $\theta\ll 1$, i.e., a weakly anisotropic system can
be close to criticality ($1-G\ll 1$) in a temperature range extending far above $\theta_c\ll 1$.
For $d\geq 2$ the Curie temperature $\theta_c$ is not small, and (\ref{gtc}) requires 
$\tau\ll 1$.

The longitudinal CFs are given by the same formulas (\ref{cfbulk}) and (\ref{kapdef}) with $\eta=1$ .
The longitudinal correlation length $\xi_{cz}\equiv 1/\kappa_z$, where
%\marginpar{kapzdef}
%
\begin{equation}\label{kapzdef}
\kappa_z^2 \equiv 2d[1/G -1] \cong 2d[1-G] \ll 1, 
\end{equation}
diverges at $\theta_c$ in different ways for the isotropic and anisotropic models according to (\ref{glin}) and (\ref{gtc}), respectively.
The critical behavior of the ASM is, for $\eta<1$, in all respects analogous to that given by the mean-field approximation.
This is due to the suppression of the singularity of the lattice Green's function $P(X)$ [see  (\ref{plims}) and (\ref{plimsd})].
For $1-\eta \ll 1$ far enough from $\theta_c$, i.e., for $\tau\gg \tau^*$, the system behaves 
isotropically and, in particular, $\xi_{cz}\cong \xi_{c\alpha}$.
Crossover at $\tau^*$ is analogous to that between the Heisenberg and Ising universality 
classes in the weakly anisotropic Heisenberg model.
Here one has the crossover between the spherical and mean-field universality classes instead.

\subsection{Continued-fraction formalism}
\label{continuous}

The linear homogeneous second-order finite-difference equation 
%\marginpar{ikeq}
%
\begin{equation}\label{ikeq}
2b_n {\cal Z}_n - {\cal Z}_{n+1} - {\cal Z}_{n-1} = 0 ,
\end{equation}
which corresponds to the CF equation (\ref{cffd}), has two linearly independent solutions, 
${\cal I}_n$ and ${\cal K}_n$.
They can be chosen so that ${\cal I}_n\to\infty$ and ${\cal K}_n\to 0$ for $n\to\infty$.
The solution of equation (\ref{cffd}) with the boundary condition (\ref{bc}) can be expressed through ${\cal I}_n$ and ${\cal K}_n$ as
%\marginpar{sigik}
%
\begin{eqnarray}\label{sigik}
\renewcommand{\arraystretch}{2.3}
&&
\sigma_{nn'} = - \frac{2d\theta}{\eta  {\cal W}_{n'} } 
\frac{ 2b_1 {\cal I}_1 - {\cal I}_2 }{ 2b_1 {\cal K}_1 - {\cal K}_2 }  {\cal K}_n {\cal K}_{n'} 
\nonumber \\
&&\qquad
{}+\frac{2d\theta}{\eta {\cal W}_{n'} } 
\left\{ 
\begin{array}{ll}
\displaystyle
{\cal I}_n {\cal K}_{n'},                      & n\leq n'                           \\
\displaystyle
{\cal I}_{n'} {\cal K}_n,                      & n'\leq n ,                       \\
\end{array}
\right. 
\end{eqnarray}
where the Wronskian ${\cal W}_n$ is given by
%\marginpar{wrdef}
%
\begin{equation}\label{wrdef}
{\cal W}_n \equiv {\cal I}_n {\cal K}_{n-1} -  {\cal K}_n {\cal I}_{n-1} .
\end{equation}
It can be shown with the help of (\ref{ikeq}) that ${\cal W}_{n+1}={\cal W}_n$, i.e.,  ${\cal W}_n$ is independent of $n$.
It is convenient to redefine ${\cal I}_n$ by replacing it by its linear combination with 
${\cal K}_n$, so that the redefined ${\cal I}_n$ satisfies the additional requirement ${\cal I}_0=0$ in the non-existing layer $n=0$.
This entails $2b_1 {\cal I}_1 - {\cal I}_2=0$, i.e., the first term in (\ref{sigik}) becomes zero.

The solution (\ref{sigik}) can be rewritten in the form of a continued fraction, which is appropriate in particular for numerical calculations.
In terms of the functions $\alpha_n$ and $\alpha'_n$ determined by
%\marginpar{aldef}
%
\begin{equation}\label{aldef}
{\cal I}_{n-1} /{\cal I}_n \equiv \alpha_n, 
\qquad {\cal K}_{n-1} /{\cal K}_n \equiv 2b_n - \alpha'_n
\end{equation}
the solution (\ref{sigik}) for $n=n'$  becomes
%\marginpar{sigrec}
%
\begin{equation}\label{sigrec}
\sigma_{nn} = \frac{2d\theta}{\eta} \frac{1}{2b_n - \alpha_n - \alpha'_n} .
\end{equation}
The functions $\alpha_n$ and $\alpha'_n$ can be found from the forward and backward recurrence relations 
%\marginpar{alrec}
%
\begin{equation}\label{alrec}
\alpha_{n+1} = \frac{1}{2b_n - \alpha_n}, \qquad
\alpha'_{n-1} = \frac{1}{2b_n - \alpha'_n},
\end{equation}
being the consequence of (\ref{ikeq}).
The initial condition for the first one is $\alpha_1=0$.
Far from the surface $\alpha_n$ approaches the bulk value $\alpha$ of (\ref{sigbulk}).
The backward relation for $\alpha'_n$ starts from $\alpha$ far from the surface.
For numerical calculations in the isotropic case for $2<d<4$, a refined asymtote is needed 
(see the end of Sec. \ref{isocrit}).
If the denominator of (\ref{sigrec}) becomes zero for $q=0$ (which is the
case for isotropic model at criticality for $2<d\leq 3$; see Sec. \ref{isocrit}), then with
the help of (\ref{alrec}) one can obtain the relation
%\marginpar{alrel}
%
\begin{equation}\label{alrel}
\alpha_n\alpha'_{n-1} = 1 \qquad (q=0).
\end{equation}

The general solution (\ref{sigik}) can be represented through the diagonal Green function  (\ref{sigrec}) via the relations
%\marginpar{signn'}
%
\begin{equation}\label{signn'}
\sigma_{n,n-m} = \sigma_{n-m,n} = \alpha_{n-m+1}\alpha_{n-m+2}\ldots\alpha_n \sigma_{nn}  
\end{equation}
or, alternatively,
%\marginpar{signn''}
%
\begin{equation}\label{signn''}
\sigma_{n+m,n} = \sigma_{n,n+m} = \alpha'_{n+m-1}\alpha'_{n+m-2}\ldots\alpha'_n \sigma_{nn}. 
\end{equation}
The consequence of these two relations is the useful formula
%\marginpar{sign'n'}
%
\begin{equation}\label{sign'n'}
\sigma_{n,n+1} = \alpha_{n+1}\sigma_{n+1,n+1} = \alpha'_n\sigma_{nn} .
\end{equation}

It is convenient to introduce the deviations from the bulk values
%\marginpar{al1def}
%
\begin{equation}\label{al1def}
\alpha_{1n} \equiv \alpha_n-\alpha, 
\qquad
\alpha'_{1n} \equiv \alpha'_n-\alpha
\end{equation}
and
%\marginpar{b1def}
%
\begin{equation}\label{b1def}
b_{1n} \equiv b-b_n = \frac d\eta \left( \frac 1G - \frac 1G_n \right) =   
\frac{ dG_{1n} }{ \eta GG_n } ,
\end{equation}
where $\alpha$, $b$, and $G_{1n}$ are defined by (\ref{sigbulk}),  (\ref{bn}),
and (\ref{g1ndef}), respectively.
The recurrence relations for the deviations $\alpha_{1n}$ and $\alpha'_{1n}$ have the form
%\marginpar{al1rec}
%
\begin{equation}\label{al1rec}
\alpha_{1,n+1} = \frac{ \alpha^2 (2b_{1n}+ \alpha_{1n} ) }
{ 1 - \alpha (2b_{1n}+ \alpha_{1n} ) }, 
\qquad
\alpha_{11} = -\alpha
\end{equation}
and
%\marginpar{al'1rec}
%
\begin{equation}\label{al'1rec}
\alpha'_{1,n-1} = \frac{ \alpha^2 (2b_{1n}+ \alpha'_{1n} ) }
{ 1 - \alpha (2b_{1n}+ \alpha'_{1n} ) },
\qquad
\alpha'_{1,\infty}  = 0 .
\end{equation}
In terms of the deviations $\alpha_{1n}$, $\alpha'_{1n}$, and $b_{1n}$ the CF (\ref{sigrec})
can be now written as
%\marginpar{sigrec1}
%
\begin{equation}\label{sigrec1}
\sigma_{nn} = \frac{d\theta}{\eta} 
\frac{1}{ \sqrt{b^2-1} + \Sigma_n }  , 
\end{equation}
where
%\marginpar{deln}
%
\begin{equation}\label{deln}
\Sigma_n \equiv -(\alpha_{1n}+\alpha'_{1n})/2 - b_{1n} 
\end{equation}
plays the role of the self-energy part for the spin CF.
At and above $\theta_c$ the constraint equation (\ref{sconstrfd}) can be rewritten as
%\marginpar{sconstr1}
%
\begin{equation}\label{sconstr1}
\int\!\!\!\frac{d^{d'}{\bf q}}{(2\pi)^{d'}} 
\left[\sigma_{nn}({\bf q}) - \sigma_{nn}^{\rm bulk}({\bf q}) \right]= 0,
\end{equation}
where the bulk result is given by (\ref{sigrec1}) without $\Sigma_n$.

Since for $n\gg 1$ the quantities $\alpha_{1n}$ and $\alpha'_{1n}$ of (\ref{al1def}) are small 
and  $n$ in the recurrence relations (\ref{al1rec}) and  (\ref{al'1rec}) can be treated as a continuous variable, these relations can be reduced to the first-order nonlinear differential equations which for $\tilde q \equiv \sqrt{\kappa^2+q^2} \ll 1$ have the form 
%\marginpar{al1difeq}
%
\begin{eqnarray}\label{al1difeq}
&&
\frac{d}{dn}\alpha_{1n} = -2\tilde q \alpha_{1n} + \alpha_{1n}^2 + 2b_{1n},
\nonumber\\
&&
-\frac{d}{dn}\alpha'_{1n} = -2\tilde q \alpha'_{1n} + (\alpha'_{1n})^2 + 2b_{1n} .
\end{eqnarray}
These Riccati equations can be transformed to linear second-order
differential equations which are equivalent to (\ref{cfdiffeq}).

\subsection{Variation of the gap parameter at low and high temperatures}
\label{lowhightemp}

The main problem with the solution of the ASM equations (\ref{cffd})--(\ref{sconstrfd}) is to find the
variation of the gap parameter $G_n$ that plays a fundamental role in the theory.
Its inhomogeneous part $G_{1n}$ defined by (\ref{g1ndef})
is analogouos, as we shall see, to the function $V(z)$ with the opposite sign, which was
considered by Bray and Moore \cite{bramoo77prl77jpa}.
The inhomogeneity $G_{1n}$ results from the deficit of  interacting neighbors in the region
near the surface and is positive.
The simplest case in which $G_n$ can be found analytically is $T=0$.
Here, for the magnetization one has $m_n=1$ everywhere, and  $G_n$ determined from
(\ref{mageqfd}) reads
%\marginpar{gnt0}
%
\begin{equation}\label{gnt0}
\renewcommand{\arraystretch}{1.5}
G_n =
\left\{
\begin{array}{ll}
\displaystyle
\frac{2d}{2d-1},                      & n=1                           \\
\displaystyle
1,		               & n\geq 2 .
\end{array}
\right. 
\end{equation}
This result also shows that the boundary layer, $n=1$, is distinguished from all other ones.
This feature that is beyond the scope of the continuous field-theoretical approaches  can be observed in the whole temperature range.
In particular, at high temperatures ($\theta\gg 1$), or in the whole region $\theta\geq \theta_c$
in the ``quasi-Ising'' limit $\eta\ll 1$, the variation of $G_n$ can be found with the help of the 
high-temperature series expansions (HTSE).
This can be most conveniently done using the diagram technique for classical spin systems
\cite{garlut84d,gar94jsp,gar96prb,gar96jpa}.
The result for the hypercubic lattice has the form ($\eta/\theta \ll 1$)
%\marginpar{gnhtse}
%
\begin{equation}\label{gnhtse}
\renewcommand{\arraystretch}{2.3}
G_{1n} \cong
\left\{
\begin{array}{ll}
\displaystyle
\frac{1}{\theta} \left( \frac{\eta}{2d\theta} \right)^2,                      & n=1                           \\
\displaystyle
\frac{4(d-1)}{\theta} \left( \frac{\eta}{2d\theta} \right)^{2(n+1)},      & n\geq 2 .
\end{array}
\right. 
\end{equation}
The terms with $n\geq 2$ are slightly different for the continuous-dimension model.
One can see, again, that the boundary layer is distinguished.
For the layers with $n\geq 2$ the expected leading diagrams of order $2n$ are cancelled, and the result is much smaller.

Calculation of $G_{1n}$ in other cases requires more specialized methods, which will be considered below.

\section{Surface-induced correlations in low and high dimensions}
\label{lowhigh}

\subsection{Solution for the ``toy'' model $d=1$}
\label{solutiond1}

In one dimension the solution of the ASM equations (\ref{cffd}) --
(\ref{sconstrfd}) is greatly simplified since $d'=0$ and there is no integration over the wave vector ${\bf q}$.
The quantity $b_n$ of (\ref{bn}) reduces in this case to $b_n = 1/(\eta G_n)$.
For the autocorrelator $s_{nn}$ one has simply $s_{nn}=\sigma_{nn}$.
Above $\theta_c$ the constraint equation (\ref{sconstrfd}) becomes $\sigma_{nn}=1$, i.e., 
all $\sigma_{nn}$ are equal to each other.
This means that for the one-dimensional ASM the {\em transverse} susceptibilities with respect 
to the layer field, $\chi_{\perp nn} = \sigma_{nn}/\theta= 1/\theta$, are the same for all layers.
Now, from the relation (\ref{sign'n'}) follows $\alpha_{n+1}=\alpha'_n$.
Using the latter together with the recurrence relations (\ref{alrec}), one can write the constraint
equation as
%\marginpar{sig1d}
%
\begin{equation}\label{sig1d}
\renewcommand{\arraystretch}{1.2}
1 = \sigma_{nn} = \frac{2\theta}{\eta}
\left\{
\begin{array}{ll}
\displaystyle
1/[2b_1-(2b_1)^{-1}],                      & n=1                           \\
\displaystyle
 1/[\alpha_n^{-1}-\alpha_n] ,    		& n\geq 2 .
\end{array}
\right. 
\end{equation}
This implies that all $\alpha_n$ and $\alpha'_n$, except for $\alpha_1=0$, are equal to 
%\marginpar{al1dbulk}
%
\begin{equation}\label{al1dbulk}
\alpha = [\sqrt{\eta^2+\theta^2} - \theta]/\eta ,
\end{equation}
which is with the bulk value given by (\ref{sigbulk}).
Then for $n\geq 2$ with the help of (\ref{sigbulk}) one can identify 
$b_n=b=[\alpha+\alpha^{-1}]/2=\sqrt{\eta^2+\theta^2}/\eta$ and $\sqrt{b^2-1}=\theta/\eta$.
The boundary-layer quantity $b_1$ can be determined directly from (\ref{sig1d}) with
the result $2b_1 = \alpha^{-1} = [\sqrt{\eta^2+\theta^2} + \theta]/\eta$.
Now the exact result for $G_n=1/(\eta b_n)$ can be written in the form
%\marginpar{gn1d}
%
\begin{equation}\label{gn1d}
\renewcommand{\arraystretch}{1.2}
G_n =
\left\{
\begin{array}{ll}
\displaystyle
 2/[\sqrt{1+\theta^2-\theta_c^2} + \theta] ,   & n=1         \\
\displaystyle
 1/\sqrt{1+\theta^2-\theta_c^2} , & n\geq 2 .
\end{array}
\right. 
\end{equation}
with $\theta_c=\sqrt{1-\eta^2}$.
One can see that $G_1>G_{n\geq 2}=G$.
In particular, in the weakly anisotropic case, $1-\eta\ll 1$, at criticality
one has $G_1\cong 2(1-\kappa)$, which is nearly two times greater than the bulk
value $G=1$ [cf. (\ref{gnt0})].
Variation of $G_n$ above belongs to the class studied at the end of Sec. \ref{ASM},
and thus $\sigma_{nn'}$ is given by (\ref{signnbc}).
In our one-dimensional model, however, one has $\alpha - 2b_{11} =0$, and the inhomogeneous
term vanishes.
Thus one arrives again at the result $\sigma_{nn}=1$, which can serve as an
independent check of the calculations.

Now we consider the longitudinal CF $\sigma_{nn'}^{zz}$ and the 
corresponding susceptibilities.
The solution of the finite-difference equation (\ref{cffd}) with $\eta=1$ and $G_n$ given by
(\ref{gn1d}) has the form (\ref{signnbc}) with
$\alpha_z^{\pm 1} = b_z \mp \sqrt{b_z^2-1}$ [cf. (\ref{sigbulk})], $b_z\equiv 1/G$, and
$b_{z11} \equiv 1/G - 1/G_1$ [cf. (\ref{b1def})].
Using (\ref{gn1d}) one can write, explicitly,
%\marginpar{alz}
%
\begin{eqnarray}\label{alz}
&&
\alpha_z = \sqrt{1+\theta^2-\theta_c^2} - \sqrt{\theta^2-\theta_c^2} ,
\nonumber\\
&&
\frac{ \alpha_z - 2b_{z11} }{ \alpha_z^{-1} - 2b_{z11} } = 
\frac{ \theta - \sqrt{\theta^2-\theta_c^2} }{ \theta + \sqrt{\theta^2-\theta_c^2} } .
\end{eqnarray}
In contrast to the transverse CF $\sigma_{nn}$ given by (\ref{signnbc}), the inhomogeneous 
term in $\sigma_{nn}^{zz}$ does not disappear.
For the weakly anisotropic ASM in the range $\theta\ll 1$, the expression for
$\sigma_{nn}^{zz}$ simplifies to
%\marginpar{sigzzas}
%
\begin{equation}\label{sigzzas}
\sigma_{nn}^{zz} \cong \frac \theta\kappa_z 
\left[ 1 +  \frac{ \kappa_z - \kappa}{ \kappa_z + \kappa} e^{ -2\kappa_z(n-1)} \right]
\end{equation}
[cf. (\ref{signnbc'})].
where $\kappa$ and $\kappa_z$ are given by 
%\marginpar{kap1d}
%
\begin{equation}\label{kap1d}
\kappa \cong \theta, \qquad \kappa_z \cong \sqrt{\theta^2-\theta_c^2} .
\end{equation}
One can see that here the extrapolation length $\lambda_e = 1/\kappa \gg 1$ is large
on the scale of the lattice spacing.
Well above $\theta_c$, where $\kappa_z\cong \kappa$, there is no difference
between the longitudinal and transverse CFs: $\sigma_{nn}^{zz} \cong \sigma_{nn} =1$.
Near $\theta_c$ one has $\kappa_z \ll \kappa$, and (\ref{sigzzas}) shows the dependence on 
the distance from the surface.
Whereas the bulk CF ($n=\infty$) diverges  with the exponent $\gamma_{nn}^{\rm bulk}=\onehalf$
(see the end of Sec. \ref{bulk}),
the semi-infinite CF (\ref{sigzzas}) does not for any finite $n$.
In the boundary layer, $\sigma_{nn}^{zz}$  takes on the exact form
%\marginpar{sigzz11}
%
\begin{equation}\label{sigzz11}
\sigma_{11}^{zz} = \frac{ 2\theta }{ \theta + \sqrt{\theta^2 - \theta_c^2} } 
\end{equation}
in the whole range of $\eta$.
It varies from 1 at $\theta\gg \theta_c$ to 2 at $\theta \cong \theta_c$.
At criticality the longitudinal surface susceptibility is two times greater
than the transverse one. 
One can see that $\gamma_{11}=-\onehalf$, as in the MFA.

\subsection{Low dimensions, $1\leq d \leq 2$}
\label{solutiond12}

For $d>1$ the ASM equations become nontrivial because of the integration over the wave vector $\bf q$ in the constraint equations (\ref{sconstrfd}).
The deviation $G_{1n}$ of the gap parameter from its bulk value is now nonzero
for all layers, $n<\infty$.
For $\theta\ll 1$, the system is close to criticality, and the inverse transverse
correlation length $\kappa$  of (\ref{kapdef}) is
small and related to $\theta$ by
%\marginpar{kapd12}
%
\begin{equation}\label{kapd12}
\theta \cong 1/P \cong \kappa^{2-d} /C_d 
\end{equation}
[see (\ref{cfhomo}) and (\ref{plimsd}), cf. (\ref{kap1d})].
The variation of  $G_n$ is for $1\leq d \leq 2$ close to that for the one-dimensional model
above [see (\ref{gn1d})] and can be searched in the form
%\marginpar{gnpert}
%
\begin{equation}\label{gnpert}
\renewcommand{\arraystretch}{1.5}
G_n =
\left\{
\begin{array}{ll}
\displaystyle
\frac{2d}{2d-1}G+ \left(\frac{2d}{2d-1}\right)^2  G_{11},                & n=1                           \\
\displaystyle
G+G_{1n},		               & n\geq 2 .
\end{array}
\right. 
\end{equation}
The correction terms $G_{1n}$ will be shown below to be proportional to $(d-1)\theta$.
[Note that the definition of  $G_{11}$ here differs from that of (\ref{g1ndef}).]

For the variation of the gap parameter above, the spin CF $\sigma_{nn}$ is
for not too small wave vectors determined by the boundary condition at the
surface and given to the zeroth order by (\ref{signnbclo}).
This makes it possible to find $G_{1n}$ perturbatively in the range $\kappa n \ll 1$.
After that one can study the corresponding corrections to $\sigma_{nn}$,
which prove to be small for $q\gg \kappa$ or $d$ close to 1.

\subsubsection{Variation of $G_n$}

The quantities $G_{1n}$ for $n\geq 2$ can be found from the constraint
equation  (\ref{sconstrfd}) written in the form 
$(d'/\Lambda^{d'})\int dq q^{d'-1} \hat\Delta^2\sigma_{nn} =0$, where
%\marginpar{sigdif}
%
\begin{equation}\label{sigdif}
\hat\Delta^2\sigma_{nn} \equiv \sigma_{n+1,n+1} - 2 \sigma_{nn} + \sigma_{n-1,n-1} .
\end{equation}
The second-order difference above can, with the help of (\ref{sign'n'}), be
rewritten as
%\marginpar{sigdif2}
%
\begin{eqnarray}\label{sigdif2}
&&
\hat\Delta^2\sigma_{nn} = 2[ -2\alpha b_{1n} +
( \sqrt{b^2-1} - b_{1n} )
\nonumber\\
&&\qquad\qquad
{}\times ( \alpha_{1n}+ \alpha'_{1n}) - \alpha_{1n}\alpha'_{1n}] \sigma_{nn}.
\end{eqnarray}
Here the first term is the only one that is important in the long-wavelength region,
$q\sim \kappa$.
By integrating it one can set $\alpha\to 1$, which yields simply 
$-4b_{1n}\cong -4dG_{1n}$ in the constraint equation.
All other terms make a contribution from $q\gg \kappa$, and the only  
important one among them is the term containing  $(b^2-1)^{1/2}\alpha_{1n}$ where 
$\alpha_{1n}$ is induced by the surface.
The latter can be found by comparing (\ref{signnbclo}) with (\ref{sigrec1}),
where the small terms $\alpha'_{1n}$ and $b_{1n}$ are neglected.
This results in 
%\marginpar{al1nsurf}
%
\begin{equation}\label{al1nsurf}
 \alpha_{1n}\cong \frac{ 2(b^2-1)^{1/2} \alpha^{2n-1} }{ 1+\alpha^{2n-1} }
\cong \frac{ 2\tilde q  }{ \exp(2\tilde q n) +1 },
\end{equation}
where the second form is valid for $q\ll 1$.
Using (\ref{signnbclo}) for $q\gg \kappa$, one finally obtains
%\marginpar{g1npert}
%
\begin{eqnarray}\label{g1npert}
&&
G_{1n} \cong  \frac{ \theta }{ 2 }
\frac{d'}{\Lambda^{d'}}\int_0^\Lambda dq q^{d'-1} (1-\alpha^2) \alpha^{2(n-1)} ,
\end{eqnarray}
where the relation $2\alpha\sqrt{b^2-1} = 1-\alpha^2$ has been employed.
This expression is explicitly small for $\theta \sim \kappa^{2-d} \ll 1$ or for
$d'\equiv d-1 \ll 1$, as was said above.
For $n\sim 1$, integration in (\ref{g1npert}) extends over the whole Brillouin zone, and 
$G_{1n}$ is nonuniversal.
It decreases with $n$ since $\alpha<1$.
In the range $1\ll n \ll 1/\kappa$ the integration is cut at $q\sim 1/n$, and one can use
$\sqrt{b^2-1} \cong q$ and $\alpha^{2(n-1)}\cong e^{-2qn}$.
Expression (\ref{g1npert}) then simplifies to
%\marginpar{g1npert1}
%
\begin{equation}\label{g1npert1}
G_{1n} \cong  \frac{ d' }{ \Lambda^{d'} } 
\frac{ \theta\Gamma(d) }{ (2n)^d }= 
\frac{ \Gamma(d) }{ dM_d } \frac{ \kappa^{2-d} }{ (2n)^d } ,
\end{equation}
where the second form is explicitly universal.
Here $M_d$ is given by (\ref{Mddef1}), or, near $d=2$, by  (\ref{Mddef1'}).
In two dimensions the result above regularizes to
 %\marginpar{g1npert12}
%
\begin{equation}\label{g1npert12}
G_{1n} \cong \frac{ 1 }{ 8n^2 \ln[1/(a\kappa )]  },
\end{equation}
where  $a\sim 1$ is a nonuniversal factor.
For $d>2$ the values of $G_{1n}$ are no longer small, and the method used above fails.
At distances $\kappa n \gsim 1$ the integral in (\ref{g1npert}) is
dominated by $q\lsim\kappa$, where $\sigma_{nn}$ and $\alpha_{1n}$ no longer have
the forms  (\ref{signnbclo}) and  (\ref{al1nsurf}).
One can write $G_{1n}$ in the whole range of $\kappa n$ in the form
%\marginpar{g1npert1gen}
%
\begin{equation}\label{g1npert1gen}
G_{1n} \cong \frac{ \Gamma(d) }{ dM_d } \frac{ \kappa^2 g(\kappa n) }{ (2\kappa n)^d  } ,
\end{equation}
where $g(\kappa n)$ is a crossover function.
The expected asymptote of $G_{1n}$ for $\kappa n \gg 1$ is
%\marginpar{g1npert2}
%
\begin{equation}\label{g1npert2}
G_{1n} \sim \frac{ \kappa^2 }{ (\kappa n)^\zeta } e^{-2\kappa n},
\end{equation}
with some exponent $\zeta$.
The analytical calculation of the prefactor here seems to be very difficult.

The value of $G_{11}$ is fixed by the constraint equation (\ref{sconstrfd}) in
the first layer.
We will see below that $\sum_{n=1}^\infty G_{1n} \sim \kappa \ll
\kappa^{2-d}$,
thus  $G_{11}$ is simply given by 
%\marginpar{g11pert}
%
\begin{equation}\label{g11pert}
G_{11} \cong -\sum_{n=2}^\infty G_{1n} \cong - \frac{ \theta }{ 2 } 
\frac{d'}{\Lambda^{d'}}\int_0^\Lambda dq q^{d'-1}  \alpha^2 ,
\end{equation}
where (\ref{g1npert}) has been used.
Although the derivation above becomes invalid for $d$ close to 1, the resulting expression has a well defined form $G_{11}=-\kappa^{2-d}/2$ in this region, which will be confirmed below by another method and is in accord with the result $G_1\cong 2(1-\kappa)$ in one
dimension [see the discussion below (\ref{gn1d})].

\subsubsection{Correlation functions near the surface}
\label{CFsurf}

In low dimensions the quantities $G_{1n}$ are small, and one can try to find the corresponding corrections to the correlation functions perturbatively.
It is clear, however, that since  $G_{1n}$ can be responsible for the gap in
the correlation function [see (\ref{signnbc'})] the direct perturbative approach to $\sigma_{nn}$ can be inefficient.
It is more convenient to use perturbation theory with respect to the self-energy part $\Sigma_n$ in (\ref{sigrec1}).
For the variation of $G_{n}$ of the form (\ref{gnpert}) in the expression for the latter one has
%\marginpar{b1npert}
%
\begin{eqnarray}\label{b1npert}
&&
b_{11} \cong  \onehalf + dG_{11}
\nonumber\\
&&
b_{1n} \cong dG_{1n}, \qquad n\geq 2 ,
\end{eqnarray}
and the quantities $\alpha_{1n}$ and $\alpha_{1n}'$
can be found from the recurrence relations (\ref{al1rec}) and (\ref{al'1rec}).
These equations simplify, since $\alpha_{1n}$ for $n\geq 2$, as
well as all $\alpha_{1n}'$, are induced by $G_{1n}$ and small.
For the first layer, taking into account $\alpha_{11}=-\alpha$ one obtains in the long-wavelength region
%\marginpar{sig11pert}
%
\begin{equation}\label{sig11pert}
\sigma_{11} \cong \frac{2d\theta}{ \tilde q + \Delta_1(\tilde q, \kappa) }  , 
\end{equation}
where
%\marginpar{Del1}
%
\begin{equation}\label{Del1}
\Delta_1(\tilde q, \kappa) \cong -2d\sum_{l=1}^\infty e^{-2\tilde q} G_{1l} 
\end{equation}
and the dependence of $\Delta_1$ on $\kappa$ is due to that of $G_{1n}$.
The longitudinal CF $\sigma_{11}^{zz}$ is given by the same expressions with 
$\tilde q_z \equiv \sqrt{\kappa_z^2 + q^2}$ instead of $\tilde q\equiv \sqrt{\kappa^2 + q^2}$.

In the sum (\ref{Del1}) only the term with $G_{11}$ is unknown, and its value
follows from the constraint equation (\ref{sconstr1}), which can, with the
help of (\ref{cfhomo1}) and (\ref{plimsd}), be put into the form 
%\marginpar{constrDel1}
%
\begin{equation}\label{constrDel1}
\int_0^{\Lambda} dq q^{d'-1} 
\left[ \frac{ 2 }{ \tilde q + \Delta_1(\tilde q, \kappa) } - \frac{ 1 }{ \tilde q } \right]  = 0.
\end{equation}
For $1<d<2$ the integral here is dominated by $q\sim \kappa$ and its upper limit can be set to $\infty$.
This implies that $\Delta_1 \sim \kappa$, i.e., the individual terms of the sum (\ref{Del1}), each being of order $\theta\sim\kappa^{2-d} \gg \kappa$, are nearly cancelled. 
The ensuing expression for $G_{11}$ is given by (\ref{g11pert}) above. 

As a result of the cancellation of the leading terms in (\ref{Del1}), taking
the next terms into account may become necessary.
The $m$th-order terms in $G_{1n}$ are proportional to $\kappa^{m(2-d)}$ and they are small in comparison to $\kappa\ll 1$ for $d< 2-1/m$.
In particular, the first-order perturbation theory written above neglects the terms starting from $m=2$ and it is, in general, valid only for $d<1.5$.
The second-order perturbation theory works for $d<1.75$, etc.
The solution of the problem seems to undergo an infinite number of crossovers for $d$ approaching 2 and it should be rather complicated.
Analytical solutions can be obtained for $d$ close to 1 and for $q\gg\kappa$,
as well as in the marginal case $d=2$, with only logarithmic accuracy.

For $d'\equiv d-1 \ll 1$ one has $M_d \cong 1/(d-1)$, and the main contribution to the integral in (\ref{constrDel1}) stems from the region $q\ll\kappa$.
For $d=1$ the obvious solution of  equation (\ref{constrDel1}) is 
$\Delta_1(\kappa, \kappa) = \kappa$.
Since for $d=1$ all $G_{1n}$ with $n\geq 2$ disappear, this leads to the one-dimensional result $G_{11} = -\kappa/2$.
In the first order in $d-1$ one can still neglect the $q$ dependence of 
$\Delta_1(\tilde q, \kappa)$.
Then the perturbative solution of (\ref{constrDel1}) leads to the simple result
%\marginpar{Del1near1}
%
\begin{equation}\label{Del1near1}
\Delta_1(\kappa, \kappa) \cong d\kappa, \qquad d-1 \ll 1 .
\end{equation}
On the other hand, the sum (\ref{Del1}) near $d=1$ consists of two contributions:
%\marginpar{Del1near1'}
%
\begin{eqnarray}\label{Del1near1'}
&&
\Delta_1(\kappa, \kappa) \cong -2dG_{11} - \frac{ \kappa^{2-d} }{ M_d } 
\sum_{l=1}^{1/\kappa} \frac{ 1 }{ l^d }
\nonumber\\
&&\qquad
\cong -2dG_{11}  - \kappa^{2-d} ( 1 - \kappa^{d-1} ) ,
\end{eqnarray}
with logarithmic accuracy.
Comparing it with (\ref{Del1near1}) yields
%\marginpar{G11}
%
\begin{equation}\label{G11}
G_{11} \cong - \frac{ \kappa^{2-d} }{ 2 } 
\cong - \frac{ \kappa }{ 2 } \left[ 1 + (d-1)\ln \frac{ 1 }{ \kappa } \right] .
\end{equation}

 For the analysis of the limit $q\gg \kappa$ and of the behavior of the longitudinal CF $\sigma_{nn}^{zz}$ for small wave vectors, it is convenient to represent the quantity $\Delta_1(\tilde q, \kappa)$ above as
%\marginpar{Del1rep}
%
\begin{equation}\label{Del1rep}
\Delta_1(\tilde q, \kappa) = \Delta_1(0, \kappa) + \bar \Delta_1(\tilde q, \kappa) ,
\end{equation}
where
%\marginpar{barDel1}
%
\begin{eqnarray}\label{barDel1}
&&
\Delta_1(0, \kappa) \equiv -2d\sum_{l=1}^\infty G_{1l} 
\nonumber \\
&&
\bar \Delta_1(\tilde q, \kappa) \equiv 2d\sum_{l=1}^\infty (1 - e^{-2\tilde q l} ) G_{1l}  .
\end{eqnarray}
The quantity $\Delta_1(0, \kappa)$ determines the gap of $\sigma_{nn}^{zz}$ at criticality 
($\tilde q_z=0$) and it is related to $\Delta_1(\kappa, \kappa)$ studied above by
%\marginpar{Del1gap}
%
\begin{eqnarray}\label{Del1gap}
&&
\Delta_1(0, \kappa) = \Delta_1(\kappa, \kappa) - 2d\sum_{l=1}^\infty (1 - e^{-2\kappa l} ) G_{1l} 
\nonumber\\
&&\qquad
{} \cong \Delta_1(\kappa, \kappa) - \frac{ \kappa \Gamma(d) }{ 2^{d-1} M_d }
\int_0^\infty \frac{ dx }{ x^d } (1 - e^{-2x} ) g(x) ,
\end{eqnarray}
where the function $g(\kappa n)$ was introduced in (\ref{g1npert1gen}).
The integral term in this formula stems from the region $x\equiv \kappa n \sim 1$ where 
$g(\kappa n)$ is unknown.
Even near $d=1$, where $\Delta_1(\kappa, \kappa)$ is given by (\ref{Del1near1}) to the first order in $d-1$, calculation of this term needed to find $\Delta_1(0, \kappa)$ would require additional efforts.
In general, there is no apparent way to calculate analytically the gap in $\sigma_{nn}^{zz}$.
On the other hand, the existence of this gap can be anticipated, since the anisotropic model shows the mean-field critical behavior in all cases. 
At criticality for $q\ll 1$ the longitudinal CF can, after the expansion of (\ref{barDel1}),
be written in the form 
%\marginpar{sigzzcrit}
%
\begin{equation}\label{sigzzcrit}
\sigma_{nn}^{zz} \cong \frac{ 2d\theta }{ \Delta_1(0, \kappa) + A q } ,
\end{equation}
where the stiffness of the longitudinal spin fluctuations is given by
%\marginpar{Adef}
%
\begin{equation}\label{Adef}
A \cong 1 + \frac{ 2^{2-d} \Gamma(d) }{ M_d } \int_\kappa^\infty \frac{ dx }{ x^{d-1} } g(x) ,
\end{equation}
which is again determined by the region $\kappa n \sim 1$, in general.
For $d<2$ the lower limit of integration can be set to zero, and $A$ is a number.
For $d \to 1$ the quantity $M_d$ diverges, and $A$ tends to 1.
For $d\to 2$ the integral in (\ref{Adef}) diverges logarithmically at the lower limit, and this divergence compensates that of $M_d$ [see (\ref{Mddef1'})].
As a result, one obtains $A \cong 2 + O[1/\ln(1/\kappa)]$.
It should be stressed, however, that in fact $A$ cannot be calculated perturbatively, as above, if $d$ is not close to 1.
This is because the term of the first-order in $G_{1n}$ gives a contribution
comparable to the zeroth-order result, and so do the terms of all orders in $G_{1n}$. 
  
The analytically tractable case is $q\gg \kappa$, in which the sum in (\ref{barDel1}) is dominated 
by $l\sim 1/q \ll 1/\kappa$ and one can use (\ref{g1npert1}) for $G_{1n}$. 
Replacing this sum by the integral and combining it with (\ref{Mddef1}) yields the result
%\marginpar{barDel1qbig}
%
\begin{equation}\label{barDel1qbig}
\bar\Delta_1(q, \kappa) \cong 
\frac{ \Gamma\left(\frac{ d }{ 2 } \right) \Gamma\left(\frac{ 3-d }{ 2 } \right) }{ \pi^{1/2} }
\frac{ 1-q^{2-d} }{ 1-\kappa^{2-d} } (1-\kappa^{d-1} ) q \left( \frac \kappa q \right)^{2-d} .
\end{equation}
Here the factor $1-q^{2-d}$  reflects the logarithmic divergence of the integral at the lower
limit for $d$ close to 2, its counterpart $1-\kappa^{2-d}$ comes from $M_d$, and the factor $1-\kappa^{d-1}$ is due to the logarithmic divergence of the integral at the upper limit for $d$ close to 1.
For $d$ not close to 1 or 2 these factors can be dropped.
For $d=2$ the expression above regularizes to
%\marginpar{barDel1qbig2}
%
\begin{equation}\label{barDel1qbig2}
\bar\Delta_1(q, \kappa) \cong 
%q \frac{ \ln(1/q) }{ \ln(1/\kappa) } 
q  \ln(1/q) / \ln(1/\kappa)  
\end{equation}
with logarithmic accuracy.
One can see that the $q$-dependent term calculated above is smaller than the
leading term $q$ in the denominator of (\ref{sig11pert}) by a factor $(\kappa/q)^{2-d}\ll 1$, 
which justifies using the perturbation theory in $G_{1n}$.
This term depends on $\kappa$ and thus signifies the gap in the correlation function.
For $2-d \ll 1$, the latter has a long-tail character, which makes the perturbation
scheme slowly convergent.
This is in contrast to the bulk behavior, where this gap tail stems simply from
the expansion of $\tilde q =\sqrt{\kappa^2+q^2}$ and has the fast-decaying
form $q(\kappa/q)^2/2$.
The most drastic situation is realized for $d=2$, where the gap tail is
logarithmic and the applicability of the perturbation theory requires
fulfillment of the very difficult criterium $\ln(1/q) \ll \ln(1/\kappa)$.

One can improve the perturbation theory by taking into account the terms of the
second order in $G_{1n}$ in the denominator of (\ref{sig11pert}).
These terms have the form of double sums over the layers and for $q\gg \kappa$
they make a contribution of order $q (\kappa/q)^{2(2-d)}$ to $\bar\Delta_1(q,\kappa)$.
For $d=2$ the formulas simplify and one obtains the contribution $qR^2/2$,
where $R$ is the ratio of logarithms in (\ref{barDel1qbig2}).
In fact, in two dimensions one can sum up (with logarithmic accuracy) all the orders of the
perturbation theory in $G_{1n}$.
This is possible because the equation
(\ref{cfdiffeq}) with  $G_{1n}$ given by (\ref{g1npert12}) can be exactly
solved in terms of the modified Bessel functions.
The corresponding calculation will be presented below; here we discuss some further
features of the semi-infinite ASM for $d\leq 2$.

In {\em each order} of the perturbation theory, $\Delta_1(q,\kappa)$ can be
represented in the form (\ref{Del1rep}), where $\bar\Delta_1(q,\kappa)$ is
determined by $l \sim 1/\tilde q \gg 1$ and is thus universal.
The quantity  $\Delta_1(0,\kappa)$ is fixed by the constraint condition
(\ref{constrDel1}), where $q\sim \kappa \ll 1$, and it is thus universal, too.
The same can be shown for all values of $n$, i.e., the spin CFs in the
semi-infinite weakly anisotropic ASM are universal in the whole half-space for
$d<2$.
In other words, in this case the {\em strong scaling} is realized, which
manifests itself in the independence of the CFs of the
lattice spacing $a_0$.
Alternatively, this can be seen from the fact that the Green function equation (\ref{cfdiffeq})
with the boundary condition (\ref{bclo}) is applicable everywhere, because
$G_{1n}$ are small and $\sigma_{nn'}$ is a smooth function of $n$ in the long
wavelength region.
The nonuniversality of $G_{1n}$ in several boundary layers does not play a
role because $\sigma_{nn'}$ is sensitive only to the cumulative action of $G_{1n}$
from a large number of remote layers, $n\sim 1/\tilde q \gg 1$.
For $d=2$ there are nonuniversal {\em logarithmic} corrections to the
strong scaling.
For $d>2$, as we shall see, the scaling is realized only in the asymptotic
region $n\gg 1$.

The expressions for all $\sigma_{nn}$ can be obtained recurrently from $\sigma_{11}$ with the 
help of relation (\ref{sign'n'}), which results in 
%\marginpar{signnrec}
%
\begin{equation}\label{signnrec}
\sigma_{nn} = \frac{ \alpha'_{n-1}\alpha'_{n-2}\ldots\alpha'_1 }
{ \alpha_{n}\alpha_{n-1}\ldots\alpha_2 } \sigma_{11} .
\end{equation}
Here, for $q\sim\kappa$ all $\alpha_n$ and $\alpha'_n $, except for
$\alpha_1=0$, are close to unity.
Thus all CFs in the layers near the surface are close to each other.
At large distances  the spin CFs should cross over to the bulk result
(\ref{cfbulk})
[cf. (\ref{sigzzas})].
For $q \sim \kappa$, this crossover occurs at $\kappa n \sim 1$ and it cannot be described analytically, since the solution depends on the values of $G_{1n}$ in this region, which
have not been determined. 
For the wave vectors $q \gg \kappa$ the equation for $\sigma_{nn}$ can be
solved perturbatively in $G_{1n}$ for all distances.
This can be seen if one rewrites (\ref{cfdiffeq}) in terms of the
dimensionless variable $z\equiv q n$ and uses (\ref{g1npert1}) for $G_{1n}$.
Then the term with $G_{1n}$ in the equation becomes of order $(\kappa/q)^{2-d}/z^d$
and can be considered perturbatively for $d<2$.
The solution is given (\ref{signnbc'}) with $\tilde q \gg c$ plus a correction
term.
Near the surface, $z\equiv q n \ll 1$,  to the lowest
order in $z$ this solution can be put into the form (\ref{sig11pert}) with 
$\Delta_n(\tilde q,\kappa)\cong\bar\Delta_1(\tilde q,\kappa) $ given by (\ref{barDel1qbig}). 
For $z\equiv q n \gg 1$ one obtains
%\marginpar{signnyxbig}
%
\begin{equation}\label{signnyxbig}
\sigma_{nn} \cong \frac{ d\theta }{ q } 
\left[ 1 + \frac{ \Gamma(d) }{ 2^dM_d } \left( \frac{ \kappa }{ q } \right)^2 
\frac{ 1 }{ (\kappa n )^d } \right] , 
\end{equation}
which is in fact the expansion of 
$\sigma_{nn} \cong d\theta/\sqrt{\tilde q^2 - 2dG_{1n}}$ for large $q$.
The latter result has a simple interpretation: For $q \gg 1/n \gg \kappa$ the
surface term in $\sigma_{nn}$ is negligible, and the correction to the bulk
result is due to $G_{1n}$  from the narrow region $|n''-n| \ll 1/q$, where $G_{1n}$ does not 
change significantly (the local correction).

For the longitudinal CF $\sigma_{nn'}^{zz}$ the corresponding effective wave vector 
$\tilde q_z \equiv \sqrt{\kappa_z^2 + q^2}$ can be smaller than $\kappa$,
especially at criticality, where $\tilde q_z =q$.
For $\tilde q_z \ll \kappa$ the exponential decrease of $G_{1n}$ for
$\kappa n \gsim 1$ [see  (\ref{g1npert2})] comes into play.
As a result, in the range $n \gsim n^* \sim 1/\kappa$ the free solution
(\ref{signnbc'}) is realized again. 
The disappearance of  $G_{1n}$  for $n \sim 1/\kappa$ and the ensuing free
solution is also characteristic for dimensions $2 < d < 4$.
As we shall see below, for $d> 4$ the value of $n^*$ is of the order of the
lattice spacing, and the free solution is realized in a wider range.
In contrast to the case $q \gg \kappa$ considered above, here the sign of the
surface term in $\sigma_{nn'}^{zz}$ is negative.
One can argue, in general, that the form of the free solution in the region 
$n^* \lsim n \lsim 1/ \tilde q_z$
is (\ref{signnbc'}) with a coefficient $f(\tilde q_z/\kappa)$ in front of the
surface-induced term.
The plausible assumption about the form of $f$ is 
$f=(\tilde q_z - c_{\rm eff})/(\tilde q_z + c_{\rm eff})$, i.e., the surface
term of the CF changes sign as a function of the wave vector.
As a justification one can stress that from the distances $n\gg n^*$ the region
$n \lsim n^*$, where the form of the CF is complicated, is seen as narrow.
Thus one can replace this region with an effective boundary condition of the
type (\ref{bclo}) set at $n \sim n^*$.
The quantity $c_{\rm eff}$ can be expected to be of order $\kappa$  in
dimensions low enough.
This is the exact result for $d=1$ [see (\ref{sigzzas})] and it will be shown
numerically to hold for $d\leq 4$.
This implies that the extrapolation length $\lambda_e = 1/c_{\rm eff}$ is of
the order of the {\em transverse} correlation length $\xi_{c\alpha}$, which 
(although nondivergent) is
much greater than the lattice spacing for the weakly anisotropic systems near criticality. 
The important implication of the negative sign of the surface term in $\sigma_{nn'}^{zz}$
at small wave vectors is the gap in $\sigma_{nn'}^{zz}$ at any finite
distances from the surface, even at criticality.
It is clear, although difficult to prove rigorously, that the existence of this gap
in the asymptotic region, where the free solution is realized, also entails
the gap in  $\sigma_{nn'}^{zz}$ near the surface.

\subsubsection{Correlation functions for $d=2$}
\label{solutiond2}

Dimension $d=2$ is the marginal one between $d<2$, where the
characteristic wave vectors in the spin CFs are $q\sim \kappa$, and $d>2$, where
$q\sim 1/n$ are important.
The solution with logarithmic accuracy for $d=2$ can be obtained if these
ranges are separated by many decades, i.e., $\ln(1/\kappa) \gg \ln n$.
In fact, the solution  (\ref{g1npert12}) for $G_{1n}$ has been obtained under
this very restrictive condition. 
With this form of $G_{1n}$, the Green function equation (\ref{cfdiffeq}) can
be solved exactly in terms of the modified Bessel functions.
This will be shown in more detail in Sec. \ref{isocrit}, and the result has
the form (\ref{sigik1}) with
%\marginpar{mud2}
%
\begin{equation}\label{mud2}
\mu \cong \frac 12 \left[ -1 + \frac{ 1 }{ \ln(1/\kappa) } \right].
\end{equation}
For $qn\ll 1$ the expression for $\sigma_{nn}$ simplifies to
%\marginpar{signnd2}
%
\begin{equation}\label{signnd2}
\sigma_{nn} \cong \frac{ 2d\theta }{ q } (qn)^{1/\ln(1/\kappa)}
\end{equation}
[cf. (\ref{sigikas2})].
This can be represented in the form of type (\ref{sig11pert}) with 
$\Delta_n = q[1-(qn)^{-1/\ln(1/\kappa)}]$.
For not too small $q$ one can expand $\Delta_n$ in powers of $\ln[1/(qn)]/\ln(1/\kappa)$
to obtain the gap tail of the spin CF.
In this way the first-order result  (\ref{barDel1qbig2}) for $n=1$ and all
other orders of the perturbation theory in $G_{1n}$ are recovered.

At small wave vectors $q$, solution for $\sigma_{nn}$ is determined by $G_{1n}$
at $\kappa n \sim 1$ where the latter are unknown and hence $\sigma_{nn}$
does not have the form above.
In fact, here the gap $\Delta_n(0,\kappa)$ in $\sigma_{nn}$ manifests itself,
and it
turns out to be much larger than the bulk gap $\kappa$.
The dependence of $\Delta_n(0,\kappa)$ can be obtained from the
constraint equation  (\ref{sconstr1}), where (\ref{signnd2}) and (\ref{cfbulk}) are used 
and the integration over $q$ is performed between $\Delta_n(0,\kappa)$ and $1/n$.
In this way one comes to an interesting formula:
%\marginpar{gapd2}
%
\begin{equation}\label{gapd2}
\Delta_n(0,\kappa) \sim \kappa^{\ln 2}/n, \qquad  \ln n \ll \ln(1/\kappa) .
\end{equation}
Here the critical index ln2 results from the fact that $\sigma_{nn}$ of
(\ref{signnd2}) is about 
2 times greater than in the bulk; thus the gap in this region
should be correspondingly greater to satisfy the constraint equation. 
The coefficient in (\ref{gapd2}) cannot be determined in the logarithmic
aproximation; it should become universal for $n\gg 1$.
This method is rather rough and it cannot distinguish detween the transverse and
longitudinal correlation functions.
One can expect that they differ by a numerical factor,  as was confirmed by
numerical calculations, which have been done, however, in a range of
$\kappa$ not small enough to confirm formula  (\ref{gapd2}) itself.

\subsection{High dimensions, $d\geq 4$}
\label{solutiond4}

As we have seen above, for the weakly anisotropic ASM in low dimensions the
problem can be solved analytically for $\kappa n \ll 1$ due to the separation
of the $q$ ranges in the transverse CF $\sigma_{nn}$, as exemplified by (\ref{sigdif2}).
The surface-induced term is important for 
$q\sim 1/n \gg \kappa$, whereas the term induced by $G_{1n}$ dominates for $q\sim\kappa\ll 1$.
In high dimensions the separation of the $q$ ranges of both terms also takes place, although in a different form.
Whereas the $q$ range of the surface term remains $q\sim 1/n$ (for $\kappa n \ll 1$), 
the $G_{1n}$-induced one dominates in the range $q\sim 1$.
Thus the ranges separate at distances $n\gg 1$, where the problem can be solved analytically.
Henceforce in this subsection we will consider the close-to-criticality case $\kappa \ll 1$; 
otherwise, at distances $n\gg 1$ deviations from the bulk values will be too small 
[cf.  (\ref{gnhtse})].

The surface term of $\sigma_{nn}$ has in high dimensions the opposite sign, as compared to
the low-dimensional result (\ref{signnbclo}), and for $n\gg 1$ its form does not depend on the details
of the behavior in the region close to the surface.
As we will see below, for $d\geq 4$ the quantity $G_{1n}$ decays fast with $n$, and for the 
calculation of the surface term it can be neglected starting from some $n^*\gg 1$.
For $n\geq n^*$ one can use for $\sigma_{nn'}$ an expression of the type (\ref{signnbc}), 
where $2b_{11}$ is replaced by some quantity determined by the region $n\leq n^*$.
In contrast to the low-dimensional case, there is no reason for the substitute for $2b_{11}$ to 
be close to unity, because the variation of $G_n$ is no longer close to (\ref{gnpert}), or
in (\ref{gnpert}) $G_{1n}$ are no longer small.
At large distances and small wave vectors one can thus use (\ref{signnbc'})
where the coefficient $f(q)$ in front of the surface term is close to $-1$,
since the quantity $c_{\rm eff}$  should be of order unity (the extrapolation length
$\lambda_e\equiv 1/c_{\rm eff}$ comparable to the lattice spacing).

Since postulating $\sigma_{nn'}$ in the form (\ref{signnbc'}) is not quite a rigorous
procedure, let us consider another derivation based on the continued-fraction formalism
described in Sec. \ref{continuous}.
For $n\gg 1$ one can employ the differential equations (\ref{al1difeq}) for the
quantities $\alpha_{1n}$ and $\alpha'_{1n}$ of (\ref{al1def}).
The boundary condition for the second equation is $\alpha'_{1\infty}=0$; the quantity 
$\alpha'_{1n}$ is generated solely by $G_{1n}$ and it is not related to the surface term of 
$\sigma_{nn}$.
For the equation for $\alpha_{1n}$ the boundary condition cannot be set on the surface, since
this equation is invalid for $n\sim 1$.
Thus we use the boundary condition $\alpha_{1n}=\alpha_{1n^*}$ at $n=n^* \gg 1$, where
$\alpha_{1n^*}$ is determined by the exact recurrence formula (\ref{al1rec}) in the region 
$n\leq n^*$.
For $n>n^*$ one can neglect $2b_{1n}$ in the differential equation for $\alpha_{1n}$, after which
it can be linearized with respect to the new variable $1/\alpha_{1n}$ and solved to give
%\marginpar{al1nd4}
%
\begin{equation}\label{al1nd4}
\alpha_{1n} = \frac{  2\tilde q\alpha_{1n^*} }
{ \alpha_{1n^*} + (2\tilde q - \alpha_{1n^*}) \exp[2\tilde q (n-n^*)] } .
\end{equation}
Since $\alpha_{1n^*}$ is generated by the boundary condition at the surface and by $b_{1n}$,
which are not explicitly small for $n\leq n^*$, one has $|\alpha_{1n^*}| \gg \tilde q$ for small wave vectors.
It can be seen that $\alpha_{1n^*}$ cannot be positive, otherwise $\alpha_{1n}$ turns to 
infinity at $n$ determined by 
$\exp[2\tilde q (n-n^*)] = \alpha_{1n^*}/(\alpha_{1n^*}-2\tilde q) \cong 1$.
Thus, $\alpha_{1n^*}<0$, and in the relevant region $n \gg n^*$, $\tilde q n \sim 1$ (\ref{al1nd4})
simplifies to the form
%\marginpar{al1nd4as}
%
\begin{equation}\label{al1nd4as}
\alpha_{1n} \cong - \frac{ 2\tilde q }
{ \exp(2\tilde q n) - 1 } ,
\end{equation}
which is independent of the behavior in the boundary region, $n\leq n^*$.
For $\tilde q n \sim 1$ one has $\alpha_{1n}\sim 1/n$; thus in the differential equation 
(\ref{al1difeq}) $\tilde q \alpha_{1n} \sim \alpha_{1n}^2 \sim 1/n^2$.
This implies that the metod used here works if $G_{1n}\propto b_{1n}$ decays faster than
$1/n^2$.
As we shall see shortly, this is the case for $d\geq 4$. 
Now the surface term of $\sigma_{nn}$ can be found from (\ref{sigrec1}) and (\ref{deln}) with
$\alpha'_{1n} = b_{1n}=0$, which results in (\ref{signnbc'}) with $f(q)$ close
to $-1$.

For $q\sim 1$ and $n\gg 1$ the quantity $\alpha_{1n}$ of (\ref{al1nd4as}), as well as the surface
term in (\ref{signnbc'}), are exponentially small.
Here the $G_{1n}$-induced term becomes dominant.
To find the values of $\alpha_{1n}$ and $\alpha'_{1n}$ in this region, one can drop the small 
terms $\alpha_{1n}^2$ and $(\alpha'_{1n})^2$ in (\ref{al1difeq}), after which the linear
inhomogeneous differential equations can be solved.
The solution at point $n$ is induced by $b_{1n''}$ from the interval of $n''$ around $n$, which
satisfies $|n''-n|\sim 1/\tilde q \sim 1 \ll n$; thus one can treat $b_{1n}$ in (\ref{al1difeq}) as a 
constant.
The solution of (\ref{al1difeq}) has the form 
$\alpha_{1n} \cong \alpha'_{1n} \cong \alpha b_{1n}/\sqrt{b^2-1}$, and the quantity $\Delta_n$
of (\ref{deln}) reads $\Delta_n \cong - bb_{1n}/\sqrt{b^2-1}$.
The resulting correction to $\sigma_{nn}$ given by (\ref{sigrec1}) is of the form
%\marginpar{sigd4loc}
%
\begin{equation}\label{sigd4loc}
\delta\sigma_{nn}^{\rm local} \cong  d\theta \frac{ b b_{1n} }{ (b^2-1)^{3/2} }
= -b_{1n} \frac{ \partial }{ \partial b } \sigma_{nn}^{\rm bulk} ,
\end{equation}
where $\sigma_{nn}^{\rm bulk} $ is given by (\ref{sigbulk}) or by the first term of 
(\ref{signnbc'}).
This correction is due to the local deviation of $b_n$ from the bulk value $b$ [see (\ref{b1def})]
and it could in fact  be written for $q\sim 1$ without calculations.

Now the value of $b_{1n}$ can be found from the constraint equation (\ref{sconstr1}), where
$ \sigma_{nn}$ is the sum of (\ref{signnbc'}) and (\ref{sigd4loc}).
One can see that $b_{1n}>0$  is needed to compensate for the negative surface term.
The integration of the local term (\ref{sigd4loc}) extends for $d>4$ over the whole Brillouin zone, $q\sim 1$, and can be accomplished with the use of (\ref{cfhomo1}).
This results in
%\marginpar{derhomo}
%
\begin{equation}\label{derhomo}
\frac{d'}{\Lambda^{d'}}\int_0^\Lambda dq q^{d'-1}
\frac{ b b_{1n} }{ (b^2-1)^{3/2} } = \frac{ b_{1n} }{ d^2 } P(\eta G) I(\eta G) ,
\end{equation}
where $I(X)$ is defined by (\ref{ieta}).
Since for $d>4$ both  $P(X)$ and  $I(X)$ do not diverge for $X\to 1$, one can set
$X=1$ for weakly anisotropic ASM near criticality.
The integration of the surface term is cut at $q\sim 1/n \ll 1$ for $\kappa n \ll 1$ and at 
$q \sim q^*= \sqrt{\kappa/n}$ ($1/n \ll q^* \ll \kappa$) for $\kappa n \gg 1$.
The resulting $G_{1n} \cong b_{1n}/d$ has for $d>4$ and $n\gg 1$ the form
%\marginpar{g1nd4+}
%
\begin{equation}\label{g1nd4+}
G_{1n} \cong \frac{d'}{\Lambda^{d'}} \frac{ d \Gamma( \frac{d-1}{2} ) }{ P(1) I(1) } 
\frac{ \kappa^{d-2} }{ (\kappa n)^{\frac{d-2}{2}} } K_{\frac{d-2}{2}} (2\kappa n) ,
\end{equation}
where $K_\nu(x)$ is the Macdonald (modified Bessel) function.
For the hypercubic lattices the first fraction in (\ref{g1nd4+}) should be replaced according to
(\ref{contint}) by $S_{d'}/(2\pi)^{d'}$ [see (\ref{Addef})].
One can see that the form of $G_{1n}$ is nonuniversal. 
The limiting forms of (\ref{g1nd4+}) are
%\marginpar{g1nd4+1}
%
\begin{equation}\label{g1nd4+1}
G_{1n} \cong \frac{d'}{\Lambda^{d'}} \frac{ d \Gamma(d-2) }{ P(1) I(1) } 
\frac{ 1 }{ (2n)^{d-2} } , \qquad \kappa n \ll 1 ,
\end{equation}
and
%\marginpar{g1nd4+2}
%
\begin{equation}\label{g1nd4+2}
G_{1n} \cong \frac{d'}{\Lambda^{d'}} \frac{ d \Gamma( \frac{d-1}{2} ) }{ 2P(1) I(1) } 
\frac{ \kappa^{d-2} e^{-2\kappa n} }{ (\kappa n)^{\frac{d-1}{2}} } , \qquad \kappa n \gg 1 .
\end{equation}

If $d$ is close to the marginal value $d=4$ the contribution to the integral (\ref{derhomo}) from small wave
vectors becomes large, and the separation of the $q$ ranges in
$\sigma_{nn}$ no longer takes the place. 
Nevertheless, the problem can be solved analytically with logarithmic accuracy at 
distances $\kappa n \ll 1$.
In this case one should integrate in (\ref{derhomo}) down to $q\sim 1/n \gg \kappa$ where the surface
term in (\ref{signnbc'}) becomes important.
This leads to the replacement
%\marginpar{PIrepl}
%
\begin{equation}\label{PIrepl}
P(1) I(1)\Rightarrow \frac{d'}{\Lambda^{d'}} \frac{ 16 }{ d-4 } 
\left[ 1 - \frac{ 1 }{ (an)^{d-4} } \right]
\end{equation}
in (\ref{g1nd4+1}), $a$ being a lattice-dependent number.  
For $G_{1n}$ near $d=4$ one obtains the result
%\marginpar{g1ndab4}
%
\begin{equation}\label{g1ndab4}
G_{1n} \cong \frac{ 1 }{ 16 n^2 } \frac{ d-4}{ (an)^{d-4}-1 },
\end{equation}
which regularizes to
%\marginpar{g1nd4}
%
\begin{equation}\label{g1nd4}
G_{1n} \cong \frac{ 1 }{ 16 n^2  \ln (an)  } ,  \qquad 1 \ll n \ll 1/\kappa,
\end{equation}
in four dimensions.

It can be seen that the applicability condition of the method used here, 
$G_{1n} = o(1/n^2)$ for $n\gg 1$, is satisfied for $d\geq 4$.
In the range $2<d<4$ an attempt to apply (for $\kappa=0$) the same method
yields for the integral (\ref{derhomo}) a value of order $\sim b_{1n} n^{4-d}$
that stems
from the region $q\sim 1/n$.
The integral of the inhomogeneous term in (\ref{signnbc'}) is, for $\kappa=0$, determined by
the same range of $q$ and it is proportional to $n^{2-d}$.
Equating both contributions according to (\ref{sconstr1}) yields
$G_{1n} \propto b_{1n} \propto 1/n^2$ with some universal coefficient.
One can see that formula (\ref{g1ndab4}) shows such a behavior for $d<4$ where
the term containing the nonuniversal number $a$ is small. 
In fact, it joins smoothly the exact solution (\ref{g1nd24}) for $2<d<4$ found by Bray and Moore \cite{bramoo77prl77jpa}, which will be considered in the next section. 

The susceptibilities of the ASM in high dimensions show the mean-field critical behavior.
In particular, at small wave vectors one can drop the local contribution (\ref{sigd4loc}) and
use (\ref{signnbc'}) for $\sigma_{nn}$ at distances far enough from the surface, 
$n\gsim n^*$.
For $q=0$ and $\kappa n, \kappa_z n \ll 1$ both transverse and longitudinal CFs
[see (\ref{signnbc'})] simplify to (\ref{sigzzext}).
CFs in the surface region, $n\lsim n^*$, should be calculated numerically.
Since  $\sigma_{11}$ can be obtained from the CFs  $\sigma_{nn}$ far from the
surface with the help of relation (\ref{signnrec}) and the quantities
$\alpha_n$ and $\alpha'_n$ are nonsingular, $\sigma_{11}$ shows the
same critical behavior as in the asymptotic region $n\gg 1$.
The latter is characterized in particular by $\gamma_{11}=-\onehalf$, which can
be found by expanding  (\ref{signnbc'}) for $q=0$ up to the second order in 
$\tilde q =\kappa \ll 1$ and using $\kappa \propto \tau^{1/2}$ following from
(\ref{kapdef}) and (\ref{gtc}).

Finally, let us transform the CF $\sigma_{nn'}(q)$ to the real-space CF $\sigma_{nn'}(\rho)$,
where $\rho$ is the distance between two points in the direction parallel to the surface.
At large distances, $n,n',\rho\gg 1$ the relevant values of $q$ are small and one can use
(\ref{signnbc'}) with $\tilde q=q\ll c$, disregarding the local contribution (\ref{sigd4loc}).
The condition for this is $e^{qn} \ll (qn)^2 n^{d-4}$, which is satisfied for
$d>4$ and $n\gg 1$.
Then one comes to the MFA result, which, at isotropic criticality,
$\kappa=0$, has the form \cite{lubrub75mfa}
 %\marginpar{sigd4rho}
%
\begin{eqnarray}\label{sigd4rho}
&&
\sigma_{nn'}(\rho) \propto 
\left\{ \frac{ 1 }{ [\rho^2 + (n-n')^2]^{(d-2)/2} } \right.
\nonumber\\
&&
\left.
\qquad\qquad {} - \frac{ 1 }{ [\rho^2 + (n+n')^2]^{(d-2)/2} } \right\} 
\end{eqnarray}
with a nonuniversal factor depending on the lattice structure.
In this expression the surface-induced term with $n+n'$, which is similar to the
``image'' term  in electrostatic problems, modifies its asymptotes at $\rho\gg n,n'$ and
$n\gg\rho,n'$.
These are \cite{lubrub75mfa,bin83ptcp}
 %\marginpar{etpar}
%
\begin{equation}\label{etpar}
\sigma_{nn'}(\rho) \propto 1/\rho^{d-2+\eta_\|}, \qquad  \eta_\| = 2 
\end{equation}
for $n,n'=\rm const$ and $\rho\gg n,n'$ and 
 %\marginpar{etperp}
%
\begin{equation}\label{etperp}
\sigma_{nn'}(\rho) \propto 1/n^{d-2+\eta_\perp}, \qquad  \eta_\perp = 1 
\end{equation}
for $\rho,n'=\rm const$ and $n\gg\rho,n'$.
One can see that near the surface correlations decay faster than in the bulk ($\eta_b=0$),
especially in the direction parallel to the surface.

\section{Dimensions between two and four}
\label{middle}

\subsection{Isotropic model at criticality}
\label{isocrit}

As we have seen above, in low and high dimensions the correlation function $\sigma_{nn}$ 
consists of two different (surface and local) terms, which are dominant in different ranges of
$q$.
This property makes possible an analytical solution of the problem for $\kappa n\ll 1$ in low
dimensions and for $n\gg 1$ in high dimensions.
In the range $2<d<4$ both terms are the most important in the range $q\sim 1/n$,
i.e., they cannot be separated from each other.
Fortunately, the problem has an exact solution for the isotropic model at
criticality for $n\gg 1$
\cite{bramoo77prl77jpa}, where the 
anticipated asymptote of $G_{1n}$ far from the surface can be written as 
[see the discussion after (\ref{g1nd4})]
%\marginpar{g1nd24}
%
\begin{equation}\label{g1nd24}
G_{1n} = \frac{ \onefourth-\mu^2 }{ 2dn^2 } , \qquad \mu = \frac{ d-3 }{ 2 } ,
\end{equation}
where the choice of the parameter $\mu$ will be justified below.
For $n\gg 1$ and $q\ll 1$ one can use the second-order differential equation 
(\ref{cfdiffeq}) for the transverse CF, in which $\tilde q=q$ at isotropic criticality.
The latter can be solved in terms of the modified Bessel functions:
%\marginpar{sigik1}
%
\begin{eqnarray}\label{sigik1}
\renewcommand{\arraystretch}{2.3}
&&
\sigma_{nn'} =  2d\theta
\left\{ 
\begin{array}{ll}
\displaystyle
\sqrt{nn'} I_\mu(qn) K_\mu(qn'),       	& n\leq n'                           \\
\displaystyle
\sqrt{nn'} I_\mu(qn') K_\mu(qn),                      & n'\leq n ,                       \\
\end{array}
\right. 
\end{eqnarray}
where the term $C(nn')^{1/2} K_\mu(qn) K_\mu(qn')$ could also be added.
This solution looks similar to the full discrete solution (\ref{sigik}), but here the constant $C$
cannot be found from the boundary condition at the surface, since equation (\ref{cfdiffeq}) is
only valid for $n\gg 1$.

There is, however, another method of finding $\sigma_{nn'}$
\cite{bramoo77prl77jpa} that avoids using the boundary condition at the surface and yields $C=0$.
The consideration starts with the eigenvalue problem
%\marginpar{eigenpro}
%
\begin{equation}\label{eigenpro}
\left( \frac{d^2}{dn^2} + \frac{\onefourth - \mu^2}{n^2} \right) \psi(a,n) = - a^2 \psi(a,n) ,
\end{equation}
whose solution $\psi(a,n) = \sqrt{an} J_\mu(an)$ satisfies
%\marginpar{ortcomp}
%
\begin{eqnarray}\label{ortcomp}
&&
\int_0^\infty \!\! dn\, \psi(a,n) \psi(a',n) = \delta(a-a') ,
\nonumber\\
&&
\int_0^\infty \!\! da\, \psi(a,n) \psi(a,n') = \delta(n-n') 
\end{eqnarray}
and thus forms an orthogonal and complete basis on the semi-infinite interval.
Then the Green function $\sigma_{nn'}$ can be expressed through its decomposition over
the set of eigenfunctions as
%\marginpar{eigenexp}
%
\begin{equation}\label{eigenexp}
\sigma_{nn'} = 2d\theta \int_0^\infty \!\! da \,\frac{ \psi(a,n) \psi(a,n') }{ a^2 + q^2} ,
\end{equation}
which results in (\ref{sigik1}) without any additional terms.

Now the value of $\mu$ can be found from the constraint condition in the form
(\ref{sconstr1}) with the bulk CF given by (\ref{cfbulk}), i.e., 
%\marginpar{sconstr24}
%
\begin{equation}\label{sconstr24}
n\int_0^\infty \!\!dz \, z^{d-2} \left[ 2 I_\mu(z) K_\mu(z) - \frac 1z \right] = 0 .
\end{equation}
This integral which can be found in Ref. \cite{bramoo77prl77jpa}, is zero for 
all $n$, if $\mu$ is given by (\ref{g1nd24}) and $2<d<4$.
There is another solution, $\mu=(d-5)/2$ for $3<d<4$,
which leads to negative values of $G_{1n}$ and it should be 
disregarded for the ordinary phase transition considered here.
The asymptotic form of the layer autocorrelation function $\sigma_{nn}$ of (\ref{sigik1})
for $qn\gg 1$ is
%\marginpar{sigikas1}
%
\begin{equation}\label{sigicas1}
\sigma_{nn} \cong \frac{ d\theta }{ q } \left[ 1 + \frac{ \onefourth - \mu^2 }{ 2(qn)^2 } \right] .
\end{equation}
Here the first term is the bulk CF and the second term in the square brackets,
$dG_{1n}/q^2$, is the local contribution analogous to that in (\ref{signnyxbig}).
This form of $\sigma_{nn}$ is responsible for the convergence of the integral in
(\ref{sconstr24}) at $z\equiv qn\sim 1$, i.e., for $q\sim 1/n \ll 1$.
The latter justifies using the long-wavelength approximation in the scaling region,
$n\gg 1$.
Here the discrete lattice structure does not show up in the long-wavelength
behavior of $\sigma_{nn'}$ and in the form of $G_{1n}$, thus $1/n$ is the only scale for $q$.
In the opposite limit, $qn\ll 1$, one can use
%\marginpar{ikzsmall}
%
\begin{eqnarray}\label{ikzsmall}
&&
I_\mu(z) \cong  \frac{ 1 }{ \Gamma(1+\mu) } \left( \frac z2 \right)^\mu [1+O(z^2)],
\qquad z \ll 1
\nonumber\\
&&
K_\mu(z) = \frac{ \pi }{ 2\sin(\pi\mu) } [ I_{-\mu}(z) - I_\mu(z) ]
\end{eqnarray}
to express $\sigma_{nn}$ in the form
%\marginpar{sigikas2}
%
\begin{equation}\label{sigikas2}
\sigma_{nn} \cong \frac{ d\theta n }{ \mu } 
\left[ 1 - \frac{ \Gamma(1-\mu) }{ \Gamma(1+\mu) } 
\left( \frac{ qn }{ 2 }  \right)^{2\mu} \right] .
\end{equation}
For $d>3$ one has $\mu>0$ and $\sigma_{nn}$ does not diverge at $q\to 0$ 
for any finite $n$.
Conversely, for $d<3$ one has $\mu<0$, and thus the second singular term in 
(\ref{sigikas2}) is dominant and it causes the divergence of $\sigma_{nn}$ at small $q$.
In the marginal case $d=3$, (\ref{sigikas2}) regularizes to 
%\marginpar{sigikas3}
%
\begin{equation}\label{sigikas3}
\sigma_{nn} \cong 2d\theta n \left[ \ln\frac{1}{qn}  + c_0 \right], 
\qquad c_0= \ln 2 - \gamma ,
\end{equation}
where $\gamma=0.5772\ldots$ is the Euler constant and $c_0 \approx 0.1159$ is rather
small.

The Fourier transformed CF (\ref{sigik1}) looks very beautiful in real space 
\cite{bramoo77prl77jpa}:
%\marginpar{sigikrho}
%
\begin{eqnarray}\label{sigikrho}
&&
\sigma_{nn'}(\rho) = \frac{ 2d\theta\Gamma(d-2) }
{ (4\pi)^{(d-1)/2} \Gamma\left(\frac{d-1}{2}\right) }
\nonumber\\
&&
\qquad {}\times
\left[ \frac{ 1 }{ \rho^2 + (n-n')^2 } - \frac{ 1 }{ \rho^2 + (n+n')^2 } \right]^{(d-2)/2} .
\end{eqnarray}
Here, in contrast to the MFA result (\ref{sigd4rho}),  the bulk term and the surface-induced image term are {\em nonadditive}.
The critical exponents $\eta_\|$ and $\eta_\perp$ determined analogously to (\ref{etpar}) 
and (\ref{etperp}) are $\eta_\| = d-2$ and $\eta_\perp = (d-2)/2$ 
\cite{bramoo77prl77jpa}.

In spite of the apparent similarity of the solution presented here and that of 
Bray and Moore \cite{bramoo77prl77jpa}, they are not completely identical.
The difference is that in the spin vector model used here the constraint $|{\bf m}_i|=1$ on 
each lattice site is obeyed, which is accounted for in the constraint equation (\ref{sconstr}).
Bray and Moore used the phenomenological $\phi^4$ field-theoretical model with the $O(\infty)$ symmetry, which has no constraint on the field $\phi$.
Accordingly, the self-consistent determination of the function $V(z)$  
\cite{bramoo77prl77jpa}, which is analogous to $-G_{1n}$ here, is more
complicated and can be done only for the single $d$-dependent magic value of the coupling 
constant $u$.

A peculiar feature of the differential equation  (\ref{cfdiffeq}) is that its solution (\ref{sigik1}) 
is twofold: for a given $\mu^2$ in (\ref{g1nd24}) solutions with both signs of $\mu =\pm |\mu|$ can be realized for $d>3$ and $d<3$.
Accordingly, the eigenvalue problem (\ref{eigenpro}) has two sets of eigenfunctions that
form two different orthogonal and complete bases.
An apparent reason for such a behavior is the singularity of $G_{1n}$ at $n=0$ in the continuous aproach, which is, naturally, not present in the original discrete formulation of the problem.
This singularity and the concomitant loss of the boundary condition at the
surface could be circumvented by Bray and Moore with the eigenfunction trick above,
which looks like a miracle 
(see other examples from quantum mechanics in Ref. \cite{cas50}).
However, the fact that the same differential equation has different solutions, e.g., for $d=2.5$ and 
$d=3.5$, contradicts the common sense and,  more importantly, 
impedes numerical solution of this equation.  
The latter would be the only possibility in situations where no general
analytical solution is available, as in the off-criticality case; in that case the loss of the boundary condition creates insurmountable
difficulties.

The key to the paradox is that in the original discrete formulation there is no singularity of 
 $G_{1n}$, and the values of $2dG_{1n}$ are {\em different} for $d>3$ and $d<3$, although they 
may coincide in the scaling region, $n\gg 1$.
The (rather essential) difference between the CFs $\sigma_{nn'}$ for $d>3$ and $d<3$ 
stems entirely from the nonscaling region, $n\sim 1$, which is not amenable to the 
field-theoretical methods.
This is most pronounced in the limits $d\to 2$ and $d\to 4$, where $G_{1n}$ of (\ref{g1nd24})
tends to zero in the scaling region but $\sigma_{nn}$ of (\ref{sigik1}) remains well defined and given by
%\marginpar{siglims}
%
\begin{equation}\label{siglims}
\sigma_{nn} \cong \frac{ d\theta }{ q }  (1 \pm e^{-2qn} ), \qquad d \to \left\{ 2 \atop 4 \right\}.
\end{equation}
The latter is nothing more than the particular forms of (\ref{signnbc'}),
the difference between the two expressions being {\em completely} determined 
by the nonscaling region near the surface.
For $d\to 2$, the parameter $c$ in (\ref{signnbc'}) disappears with $\kappa$
in the isotropic limit, and the coefficient in front of the surface terms is $f=1$.
For $d\to 4$, the parameter $c$ is of order unity, and for $q\ll 1$ the
surface term is negative, $f\cong -1$.
Thus the isotropic-criticality solution of Bray and Moore smoothly joins the
solutions obtained for $d\leq 2$ and $d\geq 4$.

To close this subsection let us look at how the continued-fraction formalism of 
Sec. \ref{continuous} works at isotropic criticality.
Here in $\sigma_{nn}$ is given by (\ref{sigrec1}), where $\sqrt{b^2-1}\cong q$ in the long-wavelength region and $b_{1n}\cong dG_{1n} \sim 1/n^2$ can be neglected for $n\gg 1$.
The quantities $\alpha_{1n}$ and $\alpha'_{1n}$ can be found from the first-order nonlinear
differential equations (\ref{al1difeq}) with $\tilde q=q$.
The latter can be reduced to the second-order linear differential equations and solved to give
%\marginpar{al1sol}
%
\begin{eqnarray}\label{al1sol}
&&
\alpha_{1n} = - \frac{ d }{dn} \ln
\left\{ e^{-qn} \sqrt{qn} \left[ I_\mu(qn) + CK_\mu(qn) \right] \right\},
\nonumber\\
&&
\alpha'_{1n} = \frac{ d }{dn} \ln
\left\{ e^{qn} \sqrt{qn} \left[ K_\mu(qn)  +C'I_\mu(qn) \right] \right\} .
\end{eqnarray}
Here the integration constant $C'$ should be set to zero, since $\alpha'_{1n}$ vanishes at
infinity.
The constant $C$ remains undefined due to the loss of the boundary
condition at the surface for the equation for $\alpha_{1n}$.
Adopting these results in (\ref{sigrec1}) and using 
$I_\mu(z)K'_\mu(z)-I'_\mu(z)K_\mu(z)=-1/z$ leads to the previously obtained expression
(\ref{sigik1}) for $n'=n$, with the same additional term containing $C$.
Thus one should set $C=0$ in (\ref{al1sol}).
Then from (\ref{ikzsmall}) it can be seen that $\alpha_{1n}$ is nonsingular at $qn\ll 1$ and 
the singular terms in $\sigma_{nn}$ [see (\ref{sigikas2})] are due solely to $\alpha'_{1n}$.
Another way of obtaining (\ref{al1sol}) is to use the definitions (\ref{al1def}) and (\ref{aldef}) 
to identify ${\cal I}_n=\sqrt{n}I_\mu(qn)$ and ${\cal K}_n=\sqrt{n}K_\mu(qn)$.
The limiting forms of $\alpha_{1n}$ and $\alpha'_{1n}$ are
%\marginpar{al1lim}
%
\begin{equation}\label{al1lim}
\renewcommand{\arraystretch}{1.5}
 \alpha_{1n} \cong \frac 1n \times
\left\{
\begin{array}{ll}
\displaystyle
-\onehalf -\mu,                      & qn\ll 1                          \\
\displaystyle
\frac{ \onefourth-\mu^2 }{2qn}\left[1+\frac{ 1 }{ qn } \right],	 & qn\gg 1 
\end{array}
\right. 
\end{equation}
and
%\marginpar{al'1lim}
%
\begin{equation}\label{al'1lim}
\renewcommand{\arraystretch}{1.5}
 \alpha'_{1n} \cong \frac 1n \times
\left\{
\begin{array}{ll}
\displaystyle
\onehalf-|\mu|,                      & qn\ll 1                          \\
\displaystyle
\frac{ \onefourth-\mu^2 }{2qn}\left[1-\frac{ 1 }{ qn } \right],	 & qn\gg 1 .
\end{array}
\right. 
\end{equation}
The results above will be used in the numerical solution of the semi-infinite ASM at the 
isotropic criticality.

\subsection{Away from anisotropic criticality}
\label{off}

The transverse correlation function $\sigma_{nn'}$ behaves similarly for the isotropic model
slightly above $\theta_c$ and for the weakly anisotropic model at or slightly above $\theta_c$.
In both cases the behavior of $\sigma_{nn'}$ is modified in comparison to that at isotropic criticality due to the finiteness of the transverse correlation length 
$\xi_{c\alpha}=1/\kappa$, where $\kappa$ is given by (\ref{kapdef}).
For $0 < \kappa \ll 1$ the function $G_{1n}$ has in the scaling region $n\gg 1$ the form 
generalizing (\ref{g1nd24}):
%\marginpar{g1noff}
%
\begin{equation}\label{g1noff}
G_{1n} = \frac{ \onefourth-\mu^2 }{ 2dn^2 } g(\kappa n) .
\end{equation}
For $\kappa n \gg 1$ one can expect, as is the case in other dimensions 
[see, e.g., (\ref{g1nd4+2})], $g(\kappa n) \propto e^{-2\kappa n}$ with some $n$-dependent prefactor.
Analytical calculation of this prefactor seems to be impossible.
In the opposite limit $g$ can be written in the form
%\marginpar{gxsmall}
%
\begin{equation}\label{gxsmall}
g(\kappa n) \cong 1 - a_d (\kappa n)^r, \qquad  \kappa n \ll 1,
\end{equation}
with $r>0$ and $a_d\sim 1$.
There is no guess about the concrete form of $g(\kappa n)$ in the intermediate
region and, moreover, even if $g(\kappa n)$ is known, one would not be able to find a general analytical solution
for the differential equation  (\ref{cfdiffeq}).
For the field-theoretical model \cite{bramoo77prl77jpa} the question of
how to generalize the choice of the coupling constant $u$ for $\kappa\ne 0$
[be it the magic value $u^*(d)$ or something else] further complicates 
the problem and makes it quite intractable. 
For the ASM, however, the situation is not so hopeless: Some features of the off-criticality
behavior can be studied analytically using available small parameters; its general properties
are well described by the scaling, and the rest can be done numerically.

\subsubsection{Scaling form of correlation functions}
\label{scalingform}

It is convenient to start the consideration with the
longitudinal correlation function $\sigma_{nn'}^{zz}$.
The latter satisfies in the scaling region $n\gg 1$ equation (\ref{cfdiffeq}) with 
$\tilde q^2 \Rightarrow \kappa_z^2+q^2$ and $G_{1n}$ 
given by (\ref{g1noff}).
In the generic case of the anisotropic criticality the longitudinal
correlation length $\xi_{cz}$ turns to infinity, and one has 
$\kappa_z\equiv 1/\xi_{cz}=0$ in  (\ref{cfdiffeq}).
In this case, for $n=n'$ there are only three length parameters
entering into the longitudinal CF: $n$, $1/q$, and the transverse correlation
length $\xi_{c\alpha} \equiv 1/\kappa$.
Thus $\sigma_{nn}^{zz}$ can be written at $\theta_c$ in the two-parameter scaling form
%\marginpar{signnsc}
%
\begin{equation}\label{signnsc}
\sigma_{nn,\theta_c}^{zz}(\kappa,q) = \frac{d\theta}{\kappa} \Phi(x,y),
\qquad
x \equiv \kappa n , \qquad y \equiv \frac q\kappa .
\end{equation}
Away from criticality one more length parameter, $\xi_{cz}$, appears, but it
does not complicate the problem.
As can be seen from  (\ref{cfdiffeq}),  $\sigma_{nn}^{zz}(\kappa,\kappa_z,q)$ 
can be represented
in the same form with $y \Rightarrow \sqrt{(\kappa_z/\kappa)^2+y^2}$.
Similarly, the transverse CF $\sigma_{nn}(\kappa,q)$ is given by (\ref{signnsc}) with 
$y \Rightarrow \sqrt{1+y^2}$.
For the isotropic model or above the anisotropic crossover temperature $\tau^*$
[see the discussion following (\ref{kapzdef})] the longitudinal CF coincides with
the transverse CF. 

It should be stressed that in the ASM the transverse correlation length plays
the main role, whereas the longitudinal one, which does not  enter into the ASM equations 
(\ref{cffd})--(\ref{sconstrfd}), is a subordinate quantity.
This feature, which should be to some extent shared by the weakly anisotropic  
classical Heisenberg model, provides a contrast to the usual scaling scheme using
the diverging $\xi_{cz}$ as the main scaling parameter
(see, e.g., Refs. \cite{bin83ptcp,bar83ptcp}).

Let us now study the limiting forms of the scaling function $\Phi(x,y)$.
The bulk limit $\Phi^{\rm bulk}(x,y)=1/y$ is clearly realized for $z \equiv xy=qn \gg 1$.
The isotropic criticality limit $\Phi^{\rm isocrit}(x,y)=2xI_\mu(xy)K_\mu(xy)$
studied above is achieved if $y\gg 1$ provided that $x\ll 1$.
Both of these conditions imply that $1/\kappa$ becomes greater than other
length scales.
For $x\gg 1$ the quantities $G_{1n}$ become exponentially small, and in the
long-wavelength region $\sigma_{nn,\theta_c}^{zz}$ is given by
(\ref{signnbc'}), as in high dimensions.
This implies the scaling function
%\marginpar{phixbig}
%
\begin{equation}\label{phixbig}
\Phi^{x\gg 1}(x,y) \cong [1-e^{-2(x+x_e)y}]/y,
\end{equation}
where $x_e \equiv \kappa \lambda_e \sim 1$ is the scaled extrapolation length.
This expression could also be written in the form of the type 
(\ref{signnbc'}), which makes no difference in the relevant region $y\ll 1$.
One can see that the longitudinal CF at criticality does not diverge for $q\to 0$
at any finite distances from the surface, as in the MFA.
We have seen above that $x_e=1$ for $d=1$ and $x_e\cong 0$ (in fact, $\lambda_e \sim 1$)
for $d>4$.
Actually, for all values of $x$ one has 
%\marginpar{Phid14}
%
\begin{equation}\label{Phid14}
\renewcommand{\arraystretch}{1.2}
\Phi(x,0) =
\left\{
\begin{array}{lll}
2(1+x),                      & d = 1                        \\
2x,		  & d > 4	,
\end{array}
\right. 
\end{equation}
and the curves $\Phi(x,0)$ for all $d$ should go between these straight lines.
For $x\ll 1$ in the wide range $y\ll 1/x$ one can write [cf. (\ref{sigikas2})]
%\marginpar{yscale}
%
\begin{equation}\label{yscale}
 \Phi(x,y)\cong \frac{ x }{ \mu } 
\left[ 1 - \frac{ \Gamma(1-\mu) }{ \Gamma(1+\mu) } 
\left( \frac{ xy }{ 2 }  \right)^{2\mu} F(y) \right] ,
\end{equation}
where $\mu = (d-3)/2$ and the scaling function $F(y)$ describes the crossover from
the zero $q$ to the isotropic criticality limit at $y\sim 1$ and satisfies
$F(\infty) =1$, as well as $F(y) \sim y^{-2\mu}$ for $y\ll 1$. 
The latter requirement serves to kill the singularity in $q$ for $\kappa\neq 0$; as a result 
$\sigma_{nn}(\kappa,0)$ behaves similarly to $\sigma_{nn}(0,q)$.
For $2<d<3$ one can simply use
%\marginpar{yscale2}
%
\begin{equation}\label{yscale2}
F(y) \cong  \Phi(x\ll 1,y)/ \Phi^{\rm isocrit}(x\ll 1,y) 
\end{equation}
to find $F(y)$.
The function $\Phi(x,0)$ shows a crossover from $\Phi(x,0)\sim x^{{\rm min} (d-2,1)}$ 
for $x\ll 1$ to $\Phi(x,0) = 2(x+x_e)$ for $x\gg 1$ [see (\ref{phixbig})].

One can see that only the second, singular, term in (\ref{yscale}) makes a  $\kappa$-dependent contribution to $\sigma_{nn}^{zz}$ in the limit $q\to 0$.
Specifically, for $2<d<4$ one has
%\marginpar{kapsc}
%
\begin{equation}\label{kapsc}
\chi_{znn}^{\rm sing} \propto  \sigma_{nn}^{zz,\rm sing}(\kappa,0)
 \propto n^{d-2} \kappa^{d-3},
\end{equation}
which in addition shows that the susceptibility increases away from the
surface, as it should.
In the isotropic case from (\ref{kapdef}) and (\ref{gtc}) it follows that $\kappa\sim \tau^{1/(d-2)}$,
thus $\sigma_{nn}^{zz,\rm sing}(\kappa,0) \sim \tau^{-\gamma_{11}}$ with
$\gamma_{11}=(3-d)/(d-2)$ \cite{bin83ptcp}.
This result means that the surface susceptibility with respect to the surface field
diverges in the ASM only for $d\leq 3$.
The leading terms of $\chi_{znn}$ near the surface are given by
%\marginpar{chizlims}
%
\begin{equation}\label{chizlims}
\renewcommand{\arraystretch}{1.2}
\chi_{znn} \sim
\left\{
\begin{array}{lll}
n,                      & d > 3                        \\
n\ln[1/(\kappa n)],		  & d =3	\\
n^{d-2} \kappa^{d-3},		  & 2< d < 3	\\
\kappa^{-1}, 			& 1\leq d <2 .
\end{array}
\right. 
\end{equation}
For the isotropic systems in $1\leq d \leq 2$ the bulk transition temperature
is zero.
With respect to the latter, $\chi_{znn}$ shows the critical behavior 
$\chi_{znn}\sim \theta^{-\gamma_{11}}$ with $\gamma_{11} = 1/(2-d)$
[see  (\ref{kapdef}) and (\ref{gtc})].
This result is complementary to that for $2<d<4$ quoted above, and it shows
 similar divergence with approaching $d=2$ from the other side. 
In the anisotropic case this low-dimensional critical behavior is realized in
the range $\tau \gg \tau^*$, where $\tau^*$ is given by (\ref{taustar}).
In the vicinity of $\theta_c$, i.e., $\tau \ll \tau^*$, the mean-field
critical behavior with $\gamma_{11} = -\onehalf$ is observed.

It should be stressed that the critical amplitudes in the nonscaling region
near the surface, $n\sim 1$, cannot be found in the continuous approximation.
Here one should numerically solve the ASM equations on the lattice.
On the other hand, it can be shown that the critical indices remain unchanged
in the nonscaling region.
The CFs in this region can be obtained from those in the region $1\ll n \ll 1/\kappa$   
with the help of the formulas of type (\ref{signnrec}).
Since the quantities $\alpha_n$ and $\alpha_n'$ are all nonsingular,
$\sigma_{11}^{zz}$ differs from the result of the continuous approximation
extrapolated to $n=1$ by a numerical factor only.

\subsubsection{The gap tail of the scaling function $F(y)$}

It turns out that the form (\ref{gxsmall}) of the function $g(x)$ for $x\ll 1$ determines
the asymptote of the scaling function $F(y)$ of (\ref{yscale}) for $y\gg 1$, and in the region
$x\ll 1$, $y\gg 1$ everything can be calculated analytically. 
For $y\gg 1$ the solution $\sigma_{nn}$ of equation (\ref{cfdiffeq}) at a
point $n\ll 1/\kappa$ 
stems from the interval $|n-n''|\sim 1/q \ll 1/\kappa$ around $n$; thus one can use $g(\kappa n)$ in the 
form (\ref{gxsmall}) and calculate the correction to $\sigma_{nn}$ perturbatively in 
$a_d(\kappa n)^r$.
The resulting expression for the scaling function $\Phi(x,y)$ of (\ref{signnsc}) has the 
form
%\marginpar{sigoff}
%
\begin{equation}\label{sigoff}
\Phi(x,y) \cong 2x [ I_\mu(z) K_\mu(z) - Q \Xi_\mu(z) ] ,
\nonumber\\
\end{equation}
where $z\equiv xy = qn$, the first term corresponds to isotropic criticality,
%\marginpar{qdef}
%
\begin{equation}\label{qdef}
Q \equiv a_d (\onefourth-\mu^2) / y^r \ll 1 ,
\end{equation}
and the function $\Xi_\mu(z)$ reads
%\marginpar{xicap}
%
\begin{eqnarray}\label{xicap}
&&
\Xi_\mu(z) =  
K_\mu^2(z) \!\! \int_{z_0}^z \!\! dt \, t^{r-1} [ I_\mu^2(t) - c_\mu^2 t^{2\mu} ]
\nonumber\\
&&
{}+ K_\mu^2(z) \frac{ c_\mu^2 z^{r+2\mu} }{ r+2\mu } + I_\mu^2(z) \!\! \int_z^\infty \!\! dt \, t^{r-1} K_\mu^2(t), 
\end{eqnarray}
with $z_0\ll 1$, and $ c_\mu \equiv [ 2^\mu \Gamma(1+\mu) ]^{-1}$ is a factor
from (\ref{ikzsmall}).
The part of the expression above without the terms containing $c_\mu$ is just what
one obtains from  the straightforward perturbative scheme using the Green
function (\ref{sigik1}).
The additional terms with $c_\mu$ in (\ref{xicap}) can be rewritten in the
form  $K_\mu^2(z) c_\mu^2 z_0^{r+2\mu} / (r+2\mu) = CK_\mu^2(z)$, i.e., they
can always be added to the solution and their amplitude should be fixed from
the boundary condition at the surface.
Since this boundary condition is lost in the continuous approximation, the
exact form of these terms in (\ref{xicap}) has been chosen above from the requirement
that the term  $CK_\mu^2(z)$, which was ruled out above with the help of the
eigenfunction trick, does not appear again in the resulting expression for $\Phi(x,y)$.  
With such a choice one can set $z_0=0$, because the first integral in (\ref{xicap}) converges
at the lower limit.
Now one can see that the terms with $c_\mu$ cancel each other, if $ r+2\mu >0$, whereas 
in the opposite case they do not.
For $ \mu <0$ (i.e., $d<3$) the function $\Xi_\mu(z)$ can be rearranged as
%\marginpar{xicap1}
%
\begin{equation}\label{xicap1}
\Xi_\mu(z) = [2\sin(\pi\mu)/\pi]^2 \bar K K_\mu^2(z)  + \tilde \Xi_\mu(z) ,
\end{equation}
where
%\marginpar{xicaptil}
%
\begin{eqnarray}\label{xicaptil}
&&
\tilde\Xi_\mu(z) =  
K_\mu^2(z) \!\! \int_{0}^z \!\! dt \, t^{r-1} \tilde I_\mu^2(t)
+ \tilde I_\mu^2(z) \!\! \int_z^\infty \!\! dt \, t^{r-1} K_\mu^2(t), 
\nonumber\\
&&
\tilde I_\mu^2(z) \equiv I_\mu^2(z) - [2\sin(\pi\mu)/\pi]^2 K_\mu^2(z)
\end{eqnarray}
and
%\marginpar{wd}
%
\begin{eqnarray}\label{wd}
&&
\bar K = \int_0^\infty \!\! dt \, t^{r-1} 
\left\{ 
K_\mu^2(t) - \left[\frac{ \pi }{ 2\sin(\pi\mu) } \right]^2 c_\mu^2 t^{2\mu}  
\right\} 
\nonumber\\
&&\qquad
{} = 2^{r-3} \Gamma(r/2+\mu) \Gamma(r/2-\mu) \Gamma^2(r/2) \Gamma^{-1}(r) .
\end{eqnarray}
In (\ref{wd}) the subtraction term with $c_\mu$ is present only for $ r+2\mu <0$; the
resulting expression is valid for both signs of $ r+2\mu $.
The representation of $\Xi_\mu(z)$ in the form (\ref{xicap1}) for  $ \mu<0$ is
convenient because of the cancellation of the divergence at $t\to 0$ terms in
$\tilde I_\mu^2(t)$.
For  $\mu \geq 0$ (i.e., $d\geq 3$) expression (\ref{xicap1}) remains valid as well,
although the subtraction makes little sense and  $\Xi_\mu(z)$ can simply be written
in the form (\ref{xicap}) with $z_0=c_\mu=0$.

The parameters $r$ and $a_d$ in (\ref{gxsmall}) should be chosen self-consistently to satisfy the spin-constraint condition. 
Here it is convenient to subtract equations (\ref{sconstr1}) at and away from 
 isotropic criticality from each other.
Thus one can write
%\marginpar{kapcon}
%
\begin{equation}\label{kapcon}
\int_0^\infty \!\! dy \, y^{d'-1} [ \Phi^{\rm isocrit}(x,y) - \Phi(x,y) ] = M_d,
\end{equation}
where $M_d$ is given by (\ref{Mddef2}).
The integral on the left-hand side of (\ref{kapcon}) is determined by $z=xy\sim 1$, i.e., 
$y\sim 1/x \gg 1$, which justifies the approximations made above.
With the use of (\ref{sigoff}) and (\ref{qdef}) one can rewrite (\ref{kapcon}) in the form
%\marginpar{kapcon1}
%
\begin{equation}\label{kapcon1}
\frac{ 2a_d(\onefourth-\mu^2) }{ x^{d-2-r} }
\!\! \int_0^\infty \!\! dz \, z^{d-2-r} \Xi_\mu(z) = M_d .
\end{equation}
This equation should be satisfied for all values of $x=\kappa n$, thus 
%\marginpar{rres}
%
\begin{equation}\label{rres}
r=d-2.
\end{equation}
Then (\ref{kapcon1}) fixes the value of $a_d$:
%\marginpar{adres}
%
\begin{equation}\label{adres}
a_d = \frac{ 2M_d }{ (d-2)(4-d) }\,  \bar \Xi ^{-1},
\qquad \bar\Xi \equiv \!\! \int_0^\infty \!\! dz \,  \Xi(z) .
\end{equation}

The scaling function $F(y)$ in (\ref{yscale}) can now be identified taking 
the limit $z\ll 1$ in (\ref{sigoff}).
This leads to $F(y)\cong 1+2Q\bar K\sin(\pi\mu)/\pi$ in the whole interval $2<d<4$.
The latter can with the use of (\ref{qdef})  be rewritten as
%\marginpar{fysc}
%
\begin{equation}\label{fysc}
F(y) \cong 1 + A_d/y^r, \qquad y\gg 1,
\end{equation}
where
%\marginpar{Adres}
%
\begin{equation}\label{Adres}
A_d = \frac{ \pi^{1/2}\tan(\pi\mu) \Gamma(d-5/2) }{ 4(d-2)\bar\Xi }.
\end{equation}
A remarkable feature of (\ref{fysc}) is that the tail of $F(y)$ is, for $d<4$,
anomalously long compared to that in the bulk, 
$F^{\rm bulk}(y) = y^2/\sqrt{1+y^2} \cong 1 - 1/(2y^2)$.
The sign of $A_d$ is determined by $\mu=(d-3)/2$, and one has  $A_d=0$ for $d=3$.
This is in accord with the structure of (\ref{yscale}); in all cases
$\Phi(x,y)$ is smaller than $\Phi^{\rm isocrit}(x,y)$, as it should.
In the limit $d\to 4$ the integral $\bar\Xi$ of (\ref{adres}) diverges at the
upper limit and $A_d$ regularizes to $A_4=\oneeights$.
For $d\to \fivehalf$ the quantity $\bar K$ given by (\ref{wd}) diverges, and thus
one can neglect $\tilde \Xi(z)$ in (\ref{xicap1}).
In this limit $A_d$ regularizes to $A_{5/2}= - 4\pi^{1/2}/\Gamma^2(1/4) \approx -0.539$.
The same situation takes place for $d\to 2$, where one obtains $A_2=-\onehalf$.
It should be noted, however, that for $d$ close to 2 the tail of $F(y)$ becomes extremely
long [see (\ref{rres})]. 
The validity of the present approximation for $F(y)$ requires, for $d\to 2$,
very large values of $y$,
which can become incompatible with the condition $z\equiv xy \ll 1$ needed to
represent $\Phi(x,y)$ in the form (\ref{yscale}). 
Actually, $d=2$ is a special case with a logarithmically decaying gap tail
(see Sec. \ref{solutiond2}).

The quantity $a_d$ given by (\ref{adres}) is positive for $\fivehalf<d<4$ and
negative for $2<d<\fivehalf$.
At $d=\fivehalf$ one has $a_d=0$ due to the divergence of $\bar K$ and hence $\bar \Xi$.
The latter could raise questions about the validity of the perturbation theory
with respect to $a_d (\kappa n)^r$ for $d=\fivehalf$ [should the higher-order
terms in (\ref{gxsmall}) be taken into account?].
As we will see below, the numerical results are in excellent agreement with the
asymptotic behavior $F(y) \cong 1 - 0.539/y^{1/2}$ for $d=\fivehalf$ and $y\gg 1$.

\begin{figure}[t]
\unitlength1cm
\begin{picture}(11,7)
\centerline{\epsfig{file=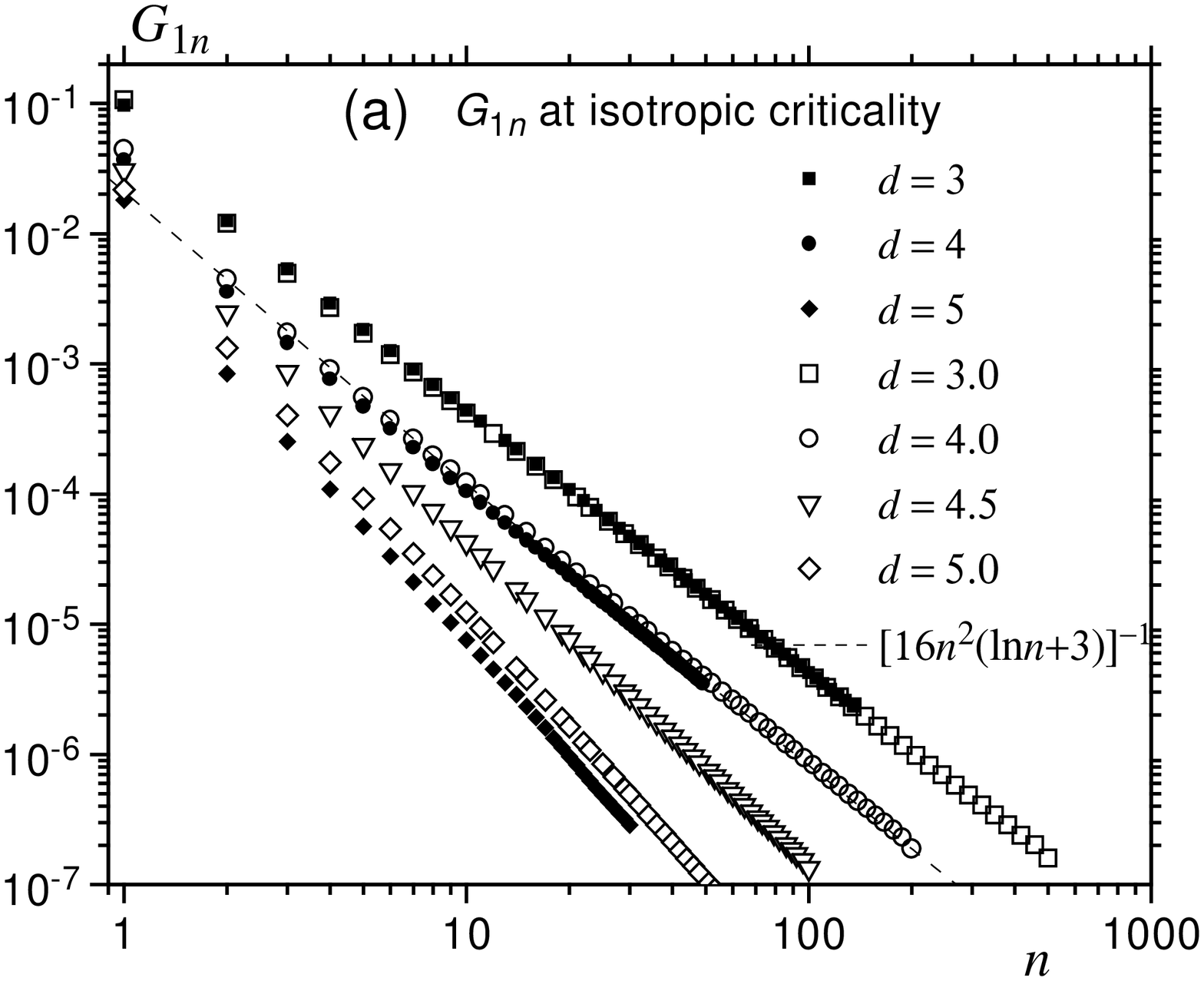,angle=-90,width=12cm}}
\end{picture}
\begin{picture}(11,6.5)
\centerline{\epsfig{file=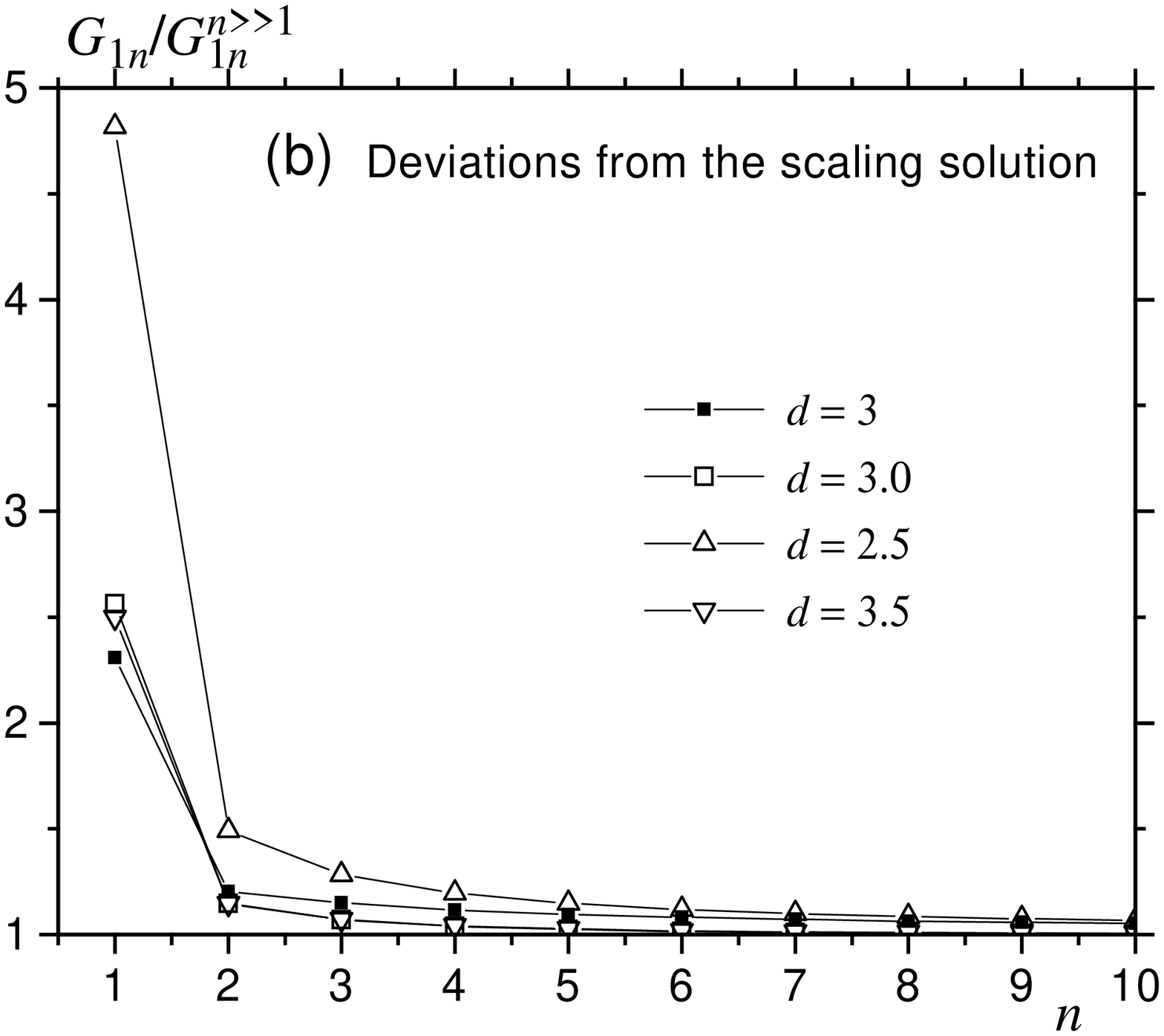,angle=-90,width=12cm}}
\end{picture}
\caption{ \label{gn_g1n} 
$G_{1n}$ at isotropic criticality for different hypercubic and 
continuous-dimension lattices:
(a) -- general view;
(b) -- surface region, deviations from scaling.
}
\end{figure}
\begin{figure}[t]
\unitlength1cm
\begin{picture}(11,7)
\centerline{\epsfig{file=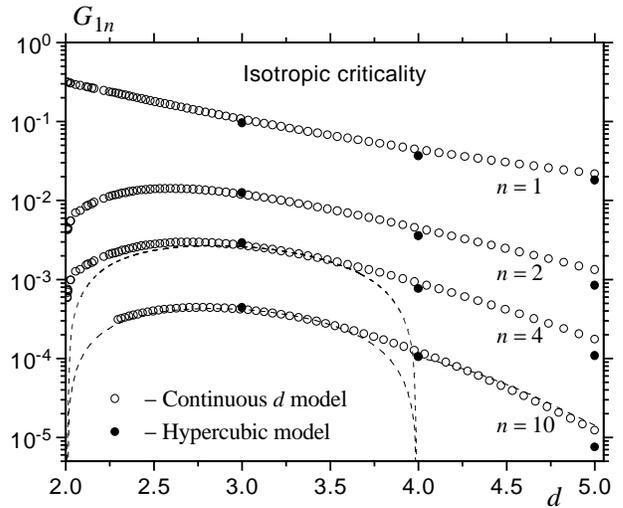,angle=-90,width=12cm}}
\end{picture}
\caption{ \label{gn_g1d} 
$G_{1n}$ at isotropic criticality vs lattice dimensionality $d$.
The asymptotic scaling result (\protect\ref{g1nd24}) for $2<d<4$ and the
asymptotic formula  (\protect\ref{g1nd4+1}) are shown by dashed lines.
}
\end{figure}
\begin{figure}[t]
\unitlength1cm
\begin{picture}(11,7)
\centerline{\epsfig{file=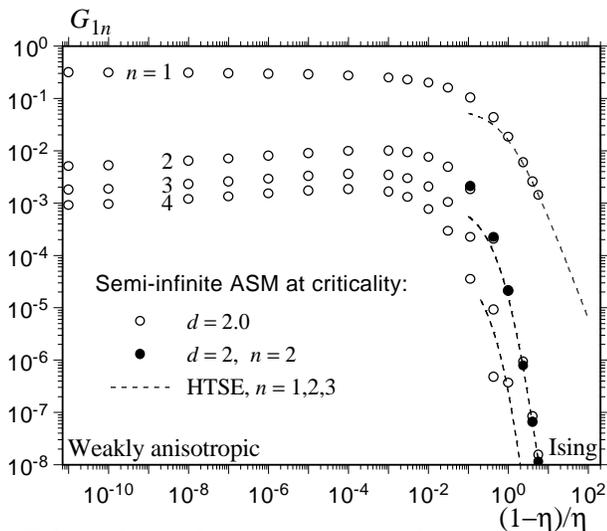,angle=-90,width=12cm}}
\end{picture}
\caption{ \label{gn_g1et} 
$G_{1n}$ at the anisotropic criticality in two dimensions vs the 
anisotropy parameter $\eta$.
The HTSE results (\protect\ref{gnhtse}) are shown by the dashed lines.
}
\end{figure}
\begin{figure}[t]
\unitlength1cm
\begin{picture}(11,7)
\centerline{\epsfig{file=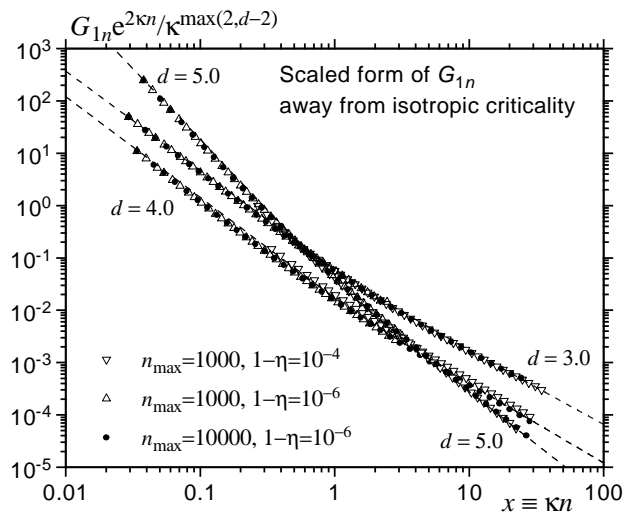,angle=-90,width=12cm}}
\end{picture}
\caption{ \label{gn_g1sc} 
Scaled form of $G_{1n}$ away from isotropic criticality.
Dashed lines are the fits describing crossovers between different power laws.
}
\end{figure}
\begin{figure}[t]
\unitlength1cm
\begin{picture}(11,7)
\centerline{\epsfig{file=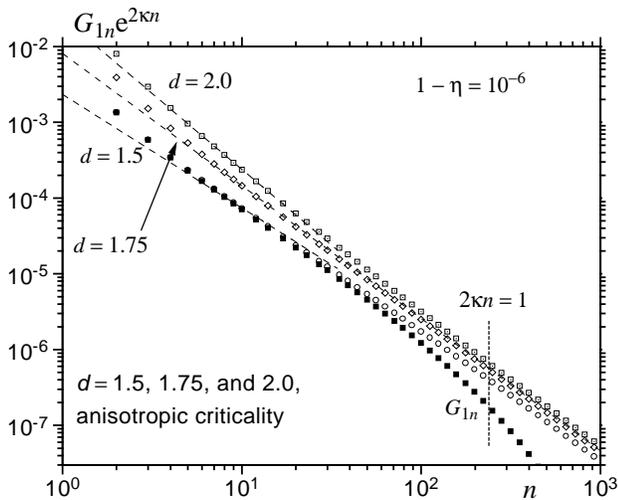,angle=-90,width=12cm}}
\end{picture}
\caption{ \label{gn_g1n_l} 
$G_{1n}$ in low dimensions at the anisotropic criticality.
Dashed lines represent the theoretical formulas (\protect\ref{g1npert1}) and
 (\protect\ref{g1npert12}), the latter with the fitting parameter $a=e=2.718$.
}
\end{figure}
\begin{figure}[t]
\unitlength1cm
\begin{picture}(11,7)
\centerline{\epsfig{file=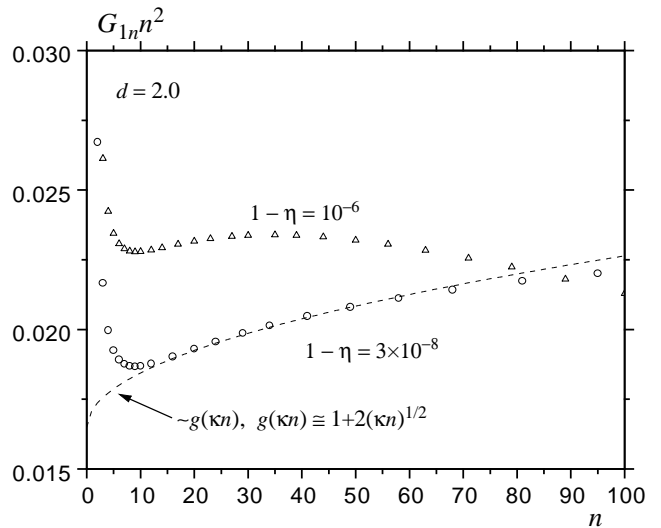,angle=-90,width=12cm}}
\end{picture}
\caption{ \label{gn_g1n_2} 
$G_{1n}$ for $d=2.0$ at the anisotropic criticality: scaling function $g(\kappa n)$ 
and deviations from scaling near the surface.
}
\end{figure}
\begin{figure}[t]
\unitlength1cm
\begin{picture}(11,7)
\centerline{\epsfig{file=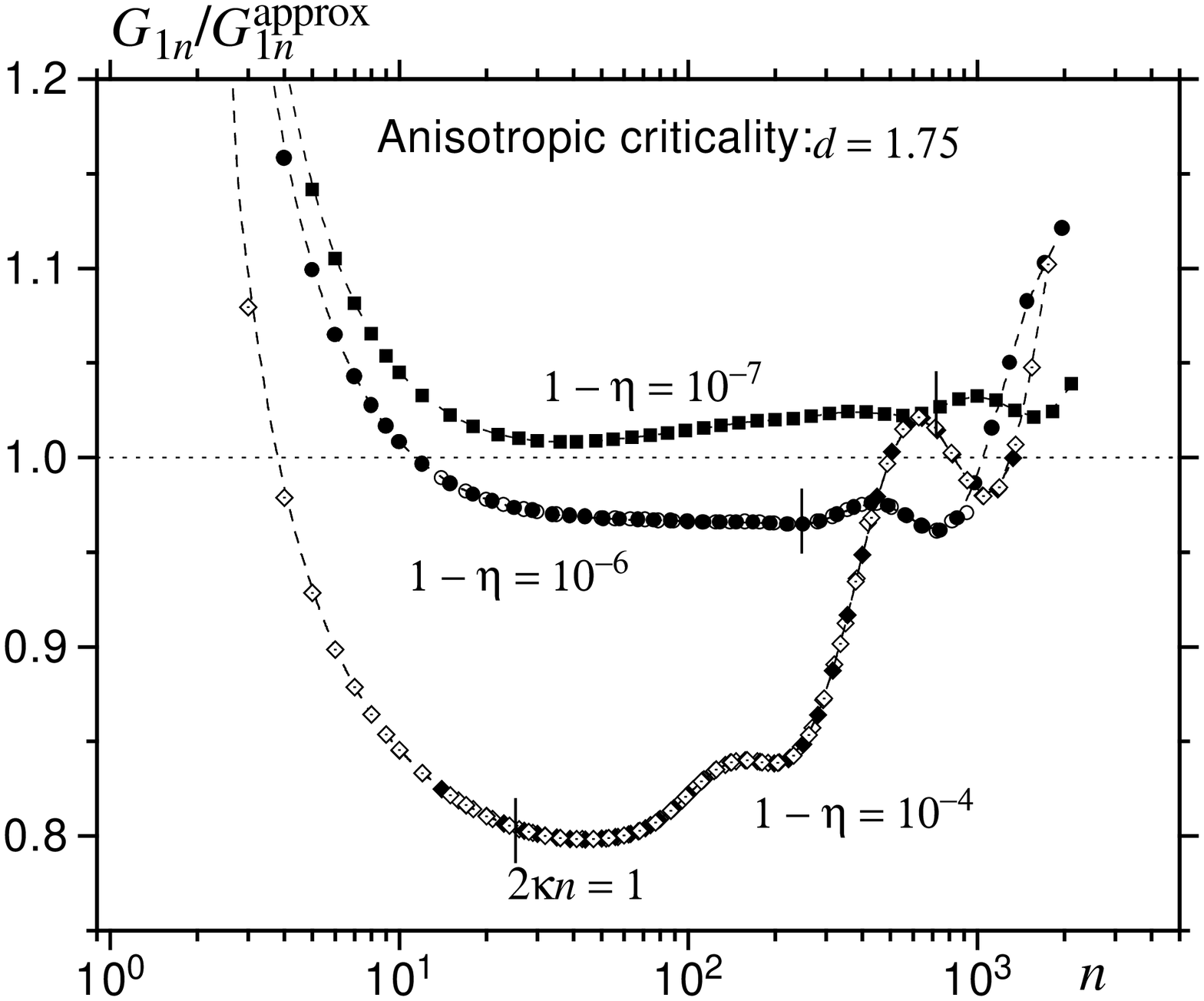,angle=-90,width=12cm}}
\end{picture}
\caption{ \label{gn_g1n_f} 
$G_{1n}$ for $d=1.75$ relative to its approximation shown by the dashed line
in Fig. \protect\ref{gn_g1n_l}.
Overlapping solid and open symbols correspond to different values of the maximal number
of layers $n_{\rm max}$ in the numerical calculation.
}
\end{figure}

\section{Numerical results}
\label{numerical}

\subsection{Variation of $G_n$}
\label{numGn}

In the symmetric phase, $m=0$, the ASM equations were solved numerically in
the following way.
For a given variation of $G_n$ and the value of the wave vector ${\bf q}$ in (\ref{bn}),
the transverse CF $\sigma_{nn}({\bf q})$ can be found from  (\ref{cffd}).
In practice, the formula  (\ref{sigrec1}) was used, where
$\alpha_{1n}$ and $\alpha_{1n}'$ were determined from the recurrence
relations  (\ref{al1rec}) and  (\ref{al'1rec}).
The result for $\sigma_{nn}({\bf q})$ can be put into the constraint
equation (\ref{sconstr}) to obtain, after the integration over  ${\bf q}$, the
system of nonlinear equations for $G_n$.
Again, it is more convenient to work with the deviations from the bulk values
and to use the constraint equation in the form (\ref{sconstr1}) where the
subtraction is done analytically to avoid the loss of accuracy.   
The integrals over $q$ have been performed in all cases over the whole
Brillouin zone, even in low dimensions.
For the continuous-dimension model [see (\ref{contint})] the range $0\leq q \leq \Lambda$ was
divided into three or four log-spaced intervals (e.g., $[0,10^{-4}\Lambda]$, 
$[10^{-4}\Lambda,10^{-3}\Lambda]$, etc.) and the Gaussian quadratures over 10
or 20 points were used in each of these intervals.
For the hypercubic lattice the products of Gaussian quadratures were used, and
the intergation was done with respect to the nonlinearly scaled $q$ components
$Q_i =q_i^{1/p}$ with $p=3$ to redistribute the contribution of the
singularity at $q=0$ more uniformly over the Brillouin zone.
In fact, similar nonlinear transformations were also done for the  continuous-dimension model.
The resulting system of equations for the deviations $G_{1n}$ was solved with a
nonlinear equation solver based on Newton method.

For the numerical solution in the semi-infinite geometry, the boundary condition at
$n=\infty$ in (\ref{al'1rec}) should be replaced by one at some 
$n_{\rm max}\gg 1$.
For the isotropic model at criticality one cannot just set $\alpha_{1n_{\rm max}}'=0$,
since $\alpha_{1n}'$ slowly decays with $n$ [see (\ref{al'1lim})].
This would spoil the behavior of correlation functions at small wave vectors and  
lead to an unphysical gap for $2<d\leq 3$.  
Fortunately, the asymptotic behavior of $\alpha_{1n}'$ in the scaling region,
$n\gg 1$, is given by (\ref{al1sol}) with $c'=0$ and it can be used as the
boundary condition at infinity.
The purpose of numerical calculations at isotropic criticality was to
check the scaling solution (\ref{g1nd24}) and to study the nonscaling effects
at $n\sim 1$.
The quantity $\mu$ was determined self-consistently as a function
of all $G_{1n}$  using the asymptotic form of $\alpha_{1n}$ at $n\gg 1$ and $q = 0$ [see the first
limit of (\ref{al1lim})].
In this way the value $\mu=(d-3)/2$ has been confirmed.

Above $\theta_c$ or at the anisotropic criticality ($\kappa> 0$) the
inhomogeneities decay as
$\exp(-2\kappa n)$ [see (\ref{g1noff})], and one can use the boundary condition 
$\alpha_{1n_{\rm max}}'=0$ for $ n_{\rm max}\gg 1/\kappa$.
Here the value of $\kappa$ should be taken rather small to study
the details of the scaling function $\Phi(x,y)$ in (\ref{signnsc}).
Indeed, to reproduce the limit $x\ll 1$ one should have 
$\kappa n \ll 1$, where $n^*\sim 10$ is the smallest
value of $n$ for which the continuous scaling solution holds.
This implies, in turn, large values of $n_{\rm max}$.
In practice, calculations were done for $1-\eta$ down to $10^{-8}$, which
corresponds to $\kappa = \sqrt{2d(1-\eta)} \approx 2.5\times 10^{-4}$ 
at the anisotropic criticality for $d=3$.
For such $\kappa$ the value $n_{\rm max}=10000$ was used, which corresponds
to $2\kappa n_{\rm max}\approx 5$.
Naturally, in this case the system of equations for $G_{1n}$ was not solved on
each of 10000 layers.
Instead, for $n\gsim 10$ only the ``representative'' layers with an
exponentially increasing spacing between them were chosen to solve the
equations.
The values of $G_{1n}$ between them were interpolated with the 
help of the formula $G_{1n}= (a/n^b)\exp(-2\kappa n)$ with the values of $a$
and $b$ determined from $G_{1n}$ at the ends of the interpolation intervals.
In all cases the number of unknowns did not exceed 50.
Computations could be performed on a 486DX 66-MHz laptop.

The results for $G_{1n}$, as defined by (\ref{g1ndef}) for {\em all} values of $n$,  are shown at isotropic criticality for different hypercubic and 
continuous-dimension lattices in Fig. \ref{gn_g1n}.
The general view,  Fig. \ref{gn_g1n}(a),  shows that the analytical result (\ref{g1nd24})
is well obeyed in three dimensions outside the surface region.
This result is universal and independent of the lattice structure; it
is the same for the simple-cubic lattice ($d=3$) and the 3$d$
continuous-dimension lattice ($d=3.0$).
Formula (\ref{g1nd24}) has also been confirmed for other values of $d$ around $d=3$; 
the results do not differ much from each other in the log scale and thus they
have not been shown.
In four dimensions the results can be fitted with formula (\ref{g1nd4}) with
$a=e^3\approx 20$, which implies significant corrections to the logarithmic approximation.
In fact, the nonuniversal number $a$ is slightly larger for the hypercubic
lattice ($d=3$), which can be seen in Fig. \ref{gn_g1n}(a).
In dimensions higher than four, $G_{1n}$ follows formula (\ref{g1nd4+1}).
The coefficient in (\ref{g1nd4+1}) depends on the lattice structure and is
clearly different for $d=5$ and $d=5.0$.

Deviations from the asymptotic solution  (\ref{g1nd24}) in the region near
the surface are shown in Fig. \ref{gn_g1n}(b).
There is a clear difference between the values of $G_{1n}$ for both
three-dimensional lattices.

The dependence of $G_{1n}$ at isotropic criticality  near the surface on $d$
is shown in Fig. \ref{gn_g1d}.
In the limit $d\to 2$ the value $G_{11}$ tends to $\onethird$, which means that the
limiting value of $G_1$ is $\fourthird$, as given by the first term of (\ref{gnpert}),
where, at criticality, $G=1$.
On the other hand, all $G_{1n}$ with $n\geq 2$ vanish in the limit $d\to 2$,
in accord with (\ref{g1npert12}), which disappears for $\kappa\to 0$.

The algorithm for solving the system of nonlinear equations for $G_{1n}$ 
based on the Newton method, which was used here, shows instability for $d\lsim 2.3$ and
$n_{\rm max}\gsim 10$.
This instability is responsible for the lack of points in the left
part of the $n=10$ curve.
The reason is that the integral in the constraint equation  
(\ref{sconstr1}) becomes more and more sensitive to the region
of small $q$ where the integrand may become infinite due to a negative gap
arising for some sets of $G_{1n}$ in the course of iterations.
Away from isotropic criticality, the gap in the spin CFs stabilizes the
algorithm.
For each dimension $d$ there is a minimal value of the anisotropy $1-\eta$,
for which the system of nonlinear equations for $G_{1n}$ does not show
instability for $n_{\rm max}$ large enough, if the starting variation of $G_{1n}$ 
is chosen sufficiently close to the actual one.
The latter is very important and necessitates using small variations of the
parameters, such as $d$,  $1-\eta$, $n_{\rm max}$, etc., in low dimensions.
The minimal values of the anisotropy  $1-\eta$ are about $3\times 10^{-8}$ for $d=2.0$,
 $10^{-7}$ for $d=1.75$, and $ 5\times 10^{-7}$ for $d=1.5$.

Contrary to the implication of the scaling solution of Bray and Moore,
 (\ref{g1nd24}), $G_{1n}$
do not go to zero and do not even show any singularity  at $d=4$ for any finite $n$.
This is due to the correction-to-scaling terms [see Fig. \ref{gn_g1n}(b)] which
become more and more pronounced as $d$ deviates from 3.
The crossover from the solution for $G_{1n}$ in the range $2<d<3$ and that for $d>4$ is
described by (\ref{g1ndab4}).

Additionally, in four surface layers for the
two-dimensional model at criticality, the dependences of $G_{1n}$ on the anisotropy parameter $\eta$ are
shown in Fig. \ref{gn_g1et}.
Calculations down to $1-\eta = 10^{-11}$ were possible here, since the value of $n_{\rm max}$
was chosen to be about 50, which is significantly smaller than the required 
$n_{\rm max}\gg 1/\kappa$.
The latter introduces a significant gap in the spin CFs, which is the artifact
of cutting the ASM equations at $n_{\rm max}\lsim 1/\kappa$.
This gap stabilizes the solution of the ASM equations.
On the other hand, the values of  $G_{1n}$ in several layers near the surface
are pretty robust and insensitive to this defect of $\sigma_{nn}$.
One can see that in the isotropic limit, $\eta\to 1$, the value of $G_{11}$ tends
to $\onethird$, whereas all other $G_{1n}$ tend to zero logarithmically in accordance
with (\ref{g1npert12}).
In the opposite limit, $\eta\ll 1$, the HTSE results (\ref{gnhtse}) are recovered.

In Fig. \ref{gn_g1sc} the calculated values of $G_{1n}$ for
continuous-dimension lattices away from the
isotropic criticality are represented in the scaled form.
The results for $d=5.0$ show crossover from (\ref{g1nd4+1}) to (\ref{g1nd4+2})
as $x\equiv \kappa n$ increases.
A similar crossover from $G_{1n}\sim 1/n^2$ to $G_{1n}\sim e^{-2\kappa n}/n^{3/2}$
takes place for $d=4.0$.
The scaling is, however, not perfect here because of the logarithmic corrections.
For $d=3.0$ the result crosses over to $G_{1n}\sim e^{-2\kappa n}/n^{1.4}$ for
$\kappa n \gg 1$.
Note that there is no analytical solution for $G_{1n}$ in this region.

The calculated values of $G_{1n}$ at the anisotropic criticality (or above criticality)
in low dimensions ($d=1.5$, 1.75, and 2) are shown in Fig.  \ref{gn_g1n_l}.
One can see that  the theoretical formulas (\ref{g1npert1}) and (\ref{g1npert12})
are obeyed starting from $n\gsim 10$, although corrections related to the
finite value of $\kappa$ and described by the function $g(\kappa n)$ in (\ref{g1npert1gen})
are quite pronounced.
For $d=2.0$ the function $g(\kappa n)$ is nonmonotonic: apart from the
exponential decrease at $\kappa n \gg 1$ it shows a singular positive
deviation from unity for $\kappa n \ll 1$.
The fit in Fig.  \ref{gn_g1n_2}
suggests $g(\kappa n) \cong 1 + 2 \sqrt{\kappa n}$ for $\kappa n \ll 1$.
One can see from Fig.  \ref{gn_g1n_l} that the correction term in 
$g(\kappa n)$ has the negative sign for $d=1.5$.
The case $d=1.75$ seems to be marginal. 
The values of $G_{1n}$ in Fig.  \ref{gn_g1n_l} nicely follow the dependence 
$G_{1n} = G_{1n}^{\rm approx}$, where $G_{1n}^{\rm approx}$ is given by
(\ref{g1npert1}) with the additional factor $\exp(-2\kappa n)$ in the whole
range of $n$.
This is, however, not an exact solution to the problem. 
The plot of  $G_{1n}/G_{1n}^{\rm approx}$ in Fig. \ref{gn_g1n_f} shows that
for $\kappa n \gsim 1$ this function begins to increase with oscillations.
These oscillations are not an artifact of cutting the ASM equations at the
maximal layer number $n_{\rm max}$.
Numerical calculations with different values of $n_{\rm max}$ give the same
results.
Although the ratio $G_{1n}/G_{1n}^{\rm approx}$  is not exactly 1, its
proximity to 1 in a wide range of $n$ is
remarkable, taking into account the strong dependence of $G_{1n}$ on $n$.

\subsection{Correlation functions}
\label{numCFs}

\begin{figure}[t]
\unitlength1cm
\begin{picture}(11,7)
\centerline{\epsfig{file=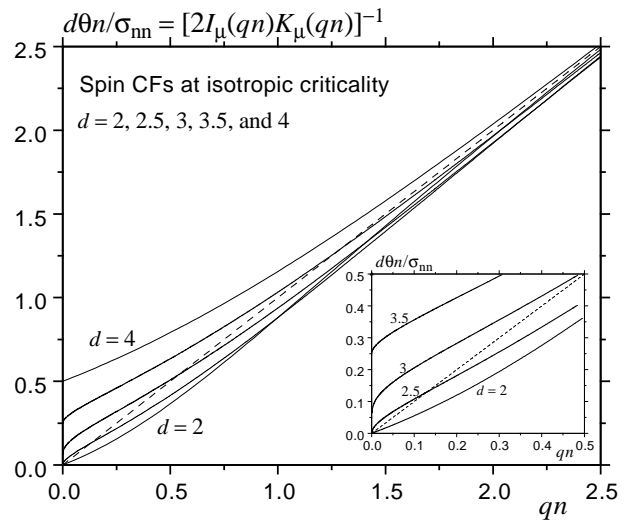,angle=-90,width=12cm}}
\end{picture}
\caption{ \label{gn_isocr} 
Reciprocal of the spin CFs at isotropic criticality in the scaling region $n\gg 1$
for different lattice dimensionalities $d$.
The dashed line is the bulk solution (\protect\ref{cfbulk}). 
}
\end{figure}

After $G_{1n}$ has been determined, the spin CFs can be found from (\ref{sigrec1})
and the recurrence relations  (\ref{al1rec}) and  (\ref{al'1rec}).
The results at isotropic criticality in the scaling region $n\gg 1$, which
illustrate the analytical solution of Bray and Moore (\ref{sigik1}) for $2<d<4$, are shown in 
Fig. \ref{gn_isocr}.
One can see that for $2<d<4$ the solution satisfies 
$\sigma_{nn } < \sigma_{nn}^{\rm bulk}$ for small wave vectors and  
$\sigma_{nn } > \sigma_{nn}^{\rm bulk}$ for large wave vectors.
For $d=4$ one has $\sigma_{nn } < \sigma_{nn}^{\rm bulk}$ everywhere, which 
contradicts the constraint equation (\ref{sconstr1}).
In fact, for $d\geq 4$ the scaling solution of Bray and Moore breaks down, and
one has to take into account the {\em positive} local contribution to $\sigma_{nn }$
at $q\sim 1$ [see (\ref{sigd4loc})], which balances the constraint equation. 
For $d=2$ one has $\sigma_{nn } > \sigma_{nn}^{\rm bulk}$ everywhere, 
and the constraint relation is violated again.
In fact, for $d\leq 2$ the form of $\sigma_{nn }$ is changed by the gap in the
region of small $q$, where $\sigma_{nn } < \sigma_{nn}^{\rm bulk}$,
thus  ensuring the constraint (see Fig. \ref{gn_invsg} below).

\begin{figure}[t]
\unitlength1cm
\begin{picture}(11,7)
\centerline{\epsfig{file=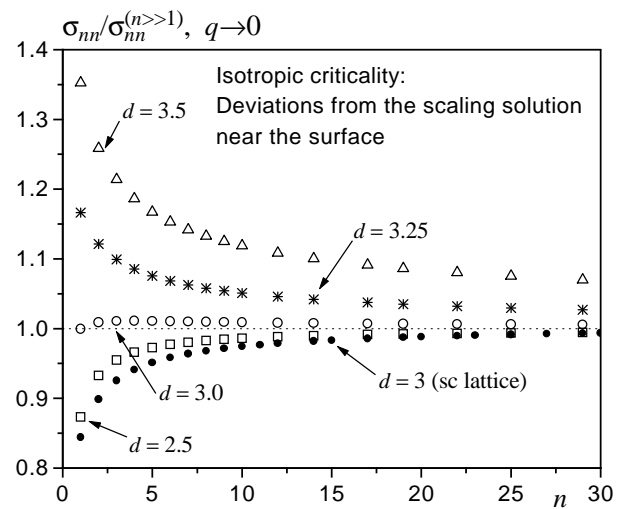,angle=-90,width=12cm}}
\end{picture}
\caption{ \label{gn_cfns} 
Deviations from the scaling solution (\protect\ref{sigik1}) for $\sigma_{nn }$
in the nonscaling region $n\sim 1$.
}
\end{figure}

Deviations from the scaling solution (\ref{sigik1}) in the region near the surface,
$n\sim 1$, are shown in Fig. \ref{gn_cfns}.
The correlation functions in different layers are related to each other by  (\ref{signnrec}), 
where the quantities $\alpha_n$ and $\alpha_n'$ are constants in the limit
$q\to 0$ and they approach the bulk value $\alpha$ of (\ref{sigbulk}) with
increasing $n$.
For $n\gg 1$ small deviations of $\alpha_n$ and $\alpha_n'$ from $\alpha$ are responsible
for the scaling form of $\sigma_{nn }$ showing only a small change when $n$
changes by one.
By contrast, in the nonscaling region, $n\sim 1$, the spin CFs change
significantly from one layer to another and they acquire in the range $q\ll 1$ nonuniversal
numerical factors, relative to the extrapolated scaling solution.
These factors, which are shown in Fig. \ref{gn_cfns}, tend to 1 as some negative
powers of $n$ far from the surface.
The accuracy of the calculations is, however,  not high enough to determine these powers precisely.
One can see that the deviations from scaling are quite large and slowly
decaying for the dimensions well above 3 [cf. (\ref{g1ndab4})].
On the other hand, for $d$ well below 3 the deviations from scaling are mainly localized
near the surface.
Note the difference between the results for the simple-cubic ($d=3$) lattice
and the three-dimensional continuous-dimension lattice ($d=3.0$).
For the latter the deviations from scaling are anomalously small, which
suggests the existence of an exact solution.
If one assumes that the scaling form of $\sigma_{nn }$ holds for all $n$, then from
 (\ref{signnrec}) and  (\ref{alrel}) it follows that $\alpha_n = \sqrt{ 1-1/n}$ and 
$\alpha_n' = \sqrt{ 1+1/n}$ for $q=0$.
Then, using the divergence of  (\ref{sigrec}) for $q\to 0$ one obtains
%\marginpar{g1ntry}
%
\begin{equation}\label{g1ntry}
G_n = \frac{ 6 }{ 4 + \sqrt{ 1+1/n} + \sqrt{ 1-1/n} }, \qquad  d=3.0,
\end{equation}
which is indeed a rather good approximation.
It has the proper behavior $G_n \cong 1 + 1/(24n^2)$ for $n\gg 1$, and the value
$G_1 = 1.1082$ is very close to 1.1067 following from numerical calculations. 
More careful analysis shows, however, that the formula above is not an exact
solution for the ASM equations, where discrepancies of the type 
$\sqrt{2}\ln 2 \approx 0.980 \ne 1$ arise.

\begin{figure}[t]
\unitlength1cm
\begin{picture}(11,7)
\centerline{\epsfig{file=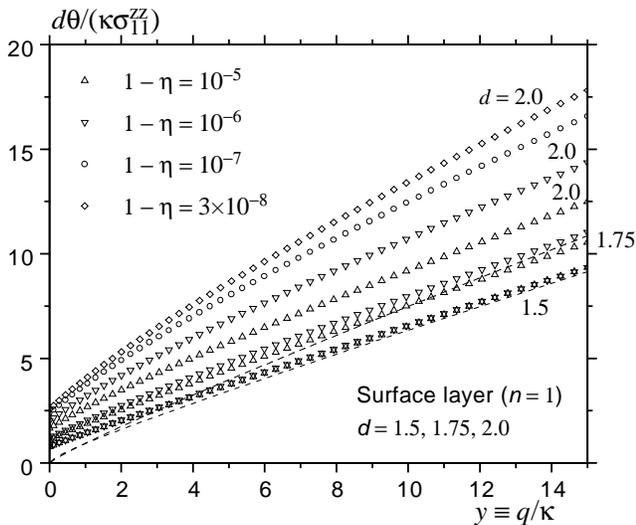,angle=-90,width=12cm}}
\end{picture}
\caption{ \label{gn_invsg} 
Reciprocal of the longitudinal CF $\sigma_{nn }^{zz}$ in the surface layer ($n=1$)
in low dimensions at anisotropic criticality.
Dashed lines represent the asymptote $[q +\bar\Delta_1(q,\kappa)]/2$ with 
$\bar\Delta_1(q,\kappa)$ given by (\protect\ref{barDel1qbig}) for $q\gg \kappa$.
}
\end{figure}
\begin{figure}[t]
\unitlength1cm
\begin{picture}(11,7)
\centerline{\epsfig{file=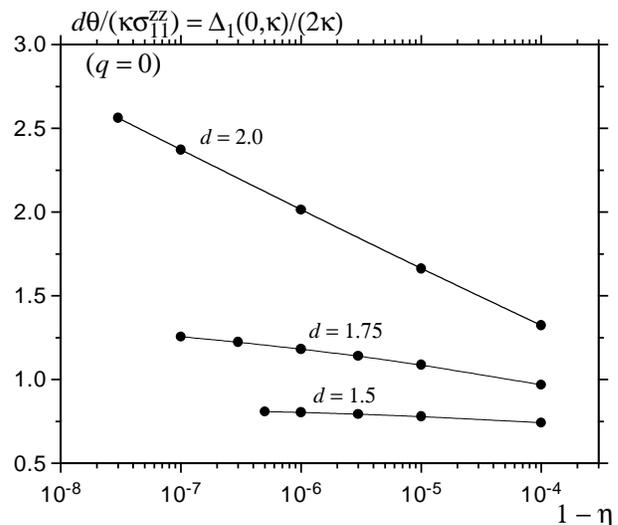,angle=-90,width=12cm}}
\end{picture}
\caption{ \label{gn_d2gap} 
Dependence of the gap in $\sigma_{11 }^{zz}$ on the anisotropy  at anisotropic criticality
in low dimensions.
}
\end{figure}
\begin{figure}[t]
\unitlength1cm
\begin{picture}(11,7)
\centerline{\epsfig{file=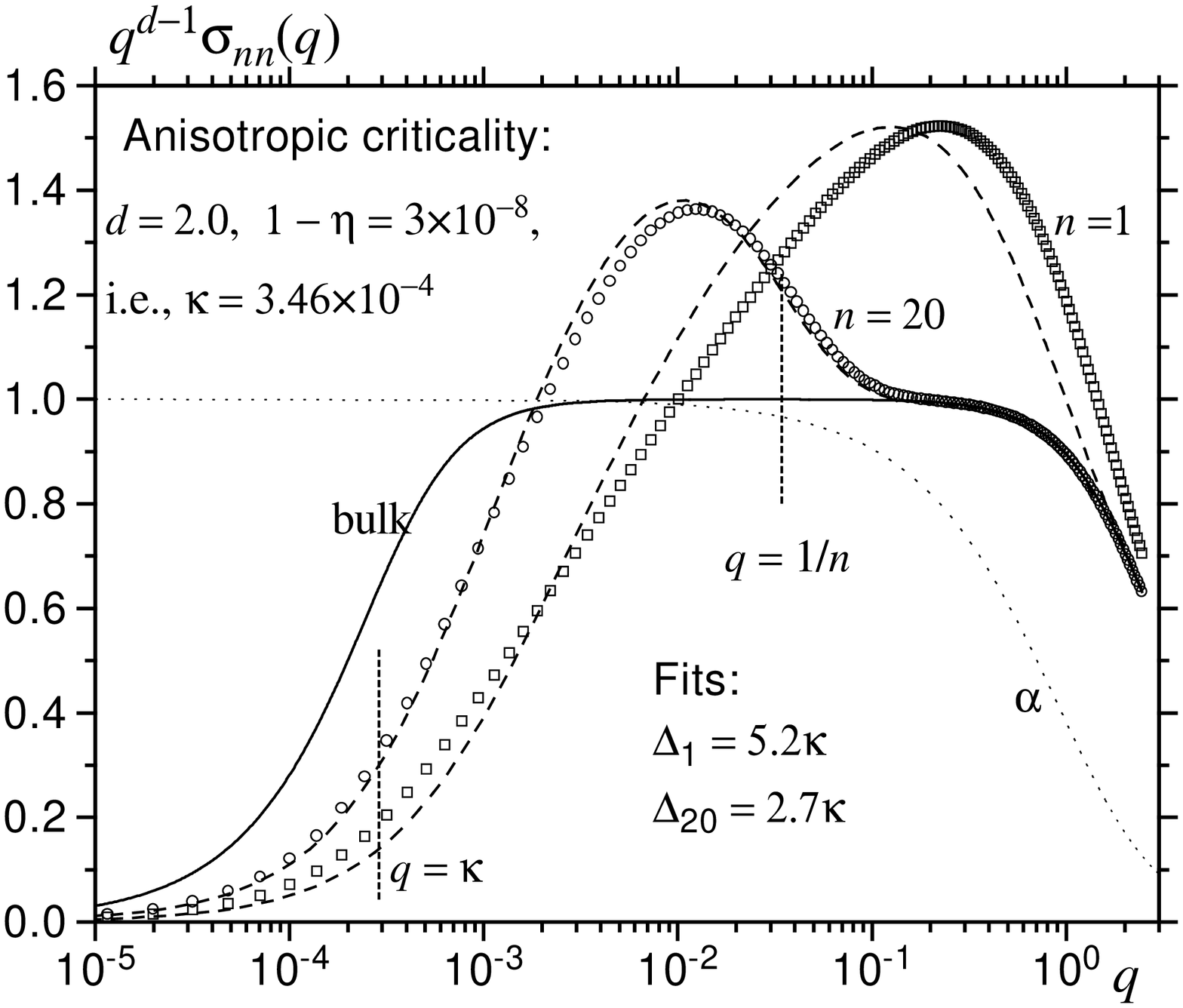,angle=-90,width=12cm}}
\end{picture}
\caption{ \label{gn_d2cs} 
Transverse CF $\sigma_{nn}$ in two dimensions.
Circles: numerical results for $n=1$ and $n=20$.
Dashed lines: Bessel-function solution with fitted gap, as explained in the text. 
}
\end{figure}

At anisotropic criticality, which is the only type of criticality in low
dimensions, the generic CF is $\sigma_{nn }^{zz}$. 
The transverse CF $\sigma_{nn }$, as well as $\sigma_{nn }^{zz}$ itself above
criticality, can be obtained from the latter by the simple change of the
wave-vector argument.
The numerical results for $\sigma_{nn }^{zz}$ in the surface layer ($n=1$) 
at anisotropic criticality in low dimensions are shown in Fig. \ref{gn_invsg}.
One can see the gap and the linear $q$ dependence at small $q$, in accord with
(\ref{sigzzcrit}).
The values of the gap $\Delta_1(0,\kappa)$ and the stiffness $A$ in (\ref{sigzzcrit})
determined from the fits of the numerical data exceed those calculated from 
(\ref{Del1gap}) and (\ref{Adef}).
The reason for this is that the first-order perturbation theory in $G_{1n}$
leading to (\ref{Del1gap}) and (\ref{Adef}) is valid for small wave vectors
only in the dimension range $d<1.5$, as was explained after (\ref{constrDel1}).
On the other hand, the asymptote $\bar\Delta_1(q,\kappa)$  for $q\gg \kappa$. 
which is given by (\ref{barDel1qbig}), works nicely for $d=1.5$ and $d=1.75$.

It can be seen that for $d=1.5$ the results for the models with different
anisotropies scale with each other.
This confirms the concept of the strong scaling (down to the surface layer)
in low dimensions, which has been suggested in Sec. \ref{CFsurf}.
For $d=1.75$ the scaling in Fig. \ref{gn_invsg} is incomplete, which can be
explained by the values of $\kappa$ not being small enough.
For $d=2.0$ the results do not scale, although either do not strongly deviate
from each other, because $d=2$ is the marginal dimension
between the strong scaling and the asymptotic ($n\gg 1$)  scaling. 
This behavior is illustrated in greater detail in Fig. \ref{gn_d2gap}, where the
gap $\Delta_1(0,\kappa)$ of $\sigma_{11 }^{zz}$ is plotted as function of the anisotropy.
Whereas the dependences for $d=1.5$ and $d=1.75$ saturate in the limit of 
$\kappa \equiv \sqrt{2d(1-\eta)}$ going to zero, which confirms 
$\Delta_1(0,\kappa) \propto \kappa $ for weakly anisotropic models, 
the almost perfectly straight line over several decades for $d=2.0$ suggests  
$\Delta_1(0,\kappa) \propto \kappa \ln (1/\kappa)$ in two dimensions. 

The numerical result for $d=2$ above is quite plausible, because logarithms
usually arise in marginal dimensions.
This would imply something like 
$\chi_{znn} \propto \kappa^{-1}/\ln [1/(\kappa n)]$ for $d=2$ in (\ref{chizlims}), 
where the corresponding position has been left empty.
It seems to be, however,  the third occasion in this work when
numerical calculations suggest some qualitative features that does not follow
from analytical considerations.
For $d=2$ the calculation with logarithmic accuracy leads to another
dependence of $\Delta_1(0,\kappa)$, which is given by (\ref{gapd2}).
The applicability of (\ref{gapd2}) requires, however, such small values of
the anisotropy that numerical calculations cannot be performed,
and for larger anisotropies no other possible analytical approximations are seen.

To shed some light on this puzzle, it is convenient to plot
$q^{d-1}\sigma_{nn}$ as a function of $\log q$ over the whole Brillouin zone.
The area under the curve is proportional to the integral over $q$ in the
constraint equation   (\ref{sconstr1}),  and the regions of $q$ making
contributions into the constraint can be well identified.
Such plots show that the integral is dominated by $q\sim 1/n$  for $d>2$ and  by
$q\sim \kappa$  for $d<2$.
In the marginal case $d=2$ the results for the lowest manageable anisotropy, 
$1-\eta =3\times 10^{-8}$, are shown in Fig. \ref{gn_d2cs}.
One can see that the area under the bulk solution (the solid line) coincides
with that under the numerical solutions for $n=1$ and $n=20$ (open circles).
The curve for $n=20$ merges with the bulk curve for $q\gsim 1/n$.
Although the distance between $q\sim 1$ or $q\sim 1/n$, on the right-hand side, and
$q\sim \kappa$, on the left-hand side, is several decades, it is not large enough
to apply the logarithmic approximation, i.e., to integrate the solution
obtained for $\kappa \ll q \ll 1/n$ between these limits.
Also, $\kappa$ is not small enough to replace $\alpha$ (dotted line in  Fig. \ref{gn_d2cs}) 
by 1 making integration in (\ref{sigdif}) to obtain $G_{1n}$ of (\ref{g1npert12}).
The applicability condition for the formula (\ref{gapd2}) is clearly not satisfied.
Nevertheless, as we have seen in Fig. \ref{gn_g1n_l}, formula (\ref{g1npert12}) 
is in reasonable agreement with the numerical results for $G_{1n}$ for small $\kappa n$.
Thus one can use the solution  (\ref{sigik1}) for $\sigma_{nn}$ in terms of
the modified Bessel functions of index $\mu$ given by (\ref{mud2}).
This solution with $\tilde q \Rightarrow \tilde q + \Delta_n(0,\kappa)$ is
plotted with dashed lines in Fig. \ref{gn_d2cs}, where the gap values $ \Delta_1(0,\kappa) = 5.2\kappa$
(taken from Fig. \ref{gn_d2gap}) and $ \Delta_{20}(0,\kappa) = 2.7\kappa$ were
used as fitting parameters.
The agreement with numerical results is rather good.
On the other hand, $1/\ln(1/\kappa)$ is not small enough to use the simplified form
(\ref{signnd2}) for $\sigma_{nn}$.
The corresponding curves deviate strongly from the numerical 
solution in Fig. \ref{gn_d2cs}, thus the final result (\ref{gapd2}) is not realized.
And analytically calculating the constraint integral with the gapped Bessel functions
 to obtain the simple empirical formula
$\Delta_1(0,\kappa) \propto \kappa \ln (1/\kappa)$ seems to be impossible.

\begin{figure}[t]
\unitlength1cm
\begin{picture}(11,7)
\centerline{\epsfig{file=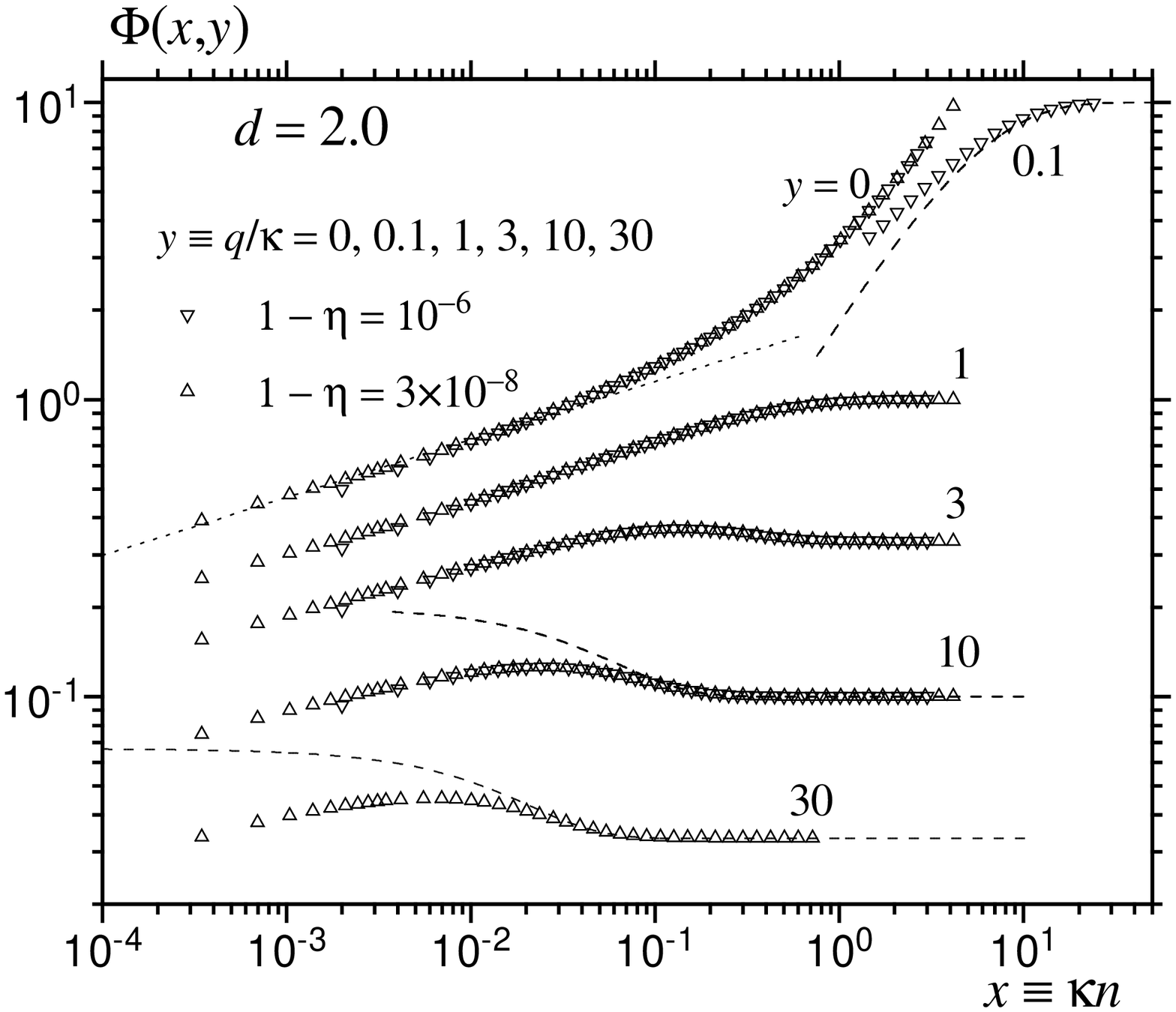,angle=-90,width=12cm}}
\end{picture}
\begin{picture}(11,6.5)
\centerline{\epsfig{file=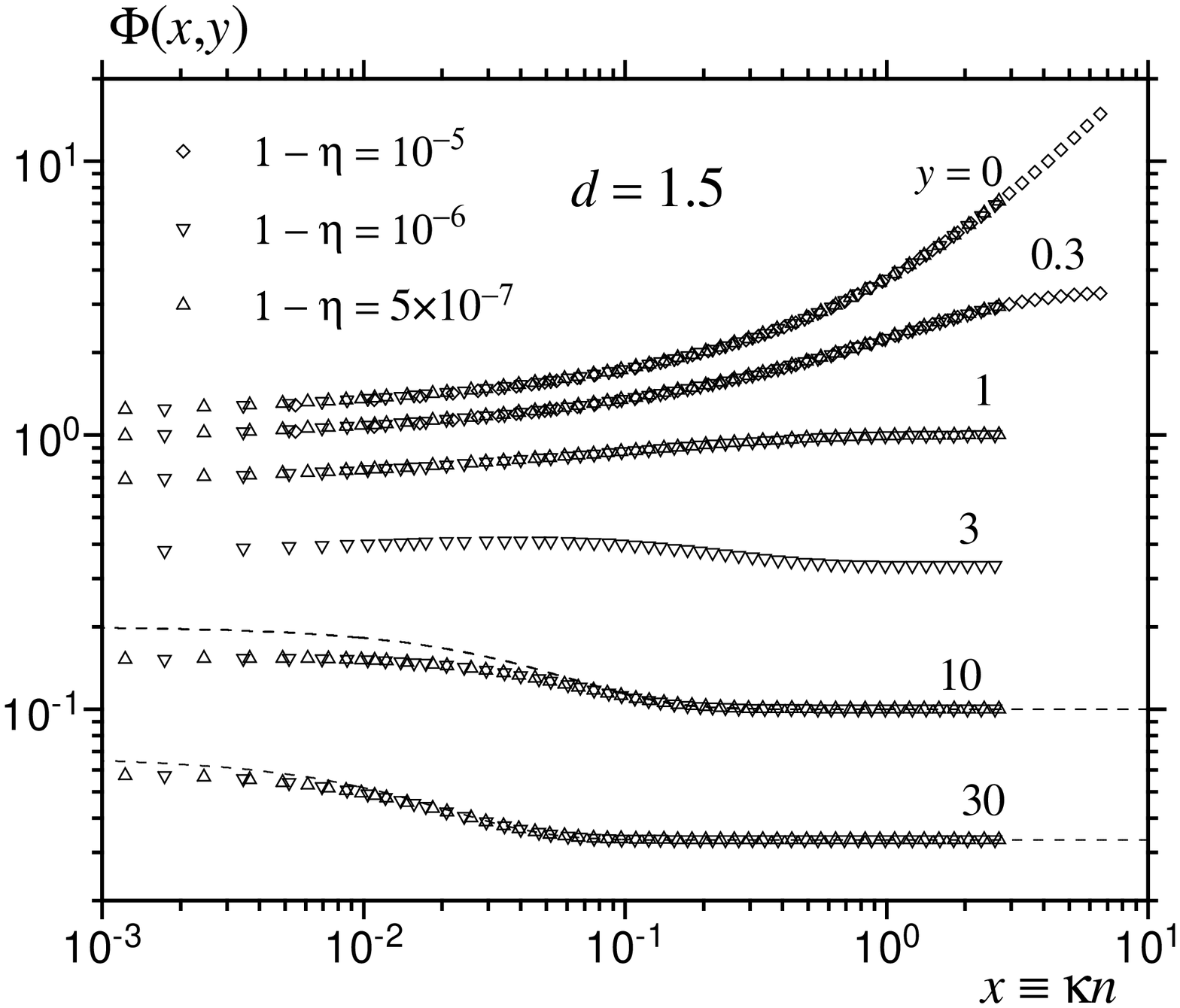,angle=-90,width=12cm}}
\end{picture}
\caption{ \label{gn_d2xy} 
Scaling representation of $\sigma_{nn}^{zz}$ for $d=2$ and $d=1.5$.
Dashed lines: free solutions $ \Phi(x,y) = (1\pm e^{-2xy})/y$ for large and small values of $y$.
}
\end{figure}
\begin{figure}[t]
\unitlength1cm
\begin{picture}(11,7)
\centerline{\epsfig{file=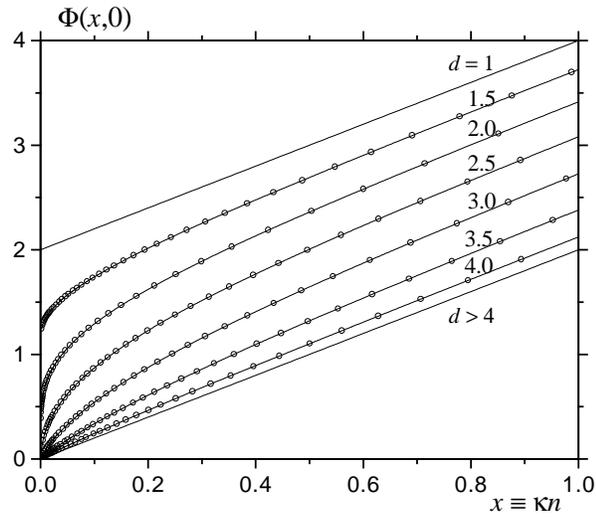,angle=-90,width=12cm}}
\end{picture}
\caption{ \label{gn_fxz} 
Scaling function at zero wave vector,  $ \Phi(x,0)$, in all dimensions. 
}
\end{figure}
\begin{figure}[t]
\unitlength1cm
\begin{picture}(11,7)
\centerline{\epsfig{file=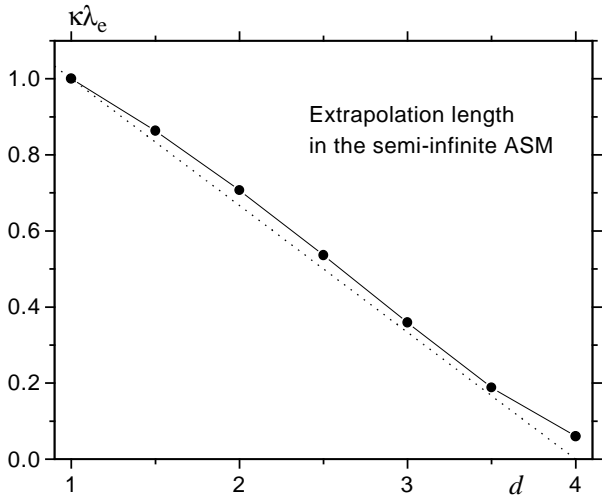,angle=-90,width=12cm}}
\end{picture}
\caption{ \label{gn_extr} 
Extrapolation length vs lattice dimensionality. 
}
\end{figure}
\begin{figure}[t]
\unitlength1cm
\begin{picture}(11,7)
\centerline{\epsfig{file=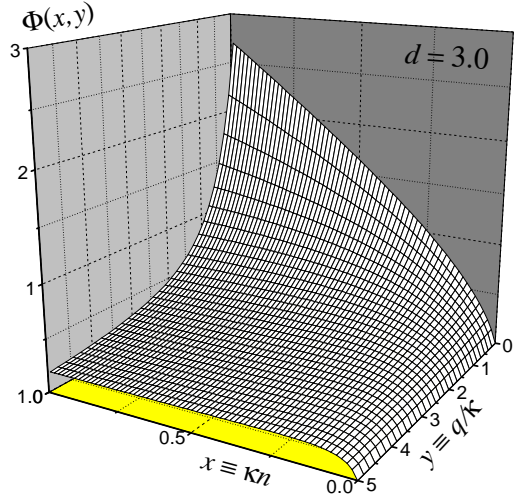,angle=-90,width=12cm}}
\end{picture}
\caption{ \label{gn_3d30} 
Scaling function $ \Phi(x,y)$ of (\protect\ref{signnsc}) for $d=3$.
}
\end{figure}
\begin{figure}[t]
\unitlength1cm
\begin{picture}(11,7)
\centerline{\epsfig{file=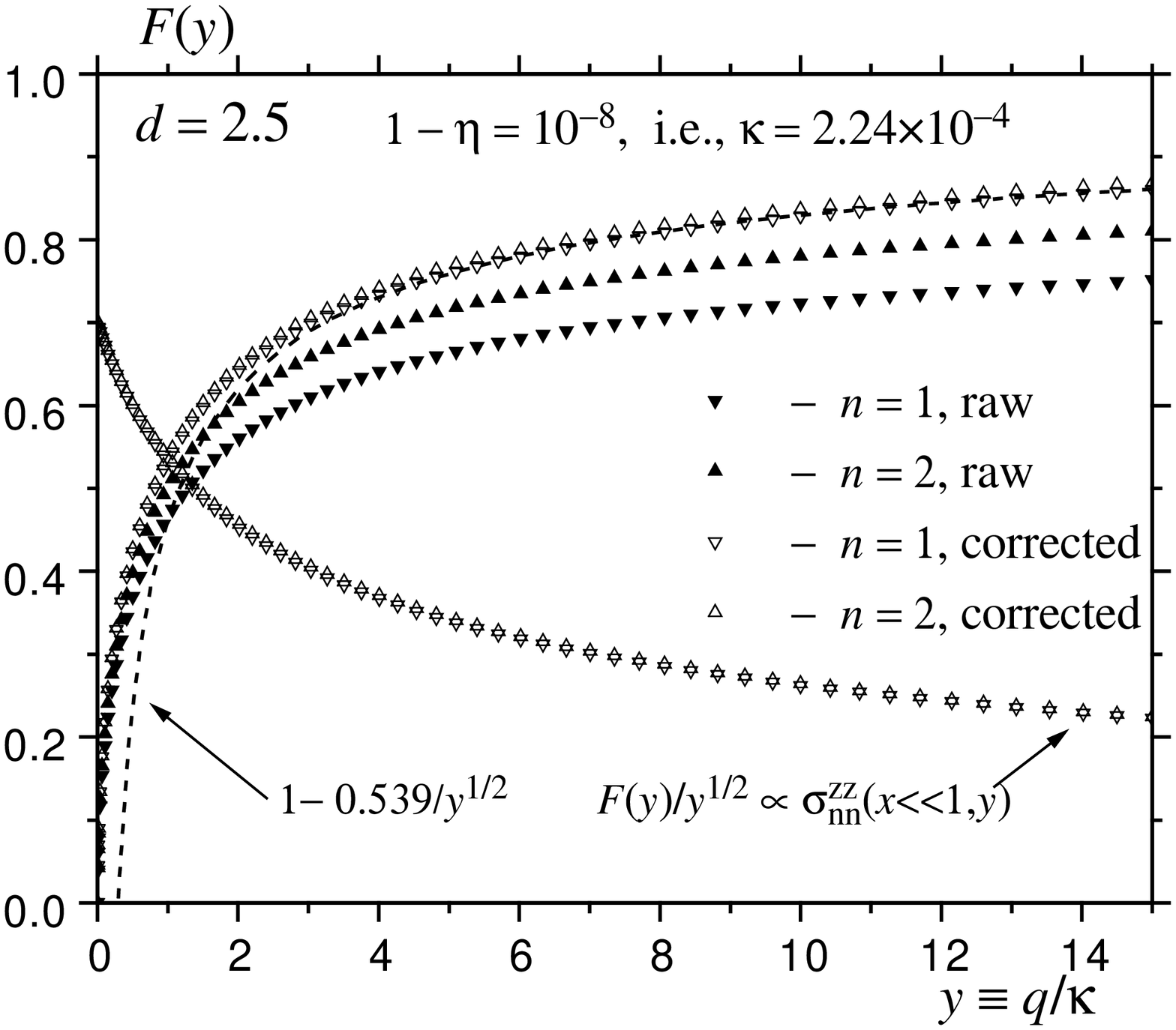,angle=-90,width=12cm}}
\end{picture}
\caption{ \label{gn_fy25} 
Scaling function $F(y)$ of (\protect\ref{yscale2}) for $d=2.5$.
}
\end{figure}

Now let us consider the numerical results for the longitudinal CF $\sigma_{nn}^{zz}$
in the scaling representation (\ref{signnsc}).
The scaling function $\Phi(x,y)$ for $d=2.0$ and 1.5 is represented in Fig. \ref{gn_d2xy}.
One can see that the asymptotic scaling (for $n\gg1$) is well obeyed.
For $d=2$ in the surface region $n\sim 1$ there are small, seemingly logarithmic
deviations from the strong scaling, as was suggested above for the
two-dimensional model.
Here, for small $x$ the results can be fitted with power-law functions, the
exponent slowly decreasing with $\kappa$.
In particular, for $1-\eta =3\times 10^{-8}$ this exponent is 0.195, which
roughly agrees with $1/\ln(1/\kappa) = 0.125$ following from (\ref{signnd2}).
For large and small values of $y$ the numerical results contain the features
of the free solution  $ \Phi(x,y) = (1\pm e^{-2xy})/y$, as was argued in Sec. \ref{CFsurf}.
In fact, $d=2$ is a marginal dimension, and for $d<2$  the free solution 
is reproduced for large and small $y$ much better, as can be
seen from the plot for $d=1.5$.
The latter also confirms the strong scaling in low dimensions. 
 
The results for $\Phi(x,0)$ in the whole range of lattice dimensionalitiess are shown
in Fig. \ref{gn_fxz}.
All the curves are bounded by the ones representing the exact expressions of (\ref{Phid14}).
For small $x$ the results are in accord with  (\ref{chizlims}).
The asymptotic form of the curves in the region $x\gg 1$, which is given by the limit
$y\to 0$ of  (\ref{phixbig}), determines the extrapolation length $\lambda_e$.
The latter is represented in Fig. \ref{gn_extr} as a function of $d$.
One can see that for $d<4$ the extrapolation length is of the order of
transverse correlation length $\xi_{c\alpha}$, which is a ``mesoscopic''
length scale between
the lattice spacing $a_0$ and the diverging longitudinal correlation length $\xi_{cz}$.

For $d>2$ the scaling form of the correlation function holds in the
asymptotic region $n\gg 1$.
The general view of the scaling function $ \Phi(x,y)$ is shown for $d=3.0$ in Fig. \ref{gn_3d30}.
One can figure out how the results look like for other dimensionalities with the help of  
Figs. \ref{gn_fxz} and \ref{gn_d2xy}.
The results for the wave-vector scaling function $F(y)$ defined by
(\ref{yscale}) or, for $2<d<3$, also by (\ref{yscale2}), are shown for
$d=2.5$ in Fig. \ref{gn_fy25}.
The dimension  $d=2.5$ is especially convenient since here the coefficient
$A_d$ in the gap tail of $F(y)$ given by (\ref{fysc}) simplifies to  
 $A_{5/2}= - 4\pi^{1/2}/\Gamma^2(1/4) \approx -0.539$.
In order to reduce the value of $z=xy=qn$, which should be small according to
the definition of $F(y)$, the solution for the CF in the first two layers has
been used.
Calculation of $F(y)$ with the help of (\ref{yscale2}) yields the curves of
solid triangles in Fig. \ref{gn_fy25}.
These curves do not scale with each other, because strong scaling does not hold for $d\geq 2$.
Nevertheless, correlation functions in the asymptotic region $n\gg 1$ differ
from those in the nonscaling region near the surface only by numerical factors,
which are represented in Fig. \ref{gn_cfns}. 
Inserting these factors into $F(y)$ makes the results for $n=1$ and $n=2$ scale.
These corrected results are in excellent accord with the asymptotic formula (\ref{fysc})
for $y\gg 1$.

The longitudinal CF $\sigma_{nn}^{zz}$ itself, which also is shown in  Fig. \ref{gn_fy25},
has the same cusplike form with a gap described by (\ref{sigzzcrit}), as in
low dimensions.
The linear $q$ dependence in the denominator of (\ref{sigzzcrit}) says that,
in spite of the gap, the correlation length of $\sigma_{nn}^{zz}$ near the
surface is infinite at the anisotropic criticality.
Actually, the correlation lengths near the surface are, in the ASM, the same as
in the bulk.
Indeed, above criticality $\sigma_{nn}^{zz}$ is a function of 
$\tilde q_z =\sqrt{\kappa_z^2+q^2}$; thus there are singularities in 
$\sigma_{nn}^{zz}$ at $q= \pm i\kappa_z$, which cause the decrease
of correlations of the type $\exp(-\kappa_z r)$ in the real space at large distances.

\section{Discussion}
\label{disc}

In this paper, a comprehensive analysis of the behavior of the  
semi-infinite anisotropic spherical model at and above the  ordinary phase
transition is presented.
The critical coupling of fluctuations, which usually necessitates application
of the $\epsilon$ expansion or purely numerical methods, dies out in this model due to the infinite
number of spin components and makes in exactly solvable.
On the other hand, the more important {\em qualitative} effects associated with Goldstone
or quasi-Goldstone modes in weakly anisotropic magnetic systems are properly
described by the ASM.
The most important of these effects is the anisotropy-induced ordering in low dimensions.
The ASM is superior with respect to the usual spherical model, which cannot
incorporate anisotropy and yields unphysical results for spacially inhomogeneous systems
because of the global spin constraint.
On the other hand, the ASM is much better defined than its phenomenological
field-theoretical analog, the infinite-component $\phi^4$ model, and it can
always be solved numerically.

Unlike the renormalization-group (RG) approach, which is based on the expansion about the
dimension $d=4$ and becomes inefficient for low dimensions, the ASM describes 
the whole range $1\leq d \leq \infty$ in  a uniform way.
The price for that is the rather complicated character of the ASM system of equations in constrained
geometries, which makes application of numerical methods necessary.
Nevertheless, there are a number of analytical solutions of the semi-infinite ASM in different limiting and particular cases.
The most important of them are the isotropic-criticality solution of Bray and Moore for $2<d<4$,
which was obtained above in an easier and more general way, the variations of the gap parameter $G_{1n}$ for $d\leq 2$ and $d\geq 4$, and the slowly decaying gap tails of the correlation functions for $q \gg \kappa$ away from isotropic criticality.

The gap parameter $G_n$, or its deviation from the bulk value, $G_{1n}$, plays a fundamental role in the theory of the ASM.
The quantity $-G_{1n}$ is similar to the function $V(z)$ used by Bray and
Moore, and it also is
proportional (and at criticality equal) to the inhomogeneity of the energy density, $\tilde U_{1n}$
[see (\ref{util1n})].
The latter has been determined in Ref. \cite{diedie81energy} using RG and scaling arguments with the result 
$\tilde U_{1n} \propto 1/n^{(1-\alpha)/\nu}$ for $2<d<4$ at isotropic
criticality, where $\nu$ and $\alpha$ are the bulk correlation length and the
heat capacity exponents.
In the ASM $\nu = 2/(d-2)$, as follows from (\ref{kapdef}) and (\ref{gtc}),
 and $\alpha=(d-4)/(d-2)$,  as follows from (\ref{utiln}) and (\ref{gtc}).
Thus the above formula reduces to
$G_{1n} \cong \tilde U_{1n} \propto 1/n^2$, as was obtained by Bray and Moore.
In these approaches, which consider $n$ as a continuous variable, the inhomogeneous part of energy shows strong and unphysical divergence at the surface.
 Although it is clear that the continuous approximation is generally invalid near the surface, this
strong singularity does not allow one to avoid the problem by replacing the surface region by some effective boundary condition, as can be done in the MFA.
As a result, a numerical solution is principally ruled out for the semi-infinite field-theoretical $O(\infty)$ model.
In contrast, no such problems arise in the ASM, which is formulated on the lattice from the beginning.
Moreover, continuous dimensionalities (in the directions parallel to the surface) can be introduced into the ASM as well, while preserving the semi-infinite dimension discrete.
The consideration in this paper shows that the nonscaling region near the surface, $n\sim 1$,
plays a very important role in the behavior of the CFs in the asymptotic region far from the surface.
So, the isotropic-criticality CFs are different for, say, $d=2.5$ and $d=3.5$, although they satisfy the same equation in the region $n\gg 1$.
The difference between them stems completely from the region $n\sim 1$.

The universality of the physical quantities in the ASM is different in different dimensionality ranges.
For $d>4$ the gap parameter $G_{1n}$ is nonuniversal and decays as $1/n^{d-2}$, 
although the CFs have the universal mean-field form for $n\gg 1$.
For $2<d<4$ both $G_{1n}$ and CFs are universal for $n\gg 1$ and nonuniversal for $n\sim 1$.
For $d<2$ the values of $G_{1n}$ are universal and decay as $1/n^d$ for $n\gg 1$.
In contrast, the correlation functions are universal in the whole range of $n$.
The reason for this strong universality and the ensuing strong scaling is that
the (transverse ) CFs satisfy the constraint equation containing the 
integral over the wave vector $q$ dominated by $q\sim\kappa\ll 1$ in low dimensions.

The next steps in studying the inhomogeneous magnetic systems with
the help of the ASM should be (i) the solution of the semi-infinite problem below
$T_c$ and in field, (ii) inclusion of surface terms into the Hamiltonian, and
(iii) numerical solution of the model in the film geometry.
The preliminary analytical investigation of the last problem can be found in Ref. \cite{gar96jpal}.

\section*{Acknowledgments}

The work on this problem continued about two years with breaks.
I would like to acknowledge the support of the Deutsche Forschungsgemeinschaft at
the University of Hamburg, which I could enjoy until the middle of this period. 
I would like to thank Hartwig Schmidt for his encouragement and many useful
discussions about this project and other plans.

%\bibliography{gar}

\end{document}